\documentclass{IEEEoj}
\usepackage{cite}
\usepackage{amsmath,amssymb,amsfonts}
\usepackage{algorithmic}
\usepackage{multirow}
\usepackage{graphicx,color}
\usepackage{textcomp}
\usepackage{xurl}
\def\BibTeX{{\rm B\kern-.05em{\sc i\kern-.025em b}\kern-.08em
    T\kern-.1667em\lower.7ex\hbox{E}\kern-.125emX}}
\AtBeginDocument{\definecolor{ojcolor}{cmyk}{0.93,0.59,0.15,0.02}}
\def\OJlogo{\vspace{-4pt}\hskip-4pt\includegraphics[height=18pt]{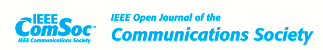}}
\begin{document}
\receiveddate{XX Month, XXXX}
\reviseddate{XX Month, XXXX}
\accepteddate{XX Month, XXXX}
\publisheddate{XX Month, XXXX}
\currentdate{11 January, 2024}
\doiinfo{OJCOMS.2024.011100}

\title{Emerging Technologies for 6G Non-Terrestrial-Networks: From Academia to Industrial Applications}

\author{Cong T. Nguyen\IEEEauthorrefmark{1,2}, Yuris Mulya Saputra\IEEEauthorrefmark{3}, Nguyen Van Huynh\IEEEauthorrefmark{4}, Tan N. Nguyen\IEEEauthorrefmark{5}, Dinh Thai Hoang\IEEEauthorrefmark{6}, Diep N Nguyen\IEEEauthorrefmark{6}, Van-Quan Pham\IEEEauthorrefmark{7}, Miroslav Voznak\IEEEauthorrefmark{8}, Symeon Chatzinotas\IEEEauthorrefmark{9}\IEEEmembership{(Fellow, IEEE)}, and Dinh-Hieu Tran\IEEEauthorrefmark{10}}
\affil{Institute of Fundamental and Applied Sciences, Duy Tan University, Ho Chi Minh City 70000, Vietnam}
\affil{Faculty of Information Technology, Duy Tan University, Da Nang 50000, Vietnam}
\affil{Internet Engineering Technology, Department of Electrical Engineering and Informatics, Vocational College, Universitas Gadjah Mada, Yogyakarta, 55281, Indonesia}
\affil{Department of Electrical Engineering and Electronics, University of Liverpool, Liverpool, L69 3GJ, United Kingdom}
\affil{Communication and Signal Processing Research Group, Faculty of Electrical and Electronics Engineering, Ton Duc Thang University, Ho Chi Minh City, Vietnam}
\affil{School of Electrical and Data Engineering, University of Technology Sydney, Sydney, NSW 2007, Australia}
\affil{Nokia Bell Labs, Murray Hill, NJ 07974, USA}
\affil{Department of Telecommunications, Faculty of Electrical Engineering and Computer Science, VSB-Technical University of Ostrava, 708 00 Ostrava, Czechia}
\affil{Interdisciplinary Centre for Security, Reliability and Trust (SnT), University of Luxembourg, 4365 Esch-sur-Alzette, Luxembourg}
\affil{Nokia, France}
\corresp{CORRESPONDING AUTHOR: Tan N. Nguyen (e-mail: nguyennhattan@tdtu.edu.vn).}
\authornote{This research is funded by the European Union within the REFRESH project - Research Excellence For Region Sustainability and High-tech Industries ID No. CZ$.10.03.01/00/22\_003/0000048$ of the European Just Transition Fund and by the Ministry of Education, Youth and Sports of the Czech Republic (MEYS CZ) through the project SGS ID No. SP 061/2024 conducted by VSB - Technical University of Ostrava.}
\markboth{Preparation of Papers for IEEE OPEN JOURNALS}{Nguyen \textit{et al.}}

\begin{abstract}
Terrestrial networks form the fundamental infrastructure of modern communication systems, serving more than 4 billion users globally. However, terrestrial networks are facing a wide range of challenges, from coverage and reliability to interference and congestion. As the demands of the 6G era are expected to be much higher, it is crucial to address these challenges to ensure a robust and efficient communication infrastructure for the future. To address these problems, Non-terrestrial Network (NTN) has emerged to be a promising solution. NTNs are communication networks that leverage airborne (e.g., unmanned aerial vehicles) and spaceborne vehicles (e.g., satellites) to facilitate ultra-reliable communications and connectivity with high data rates and low latency over expansive regions. This article aims to provide a comprehensive survey on the utilization of network slicing, Artificial Intelligence/Machine Learning (AI/ML), and Open Radio Access Network (ORAN) to address diverse challenges of NTNs from the perspectives of both academia and industry. Particularly, we first provide an in-depth tutorial on NTN and the key enabling technologies including network slicing, AI/ML, and ORAN. Then, we provide a comprehensive survey on how network slicing and AI/ML have been leveraged to overcome the challenges that NTNs are facing. Moreover, we present how ORAN can be utilized for NTNs. Finally, we highlight important challenges, open issues, and future research directions of NTN in the 6G era.
\end{abstract}

\maketitle
\section{INTRODUCTION}
\IEEEPARstart{T}{errestrial} networks, encompassing land-based infrastructures such as fiber optics, coaxial cables, and wireless transmission, are the backbone of modern communication systems, enabling the seamless exchange of information in our interconnected world. However, with the ever-increase of demands, e.g., nearly 5 billion Internet users worldwide~\cite{adinoyi2022future}, terrestrial networks are facing a wide range of challenges, from coverage and reliability to interference and congestion. Particularly, as reported by Ericsson, only 10\% of the global population has access to mobile broadband services~\cite{intro_1}. The main reason for this is due to the difficulties in establishing infrastructure in underserved regions, such as rural areas, islands, and isolated communities. Moreover, the reliability and resilience of terrestrial networks are required to be improved, especially in case of natural disasters, accidents, or attacks~\cite{sec1_feltrin2021potential}. Furthermore, current terrestrial networks are not suitable for new use cases and services that require high bandwidth, low latency, or global coverage~\cite{rinaldi2020non}.

Non-terrestrial Network (NTN), encompassing satellites, Unmanned Aerial Vehicles (UAVs), and High-Altitude Platform Stations (HAPS) networks, is a promising solution to address the challenges that terrestrial networks are facing. For instance, through the utilization of satellites and HAPS, NTN can effectively overcome coverage gaps in terrestrial networks, facilitating broadband connectivity across expansive distances. NTN can also play a vital role in enhancing the resilience of terrestrial infrastructure during crises by utilizing UAVs to provide emergency access \cite{HieuTWC}. Furthermore, by integrating with terrestrial networks, NTN can unlock new possibilities for applications requiring high bandwidth, low latency, and global coverage, including autonomous driving, smart cities, Internet-of-Things (IoT), Augmented Reality, Virtual Reality, cloud gaming, and video conferencing, in the 6G era~\cite{sec1_azari2022evolution,sec1_vaezi2022cellular}. Table~\ref{tab:com2}~\cite{azari2022evolution} summarizes the main differences between NTNs and terrestrial networks.
\begin{table*}[!]
	\caption{Comparison between NTNs and Terrestrial Networks~\cite{azari2022evolution}}
	\centering
	\begin{tabular}{|l|l|l|}
		\hline
		\textbf{Aspect} & \textbf{Non-Terrestrial Networks (NTN)} & \textbf{Terrestrial Networks} \\ \hline
		Coverage Area & Global (up to 3500 km) & Regional (up to 100km) \\ \hline
		Communication Rates & Up to 150 Mbps (Starlink) & More than 100 Mbps on average \\ \hline
		Latency & Higher latency (more than 500ms for GEO satellite) & Lower latency (lower than 40ms ) \\ \hline
		Infrastructure Cost & Very high for satellites & Lower for ground infrastructure \\ \hline
		Reliability & Robust to disasters & Vulnerable to disasters, single points of failure \\ \hline
		Accessibility & Can reach remote/underserved areas & Difficulty in remote area coverage \\ \hline
		Mobility Support & Seamless mobility across wide areas & Frequent handovers for mobility \\ \hline
	\end{tabular}
	\label{tab:com2}
\end{table*}

\begin{table*}[!]
	\caption{Comparison with existing surveys}
	\centering
	\begin{tabular}{|l|l|l|}
		\hline
		\textbf{Ref.}&\textbf{Focus}  &\textbf{Differences}  \\ \hline 
		\cite{sec1_araniti2021toward} & Integration of New Radio in NTN  & Not focus on network slicing, AI/ML, and ORAN    \\ \hline
		\cite{rinaldi2020non} & Role of NTN in 5G systems & Not focus on network slicing, AI/ML, and ORAN   \\ \hline
		\cite{sec1_azari2022evolution} & Integration of NTN with different types of networks & Not focus on network slicing, AI/ML, and ORAN  \\\hline 
		\cite{sec1_vaezi2022cellular} & Utilization of cellular, Wide-Area, and NTN for IoT & Not focus on NTN  \\\hline 
		\cite{sec1_iqbal2023empowering}  & Application of AI/ML in NTN & Not focus on network slicing and ORAN  \\\hline 
		\cite{sec1_wang2019convergence} & Architecture, performance evaluation, and standardization of NTNs & Not focus on network slicing, AI/ML, and ORAN  \\\hline 
		\cite{sec1_zhang2020survey}  & Architecture, performance evaluation, and standardization of NTNs & Not focus on network slicing, AI/ML, and ORAN  \\
		\hline
		\cite{khalid2024artificial}  & Integration of AI/ML for NTN & Not focus on network slicing and ORAN  \\\hline 
	\end{tabular}
	\label{tab:com}
\end{table*}

\begin{figure*}[!]
	\centering
	\includegraphics[width=1.5\columnwidth]{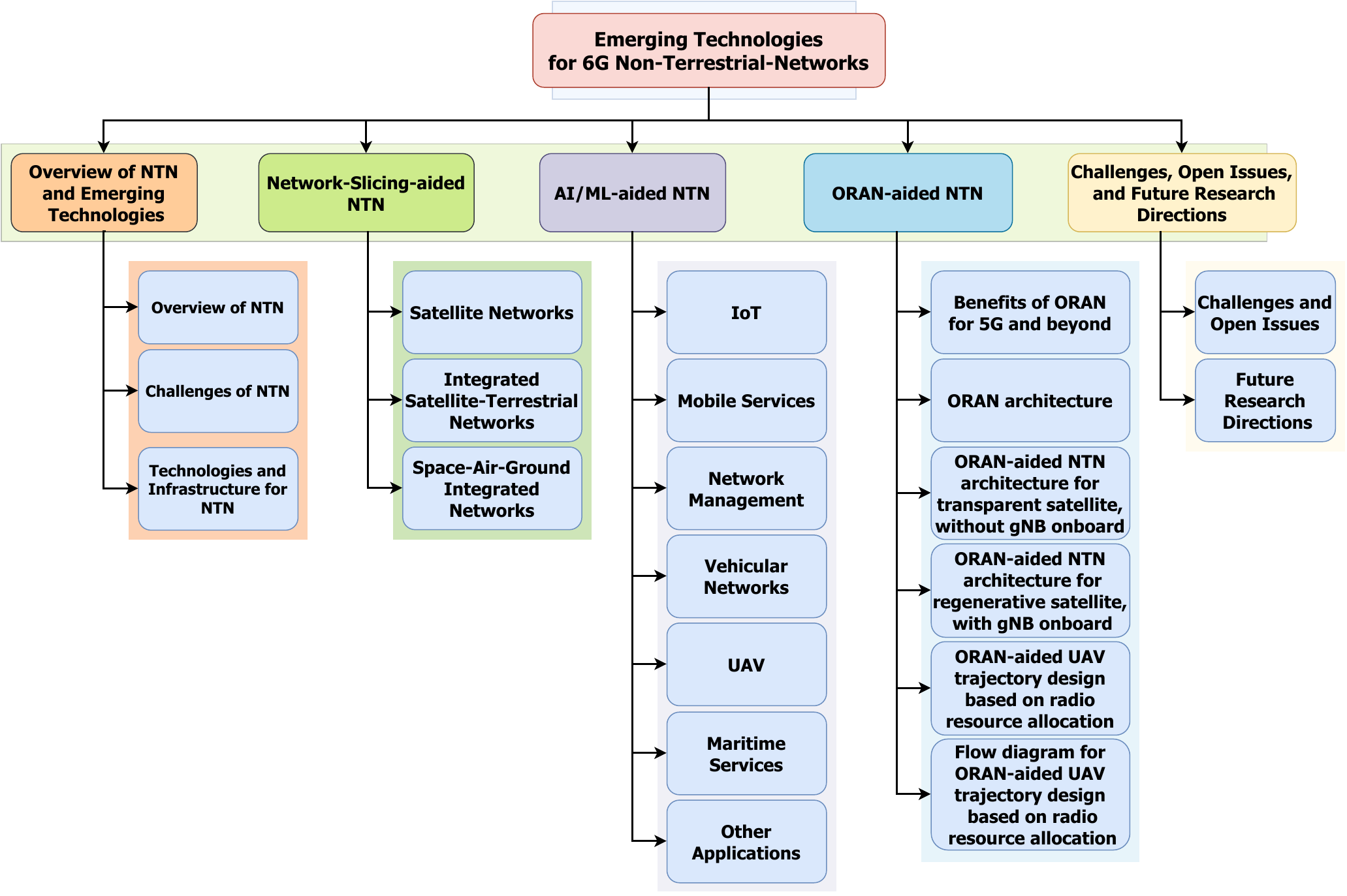}
	\caption{Organization of this paper.}
	\label{orga}
\end{figure*} 

Despite its potential, the development of NTN is also facing serious challenges. Particularly, NTN often faces significant propagation delay and path loss due to high altitudes. This can significantly hinder the effectiveness of NTN in time-sensitive applications. Moreover, the challenging nature of channel estimation is exacerbated by the time-variant characteristics of NTN. The high moving speed of satellites also causes severe Doppler effects and serious challenges in mobility management.~\color{black}Another limitation of NTN is the high initial investment cost compared to that of terrestrial networks, especially for satellites, e.g., more than \$2 million launch cost per satellite~\cite{intro_2}.~\color{black}Additionally, the complex integration of NTN and existing terrestrial network infrastructure necessitates careful design and considerations to optimize. 

To address these challenges, various solutions have been developed in recent years, leveraging technologies such as network slicing, Artificial Intelligence/Machine Learning (AI/ML), and Open Radio Access Network (ORAN). Particularly, with the powerful ability to create multiple virtual networks from a shared physical network architecture, network slicing enables a single NTN to serve multiple types of applications and users with different demands, e.g., low-latency communication for UAVs and high bandwidth for satellite Internet services. Moreover, AI/ML can help NTNs to overcome critical challenges of NTNs. Specifically, AI/ML techniques are very effective in handling non-linear effects and general impairments of channels in NTNs. Those techniques can
also enable the autonomous operation of numerous wireless applications, thereby reducing the need for constant human intervention.  Additionally, by utilizing the current diverse vendors ecosystem, ORAN can help to significantly improve NTNs' resiliency, scalability, and flexibility. 

This paper aims to provide an in-depth and comprehensive survey on the utilization of those technologies to address the diverse challenges of NTNs from the perspectives of both academia and industry. Particularly, we first provide an in-depth tutorial on NTN and the enabling technologies including network slicing, AI/ML, and ORAN. Then, we provide a comprehensive survey on how network slicing and AI/ML have been leveraged to overcome the challenges that NTNs are facing. Moreover, we present how ORAN has been utilized for NTNs from an industry standpoint. Finally, we discuss the current challenges and open issues and introduce potential research directions for NTN in the 6G era.

As summarized in Table~\ref{tab:com}, there are a few surveys on the development of NTN in the literature, such as~\cite{sec1_azari2022evolution,sec1_vaezi2022cellular,rinaldi2020non,sec1_iqbal2023empowering,sec1_wang2019convergence,sec1_araniti2021toward,sec1_zhang2020survey}. Particularly,~\cite{sec1_araniti2021toward} focuses on the integration of New Radio in NTN, whereas~\cite{rinaldi2020non} elaborates on the role of NTN in 5G systems. Moreover,~\cite{sec1_azari2022evolution} surveys the integration of NTN with different types of networks such as IoT, Mobile Edge Computing (MEC), and mmWave. Taking another approach, \cite{sec1_vaezi2022cellular} discusses the utilization of cellular, Wide-Area, and NTN for IoT applications.~\color{black}Additionally, although~\cite{sec1_iqbal2023empowering} and~\cite{khalid2024artificial} provide comprehensive surveys on the application of AI/ML in NTN, network slicing and ORAN are not the focus of these surveys.~\color{black}Taking another approach,~\cite{sec1_wang2019convergence} and \cite{sec1_zhang2020survey} discuss the architecture, performance evaluation, and standardization aspects of NTNs and terrestrial networks integration. To the best of our knowledge, there exist no comprehensive studies on how the abovementioned technologies are employed to address the challenges in NTN, especially from an industry perspective. Given the rapid increase in user demands and the emergence of new services, there is an urgent need for new solutions to address the limitations of NTNs. As a result, this paper is expected to fill the gap in the literature and contribute to the future development of 6G networks.

As illustrated in Fig.~\ref{orga}, the rest of this paper is organized as follows. Section~\ref{sec:tutorial} provides a tutorial on NTN and its core enabling technologies including network slicing, ORAN, and AI/ML. Then, the applications of network slicing and AI/ML in NTN are discussed in detail in Section~\ref{sec:network_slicing} and~\ref{sec:aiml}. Next, the industry applications of ORAN are presented in Section~\ref{sec:oran}. Open issues, challenges, and future research directions are presented in Section~\ref{sec:challenge}, and conclusions are given in Section~\ref{sec:Summary}.


\section{Overview of NTN and Enabling Technologies }\label{sec:tutorial} 
In this section, we first provide an overview of NTN, including its architectures, advantages, and use cases. Then, the challenges in NTNs are discussed thoroughly. Finally, we provide a brief background on advanced technologies and infrastructures that can be adopted to efficiently address these challenges, including network slicing, AI/ML, and ORAN.

\subsection{Overview of NTN}

\begin{figure*}[!]
	\centering
	\includegraphics[scale=0.36]{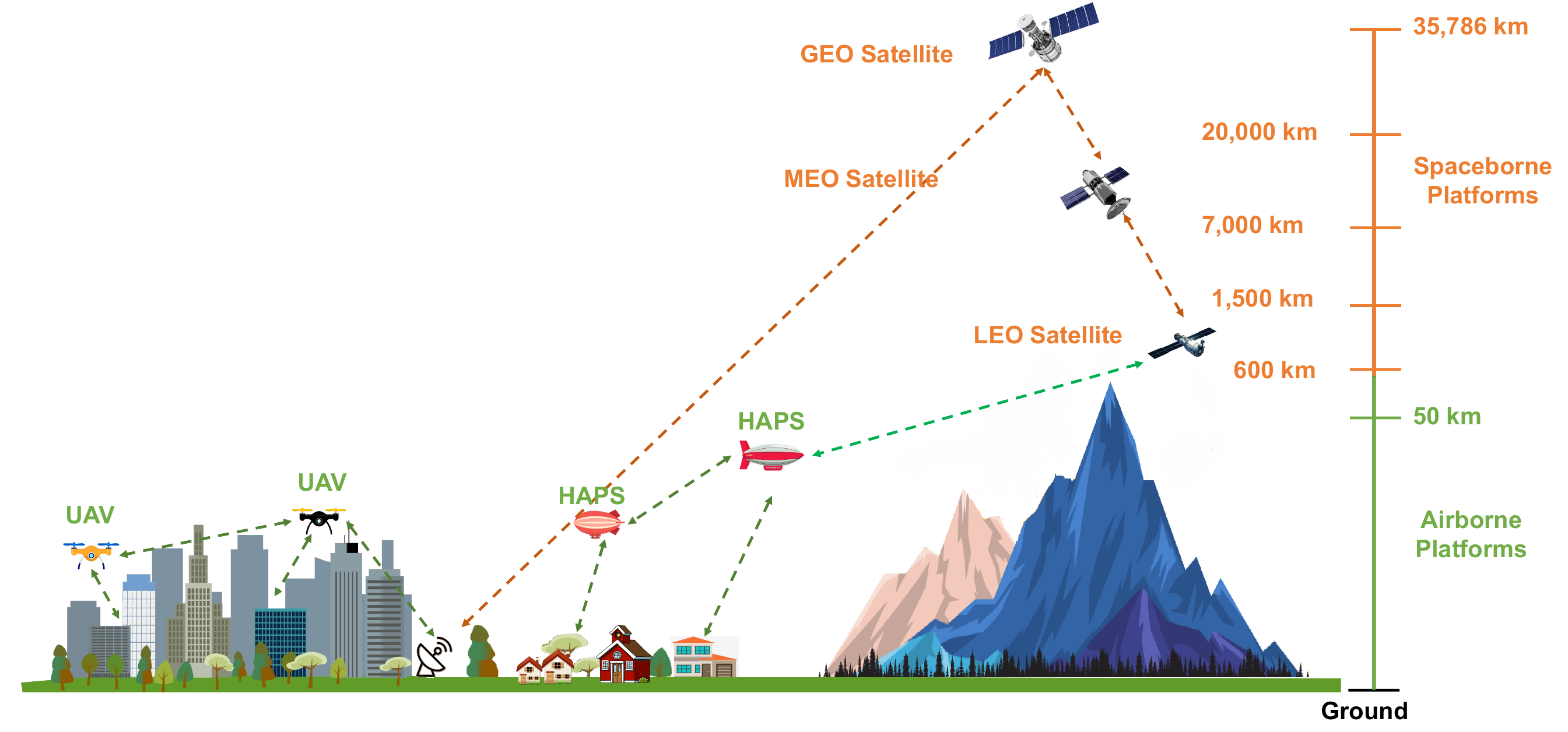}
	\caption{Overview of NTN~\cite{mahboob2023tutorial}.}
	\label{fig_tutorial:overview_NTN}
\end{figure*}

NTNs are communication networks that partially or fully operate through the airborne or spaceborne vehicle(s)~\cite{TanNTN2023,azari2022evolution, HieuTVT2020,mahboob2023tutorial}. With connections from the air or space, NTNs can empower ultra-reliable communications when terrestrial infrastructures are not available, such as in remote and unreachable areas or during natural disasters~\cite{geraci2022integrating}. In addition, NTNs can easily provide multicast connectivity with high data rates and low latency over a large region, enabling massive Machine Type Communication (mMTC) and enhanced Mobile Broadband (eMBB) communications. Moreover, multiple NTNs can be connected together and/or connected to existing terrestrial networks to provide continuous and ubiquitous wireless coverage, thus bringing us closer to the era of anything, anytime, and anywhere communications~\cite{azari2022evolution}.

\textcolor{black}{The development of NTNs can be traced back to the 1990s with the commercialization efforts of the two companies Globalstar and Iridium Communications~\cite{NTN_history}. By deploying several satellites, they can provide low-bandwidth connections to specialized handsets. Since then, NTN technologies have been studied extensively by both academia and industry with a vision of creating a network of satellites and providing Internet services to users anywhere on Earth. The potential features, requirements, and protocols for NTNs have been standardized by 3GPP since 2017 with a study item on deployment scenarios and channel models in Release-15~\cite{geraci2022integrating}. After that, the required features to enable New Radio (NR) support for NTN are determined by a study in the Radio Access Network (RAN) working group of Release-16. This is then developed further in Release-17 with a set of potential features to enable 5G NR to operate over NTN at frequencies up to 7.125 GHz~\cite{geraci2022integrating}. Potential use cases of IoT over NTN with link budget and parameters are also discussed in this Release-17. Release-18 is under development and will focus on 5G systems with satellite backhaul architecture as well as addressing the mobility and continuity problems when connecting different NTNs and also connecting NTNs to terrestrial networks~\cite{geraci2022integrating}.}

NTN has been attracting great attention from both industry and academia recently due to its potential to complement and enhance existing terrestrial networks in terms of coverage, capacity, and mobility. It is expected to play an essential role in the development of 5G-advanced and 6G networks. For that, big tech companies like SpaceX, OneWeb, Amazon Kuiper, and SoftBank are investing billions of dollars in this field to provide global connectivity and expand their business models~\cite{starlink, kuiper}. In general, NTNs can be categorized into airborne and spaceborne platforms as illustrated in Fig.~\ref{fig_tutorial:overview_NTN}.

\subsubsection{Airborne Platforms}

This category includes UAVs and HAPSs \cite{TranTVTbackscatter}. While UAVs operate at low altitudes, (e.g., a few hundred meters), HAPSs, such as airplanes, balloons, and airships, can reach the stratosphere region with altitudes ranging from 20 kilometers to 200 kilometers~\cite{mahboob2023tutorial}. One example of airborne NTN is Project Loon~\cite{Lardinois_2013, NZHerald_2020} from Google which deploys balloons at high altitudes from 18 kilometers to 25 kilometers to provide connectivity to remote and rural areas. Facebook is also involved in airborne NTN by developing solar-powered drones that operate at an altitude of up to 27 kilometers and provide Internet services to an 80-kilometer-radius area below its flying path through Project Aquila~\cite{CNNMoney}.

Airborne NTN can be deployed quickly at a lower cost and has a much smaller propagation delay compared to spaceborne NTN. However, airborne platforms face critical challenges of stabilization on air and refueling. In particular, operating at low altitudes makes them vulnerable to environmental conditions such as strong winds or storms that can change the flying path of balloons and drones or even destroy them. It is also difficult and inefficient to refuel airborne vehicles while maintaining connections for users. These challenges could be the reason for the termination of projects Aquila and Loon in 2018 and 2021, respectively. While airborne platforms may not be effectively used to provide Internet services, they can be used in emergency scenarios such as during natural disasters and rescue missions in rural areas where cellular coverage is not available.

\subsubsection{Spaceborne Platforms}
Recently, spaceborne NTN has been emerging as a promising technology for future communication networks (e.g., 5G-Advanced and 6G) where satellites are placed in space to provide communications for users on Earth. These satellites fly around the Earth in specific orbits and can be categorized into: (i) Geostationary Earth Orbit and (ii) Non-Geostationary Earth Orbit~\cite{TanSatLetter,TanSatSystem,mahboob2023tutorial,HieuOJVT2022}.

\paragraph{Geostationary Earth Orbit (GEO)}
Satellites operating in GEO (i.e., GEO satellites) fly around the Earth above the equator from west to east following the Earth's rotation. GEO satellites travel at the same rate as the Earth and take 23 hours 56 minutes and 4 seconds to complete one orbital period. As a result, these satellites are ``stationary'', i.e., appear motionless, at a fixed position in the sky to observers on Earth~\cite{type_orbits}. To exactly match the Earth's rotation, GEO satellites must travel at the speed of about 3 kilometers per second at an altitude of 35,786 kilometers. Due to the high altitude, GEO satellites can cover a large portion of the Earth's surface. Theoretically, three GEO satellites can provide near-global coverage. The beam footprint of GEO satellites can range from 200 kilometers to 3,500 kilometers~\cite{azari2022evolution}. There are currently hundreds of GEO satellites in the orbit~\cite{list_satellites}, and most of them are used to provide services such as weather monitoring, TV broadcasting as well as remote sensing and positioning. GEO satellites can also be used for communication services. However, due to the high altitude, the communication latency is significantly high (around 600 milliseconds according to Starlink~\cite{starlink}). To provide low latency connections, satellites operating at low orbits have been gaining great attention recently.

\paragraph{Non-Geostationary Earth Orbit (NGEO)}
Different from GEO satellites, NGEO satellites operate at lower altitudes with orbital periods of less than 24 hours. In addition, their positions can be always changed with respect to observers on Earth. There are two types of NGEO satellites according to their altitudes: (I) Medium Earth Orbit (MEO) satellites and (ii) Low Earth Orbit (LEO) satellites.

\begin{itemize}
	\item \textit{MEO satellites} usually operate at 2,000 kilometers to 25,000 kilometers above the Earth's surface~\cite{mahboob2023tutorial}. Flying at these altitudes allows MEO satellites to create beams with diameters from 100 kilometers to 500 kilometers. 
	
	\item \textit{LEO satellites} can be deployed at altitudes from 200 kilometers to 2,000 kilometers~\cite{mahboob2023tutorial} which is lower than other orbits but still very far from the Earth's surface. LEO satellites can create beam footprints with diameters from 5 kilometers to 200 kilometers.
\end{itemize}

MEO and LEO satellites can be used for sensing, positioning, and communication systems due to their low propagation delay compared to GEO satellites. For example, the US Global Positioning System (GPS) uses at least 24 MEO satellites for its global positioning services. As of June 26, 2022, the GPS constellation consists of 31 operational satellites~\cite{gps}. Although MEO satellites can be used for communication services, the communication latency is still high and may not be feasible for today and future communication applications. Recently, LEO satellites have been emerging as promising platforms to provide Internet services to users on Earth anywhere and anytime. For example, Starlink, the world's first and largest constellation, uses thousands of LEO satellites to deliver broadband Internet for services such as streaming, online gaming, and video calls to users around the world~\cite{starlink}. Starlink's satellites fly at an altitude of about 550 kilometers with much lower propagation delay compared to those of GEO and MEO satellites. As a result, Starlink can achieve a roundtrip delay of around 25 milliseconds for its services~\cite{starlink}. The current biggest competitor of Starlink is Amazon's Project Kuiper which aims to launch over 3,236 LEO satellites to provide low latency and high-speed Internet services to users on a global scale. In addition, this satellite system is also integrated into resilient communication infrastructure powered by a global network of ground stations and Amazon Web Services for better services~\cite{kuiper}.

\subsection{Challenges of NTN}

Although providing various promising applications and use cases, NTN faces several technical challenges that need to be fully addressed to ensure its success in future communication networks.

\begin{itemize}
	\item \textit{Propagation delay and path loss:} NTN platforms face a particular challenge of propagation and path loss due to their high altitudes. For example, GEO satellites experience a round-trip latency of around 600 milliseconds~\cite{starlink} which is considered significantly high. This may not be feasible for communications that require low or ultra-low latency such as online gaming, video streaming, and VR/AR. Compared to terrestrial communications, the propagation path losses of NTN platforms are much higher due to long distances to UEs. In addition, NTN platforms at high altitudes can cover large areas to serve a massive number of users with different propagation delays and path losses in different regions~\cite{azari2022evolution}. Consequently, ensuring good communications for all users is very challenging. Moreover, managing initial accesses and synchronizations for these diverse users also poses another challenge to NTNs.

	\item \textit{Channel estimation:} Channel estimation is an essential task in every communication system. For NTNs, it is even more challenging due to the inherent time-variant property of NTN platforms. In particular, at high altitudes, NTN platforms fly from horizon to horizon very fast (e.g., around 5-10 minutes for LEO satellites~\cite{mahboob2023tutorial}). As a result, users on Earth remain in the coverage of a particular NTN platform for a very short period. In addition, the long propagation delay may make estimated channel state information outdated quickly~\cite{3gpp_38821}. Hence, traditional estimation methods in terrestrial networks may not be feasible for NTNs, and advanced approaches are required to ensure the success of NTNs.

	\item \textit{Doppler effect:} Due to their high movement speeds, NTN platforms introduce significant Doppler effects on communication links between them and users on Earth. In particular, the Doppler effect is the shift in the frequency of signals during the relative motion between transceivers. The Doppler effect also happens in terrestrial networks, e.g., users on high-speed trains or cars. However, in NTNs, this effect is more serious as NTN platforms fly at very high speeds. For example, a user communicating with an LEO satellite operating at 600 kilometers above the ground may experience a Doppler shift of up to 48 kHz given the carrier frequency of 2 GHz~\cite{lin20215g}. This Doppler shift is significantly larger compared to those of users in terrestrial networks.

	\item \textit{Mobility management:} Another challenge in NTNs is mobility management due to the high speeds of NTN platforms. For example, satellites operating at NGEO have short orbital periods (around 2-10 hours~\cite{mahboob2023tutorial}). Consequently, users on the ground can only observe a particular NGEO satellite over a very short period, typically several minutes~\cite{mahboob2023tutorial}. In this case, the ground users will need to perform handovers frequently, especially when these satellites use multiple beams to cover an area on Earth.

	\item \textit{Resource management:} Compared to terrestrial terminals, NTN platforms need to transmit signals with much higher power to deal with the high path loss and ensure users on Earth can successfully decode the transmitted signals. This introduces a new challenge for NTN platforms as they are not equipped with stable power sources like terrestrial terminals. Moreover, the frequency bands assigned to NTN communications are limited and already crowded. In particular, the S-band and Ka-band are the target bands for NTNs~\cite{mahboob2023tutorial}. However, 4G LTE devices are using the S-band, and millimeter wave-enabled devices in 5G are using the Ka-band. As a result, users in NTNs may experience co-channel interference from these terrestrial devices. This demands novel spectrum-sharing solutions to intelligently and efficiently utilize the limited frequency bands.
\end{itemize} 
\subsection{Technologies and Infrastructures for NTN}
\textcolor{black}{Solutions to address the aforementioned challenges of NTNs have been actively developing in the past few years, including beamforming designs, efficient resource allocation, dynamic routing, and intelligent operation management~\cite{zhang2019robust, niephaus2016qos}, by leveraging advanced technologies in AI/ML, communications and networking, and computing. Among them, network slicing, ORAN, and AI/ML are promising technologies that are expected to play vital roles in NTNs.} In this section, we will provide the fundamentals and advantages of these technologies for NTNs.

\subsubsection{Network Slicing}
\begin{figure}[h]
	\centering
	\includegraphics[scale=0.34]{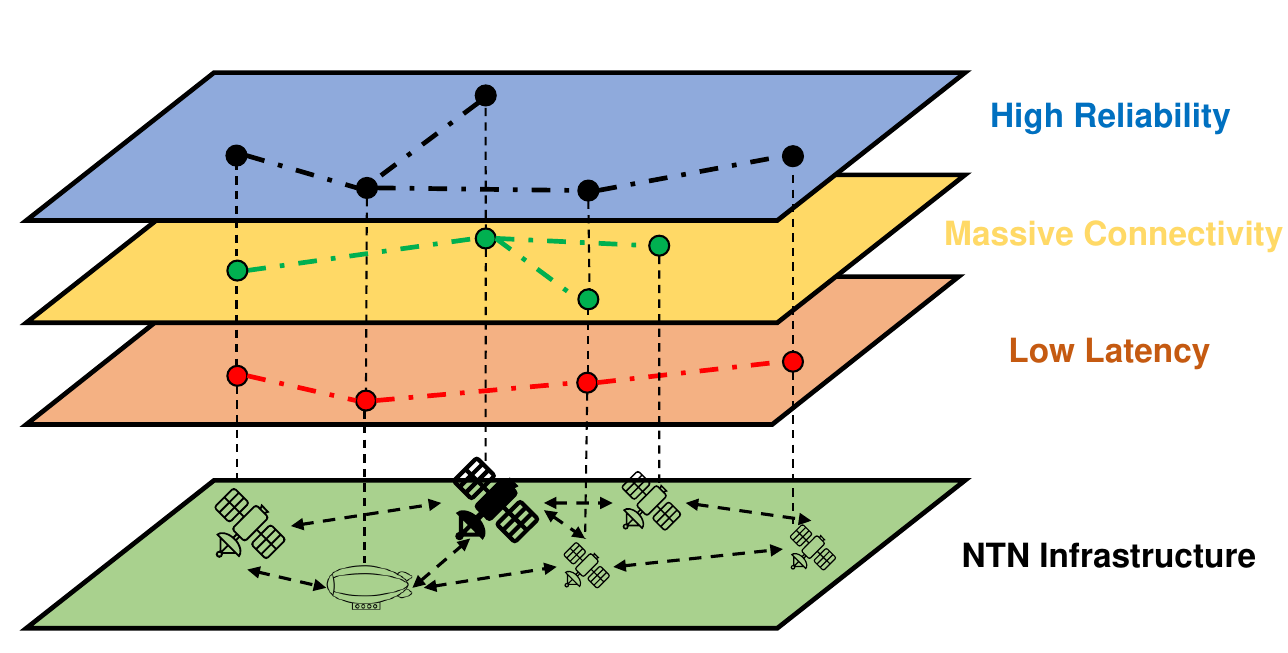}
	\caption{General architecture of network slicing-powered NTN~\cite{drif2021slice, van2019optimal}.}
	\label{fig_tutorial:networkslicing_NTN}
\end{figure}

Network slicing is a new technology that allows us to create multiple unique logical and virtualized networks, called slices, on top of a physical infrastructure~\cite{foukas2017network,afolabi2018network, van2019optimal}. With network slicing, service providers can quickly serve diverse services according to their requirements such as resources, quality of service (QoS) demands, and network functionalities, simultaneously. To do that, network slicing deploys various technologies such as Software-Defined Networking (SDN), Network Function Virtualization (NFV), and Network Orchestration. In particular, SDN is an approach to decouple the control plane from the data plane of network devices and make the control plane programmable~\cite{xia2014survey}. In this way, SDN can dynamically perform network configurations through the centralized controller, resulting in better network performance compared to conventional networking. NFV is a novel technology to virtualize network services such as firewalls, routers, and load balancers by using virtualization technologies and commercial off-the-shelf programmable hardware~\cite{han2015network}. In such a way, NFV brings various benefits to network operators such as decoupling software from hardware, flexible network function deployment, and dynamic scaling~\cite{mijumbi2015network}. Finally, network orchestration refers to the automated rules to dynamically control and automatically program the network, allowing it to ensure the service level agreements for the services. With these advanced technologies, network slicing is expected to play an essential role in 5G and beyond due to its advantages as follows~\cite{afolabi2018network,wijethilaka2021survey}:

\begin{itemize}
	\item \textit{Enabling Scalability and Flexibility:} In practice, network services may require different amounts of resources and network functions at different times. To better utilize the system's resources, network slicing allows network operators to dynamically allocate resources from a particular slice to other slices that demand higher resource requirements without affecting services in the source slice. Network slicing also can quickly create a slice for new services when requested, e.g., critical applications, and adjust the allocated resources and network functions based on users' demands.
	
	\item \textit{Ensuring Security and Privacy:} Although using the same physical infrastructure, slices in network slicing networks are independent. As such, security attacks in one slice cannot affect other slices in the system. This is called slice isolation, an important design principle of network slicing.
	
	\item \textit{Improving QoS:} Network slicing can allocate different types of resources and network functions to diverse services based on their QoS requirements. In addition, during congestion situations, network slicing can still guarantee the QoS for users by dynamically allocating resources and prioritizing traffic between slices in the network.
\end{itemize}

\textcolor{black}{With the aforementioned advantages, network slicing has emerged as a potential enabler that can effectively address various challenges of NTNs. A general network slicing-powered NTN system is illustrated in Fig.~\ref{fig_tutorial:networkslicing_NTN}. In particular, like traditional network components, NTN platforms can also be virtualized both in terms of resources and network functions by using SDN and NFV to support different types of services simultaneously. With network slicing, service providers can quickly establish new slices for NTN services such as Earth observation, broadcast services, and broadband services to serve different groups of users~\cite{drif2021extensible}. This can efficiently address the problem of high cost and high delay in building satellite constellations in orbits. In addition, network slicing also allows NTNs to be effectively and flexibly integrated into existing terrestrial networks. Nevertheless, network slicing-powered NTN systems still face several challenges in designs, protocols, and optimizations. In particular, the high mobility of satellites can create frequent topology changes and handovers that may not be fully addressed by using existing network slicing technologies. In addition, satellite communications have higher latency and energy consumption compared to traditional communications. As such, managing and allocating different types of virtual resources to maintain efficient network slicing operations are more complex in NTNs than in terrestrial networks. These challenges and existing approaches in the literature will be discussed in detail in Section~\ref{sec:network_slicing}.}

\subsubsection{AI/ML}
AI/ML has been developing significantly since the deep neural network (DNN) architecture was re-invented and trained over a large amount of data by powerful computational computers. It has been successfully applied to various areas such as computer vision, gaming, and natural language processing. In the fields of communications and networking, AI/ML has been emerging as a promising solution to efficiently and significantly improve communication performance. AI/ML is a potential technology for NTNs to overcome their critical challenges that conventional approaches cannot handle well. For instance, Deep Reinforcement Learning (DRL), an advanced AI/ML algorithm, can be used to address resource allocation, routing, handover, and beamforming problems in NTNs~\cite{deng2019next,wu2021learning, zhang2020towards} due to its capability in dealing with the dynamics and uncertainty of the system. Deep learning (DL) with advanced neural network architectures such as Convolutional Neural Network (CNN) and Recurrent Neural Network (RNN) is very effective in handling non-linear effects and general impairments of channels in NTNs. In this section, we will present the fundamentals of common AI/ML techniques that can be used to improve the performance of NTNs. The details of existing AI/ML applications for NTNs will be presented in Section~\ref{sec:aiml}.

AI refers to a field of computer science that focuses on building machines (especially computer systems) to perform tasks that require human intelligence such as learning, problem-solving, decision-making, and reasoning. ML is a special subset of AI in which a machine tries to learn a specific task (e.g., image classification, voice recognition, or signal classification and resource allocation in communication networks) and performance metrics such as classification accuracy and performance loss by using only the data collected from the task. ML can be generally categorized into three subsets: (i) supervised learning, (ii) unsupervised learning, and (iii) Reinforcement Learning (RL)~\cite{bonaccorso2017machine}.

\begin{itemize}
	\item \textit{Supervised learning:} In this type of ML, the learning model is trained with labeled datasets to classify data or predict outcomes accurately. For example, an ML model can be trained with a dataset of images of different animals that are labeled by humans. Over time, the model can learn ways to identify these animals. Supervised learning is the most common type of ML used today. The standard supervised learning techniques include Naive Bayes, Liner Regression, Decision Tree, Support Vector Machine, and Logistic Regression~\cite{bonaccorso2017machine}.
	
	\item \textit{Unsupervised learning:} Different from supervised learning, unsupervised learning models are trained on unlabeled data. To do that, unsupervised learning algorithms scan through training data to identify patterns or trends without human intervention. For example, unsupervised learning can be used to learn from online sales data to look for different types of clients making purchases. The standard techniques in unsupervised learning include K-means clustering, Self-Organization Map, and Principal Component Analysis~\cite{bonaccorso2017machine}.
	
	\item \textit{Reinforcement learning:} This type of ML does not require a prior dataset for training. In particular, RL is trained through trial and error by interacting with an external environment. Given a particular state, the RL agent makes an action based on its current policy. After that, it observes the next system state and the immediate reward calculated by a predefined reward function. All these observations will be learned by the agent to gradually obtain the optimal policy. The standard techniques in RL include Q-learning, deep Q-learning, and multi-armed bandit~\cite{bonaccorso2017machine, goodfellow2016deep, hoang2023deep,hou2022environment}.
	
\end{itemize}

\textcolor{black}{To further improve the prediction performance of ML models, DL has been proposed by leveraging the capabilities of DNNs in learning from a large amount of data.} Compared to conventional ML algorithms, DL has several advantages such as no need for system modeling, supporting parallel and distributed algorithms, and being reusable. A typical DNN consists of four main components: (i) neurons, (ii) weights, (iii) biases, and (iv) activation functions. In particular, layers of a DNN are connected to each other by neurons, also known as nodes. Each neuron has an activation function such as tanh, sigmoid, and relu~\cite{goodfellow2016deep}. The activation function is used to calculate the output of each neuron given its weight and bias. During training, the weights of the DNN are updated by calculating the gradient of the loss function. There are three main types of DNNs including (i) Artificial Neural Network (ANN), (ii) Recurrent Neural Network (RNN), and (iii) Convolutional Neural Network (CNN).
\begin{itemize}
	\item \textit{Artificial Neural Network:} ANN, also known as feed-forward neural network, is the most common type of DNNs. Typically, an ANN consists of nonlinear processing layers, including an input layer, several fully connected hidden layers, and an output layer as shown in Fig.~\ref{fig_tutorial:ANN}. As a hidden layer takes the outputs of its previous interconnected layer as its inputs, ANN processes information in one direction from the input layer through the hidden layer to the output layer. Generally, ANN can work well with nonlinear functions, and thus it can be considered a universal function approximation. \textcolor{black}{Due to its simple architecture and ability to extract useful information from training data, ANN has been commonly adopted to address emerging issues in communications and networking. For example, the authors in~\cite{ye2017power} propose to use an ANN architecture with only three hidden layers for channel estimation and signal detection in orthogonal frequency-division multiplexing (OFDM) systems. Simulation and experimental results then reveal that ANN is a promising tool for channel estimation and signal detection in wireless environments under complex channel distortion and interference. Moreover, in~\cite{chen2019artificial}, the authors present several use cases of applying ANNs to different problems in wireless communications such as UAV-based wireless networks, radio access, and mobile edge caching and computing.}
	\begin{figure}[h]
		\centering
		\includegraphics[scale=0.25]{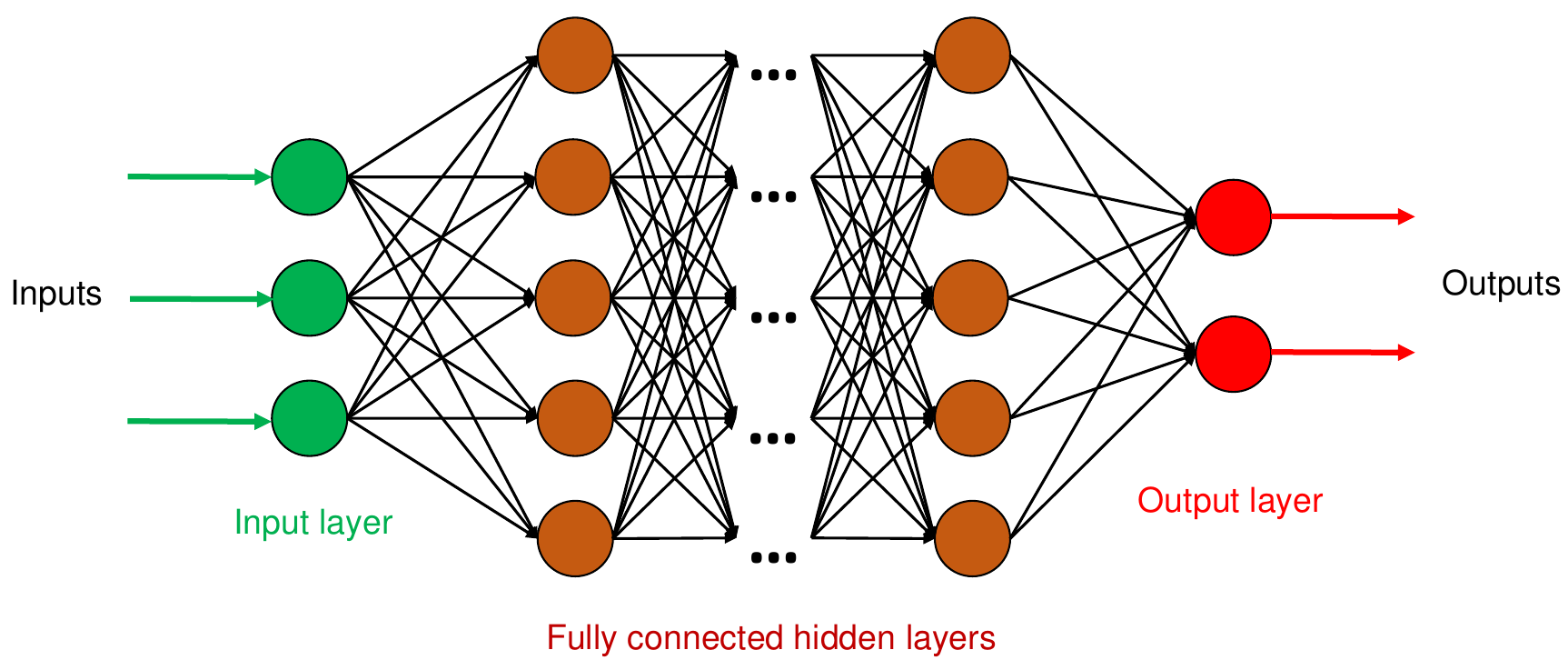}
		\caption{General architecture of ANN.}
		\label{fig_tutorial:ANN}
	\end{figure}
	\item \textit{Recurrent Neural Network:} RNN is a special architecture of DNNs that is widely used for time series data or data that involves sequences. To do that, RNN deploys a feedback loop and hidden states to store information of previous inputs to improve the learning processes of the next data, as illustrated in Fig.~\ref{fig_tutorial:RNN}. In particular, the output of the RNN cell at time $t-1$ will be stored in the hidden state $h_t$. This stored information will be used for learning the next sequence at time $t$. \textcolor{black}{In complex systems, RNN may not perform well due to the ``vanishing'' or ``exploding'' gradient problem during the backpropagation operation. To overcome this issue, an extended version of RNN, namely Long Short-Term Memory (LSTM), is proposed. In particular, LSTM employs additional gates to determine the amount of previous information in the hidden state that will be used for the output and the next hidden state. In this way, LSTM can efficiently learn the long-term dependencies in training data while mitigating the issue of the backpropagation process. RNN, especially LSTM, has emerged as a promising architecture for signal classification in wireless communications due to the fact that signals are naturally sequential and collected over multiple antennas~\cite{zhang2021signal, van2022defeating}. In addition, LSTM can be used for resource allocation, modulation classification, intrusion detection, and beamforming~\cite{gupta2021lstm, xu2021generative, diro2018leveraging}.}
	\begin{figure}[h]
		\centering
		\includegraphics[scale=0.25]{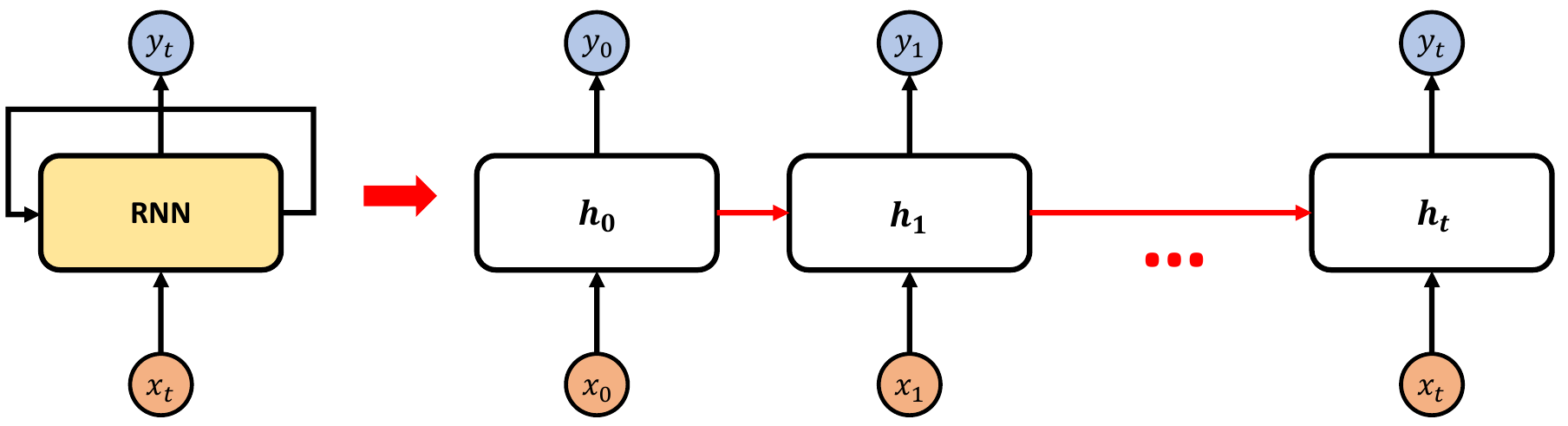}
		\caption{General architecture of RNN.}
		\label{fig_tutorial:RNN}
	\end{figure}
	\item \textit{Convolutional Neural Network:} CNN is designed mainly for training over image data. The general architecture of a CNN is illustrated in Fig.~\ref{fig_tutorial:CNN}. Specifically, a CNN consists of convolution layers that have a set of convolutional filters. Each convolutional filter extracts specific features from image data. After each convolution operation, CNN uses a Rectified Linear Unit (ReLU) transformation to maintain positive values during training. This helps the training process to be faster and more effective. The pooling layer, also known as downsampling, is used to reduce the number of training parameters by performing dimensionality reduction. It has been widely demonstrated that CNN can handle image data much more effectively than ANN. This is because CNN does not need to convert images to 1-dimensional data before training, which increases the number of training features as well as removes the correlations of features in images. CNN, in contrast, can learn these features directly from image data by using convolutional layers. \textcolor{black}{As a result, CNN has been widely applied in communications and networking to handle data in the form of images or high-dimensional matrices. For example, the authors in~\cite{liu2019deep} propose to use CNN for spectrum sensing in cognitive radio networks. In addition, CNN can be used for automatic modulation classification as demonstrated in~\cite{hermawan2020cnn} by taking IQ time-domain vectors of modulated signals as its inputs. Due to the ability to handle high-dimensional input data, CNN is a promising architecture for channel feature extraction as studied in~\cite{li2017wireless}.}
	\begin{figure}[h]
		\centering
		\includegraphics[scale=0.25]{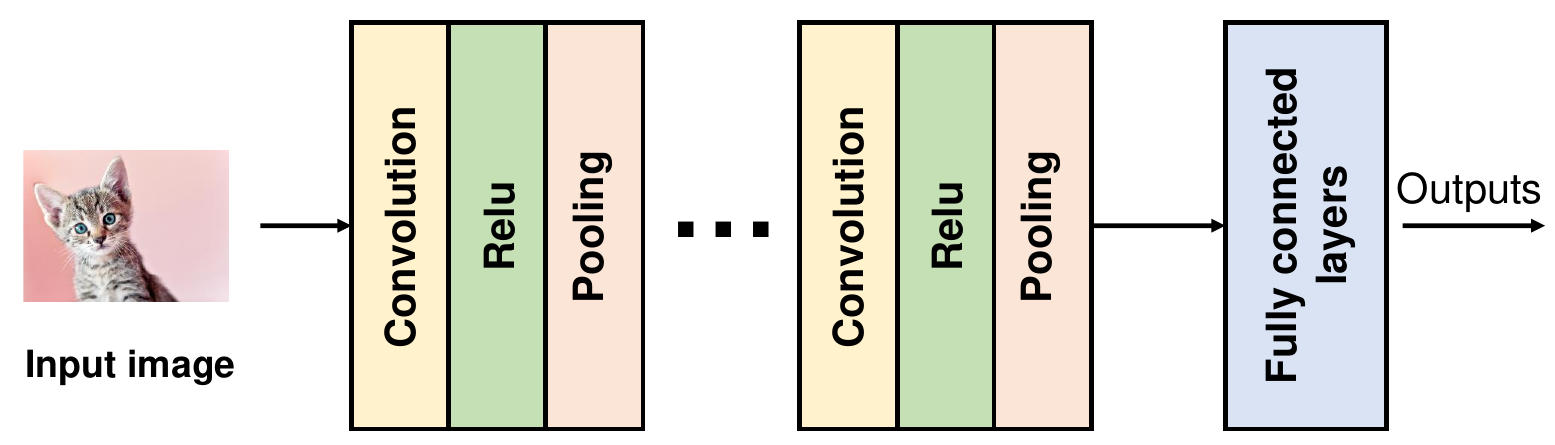}
		\caption{General architecture of CNN.}
		\label{fig_tutorial:CNN}
	\end{figure}
\end{itemize}

\textcolor{black}{As mentioned, DL approaches usually require sufficient training data from users to achieve good performance. However, in practice, users may not be willing to share their data with the centralized server due to privacy and security concerns. To address this issue, federated learning (FL) was introduced in~\cite{mcmahan2016federated}. In FL, each user, i.e., client, uses their local data to train a DL model. After that, they send their model updates, i.e., their models' weights, to a centralized server for aggregation. The server then sends the aggregated global model to all the clients. Based on this global model, the clients then continue to train it with their local data. This process is repeated until a desirable accuracy is obtained. In this way, FL can effectively address the privacy problem as well as reduce the required bandwidth as only models' weight is transmitted to the server instead of raw training data. Due to these advantages, FL has been widely adopted in wireless communications and networking to address a wide range of problems such as spectrum management, caching, and IoT~\cite{niknam2020federated,hou2023efficient}. Applications of FL in NTNs will be also discussed in detail in Section~\ref{sec:aiml}.}

\subsubsection{ORAN}

ORAN is the disaggregation of the traditional RAN, allowing cellular equipment provided by different vendors can be interoperated by using open and standards-based protocols~\cite{Cisco_2021, openran_uk}. In the following, we first provide the fundamentals of RAN and then discuss the architecture and advantages of ORAN over RAN.

\begin{figure}[h]
	\centering
	\includegraphics[scale=0.27]{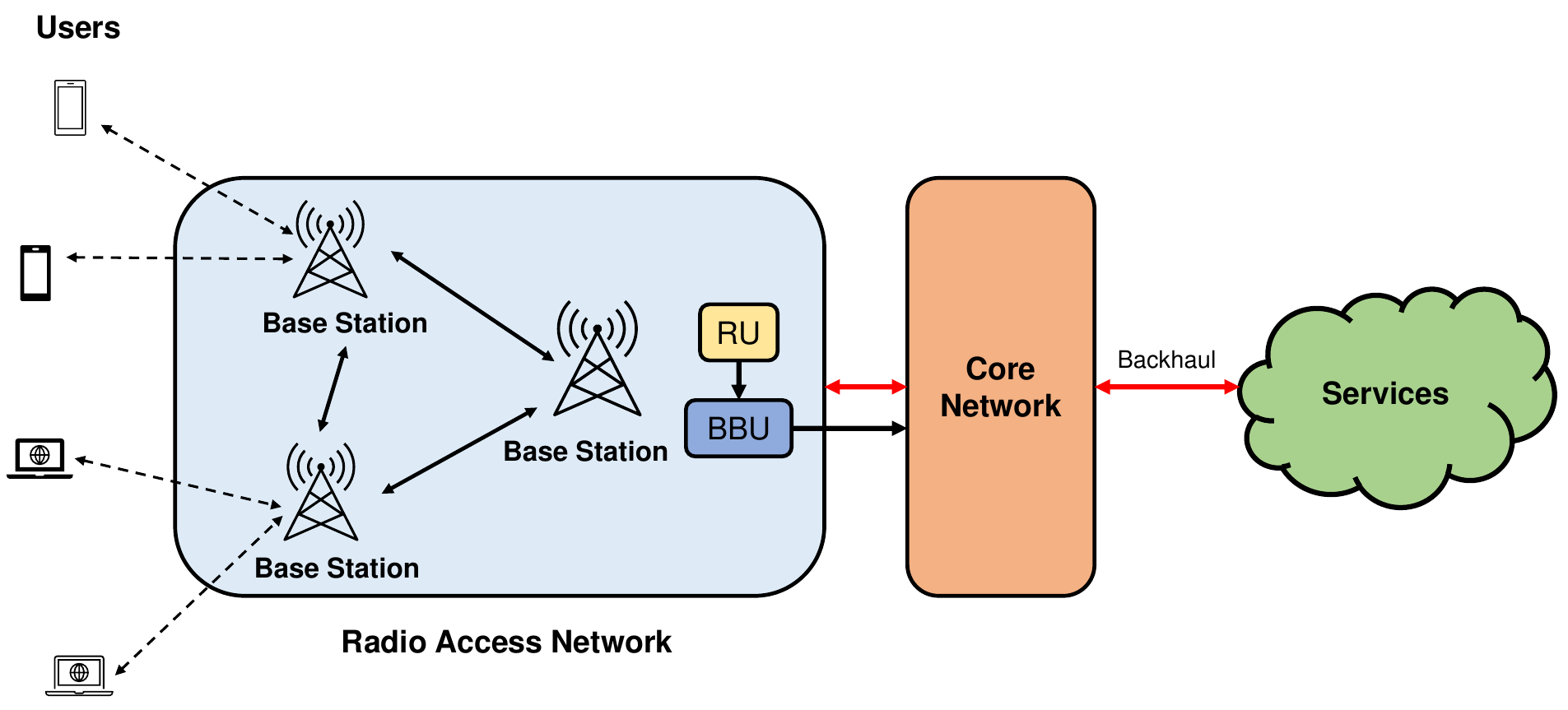}
	\caption{General architecture of mobile networks.}
	\label{fig_tutorial:RAN}
\end{figure}

Fig.~\ref{fig_tutorial:RAN} illustrates a general architecture of mobile networks, including RAN, core network, and services. In particular, the purpose of RAN is to connect user devices (e.g., mobile phones and computers) to the core network to access services provided by network operators. The main component of RAN is base stations. In general, a base station consists of two main units: (i) radio unit (RU) and (ii) baseband unit (BBU). RU receives signals from users and sends them to BBU for processing before transmitting them to the core network. Traditionally, a single vendor provides all the units, software, and connections between them of the base station. As a result, it is difficult if not impossible to change any component of RAN, making it costly and less flexible in deployment and operation. To address all these drawbacks of conventional RAN, ORAN was proposed recently with a more flexible design.

\begin{figure}[h]
	\centering
	\includegraphics[scale=0.25]{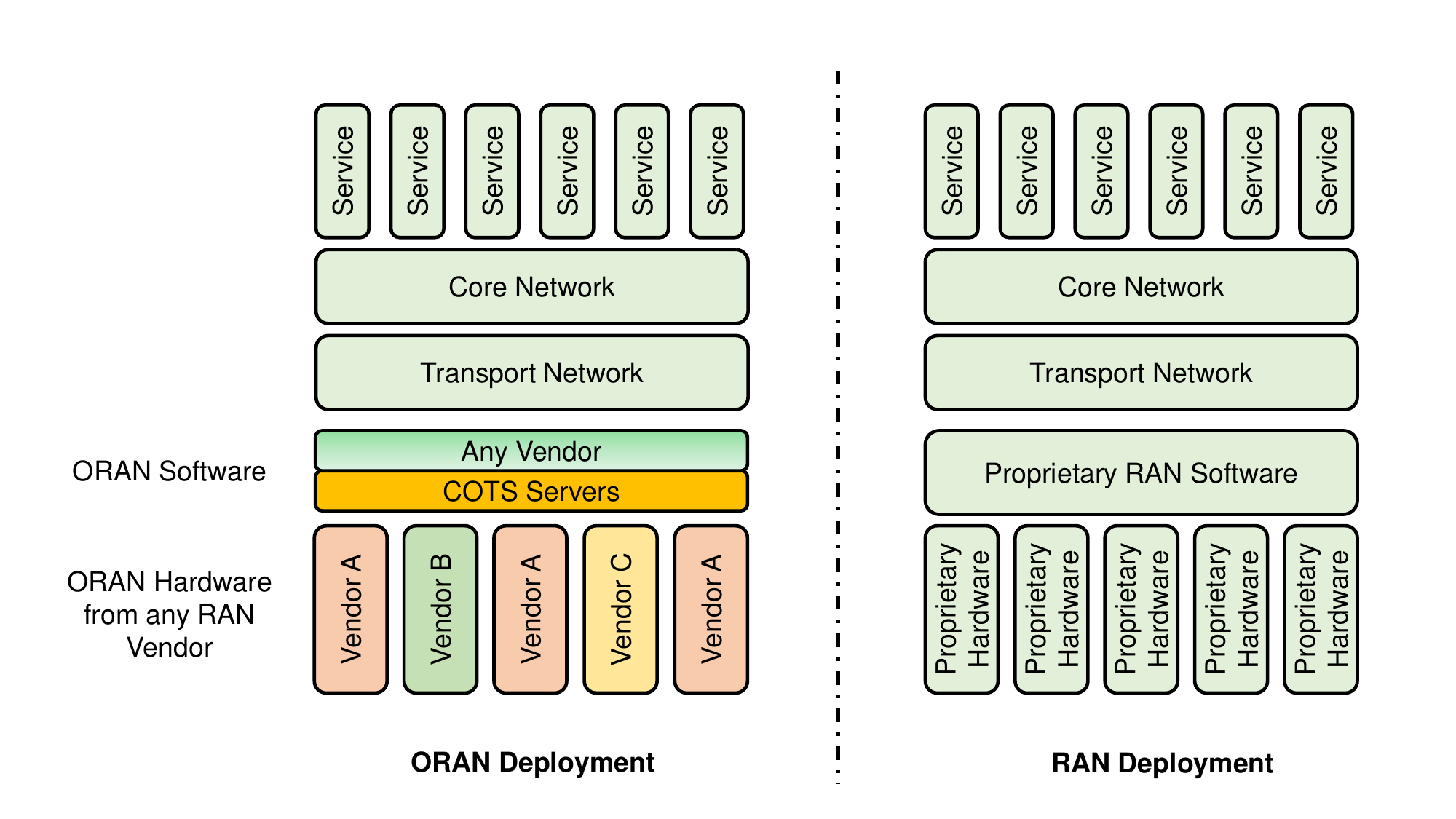}
	\caption{RAN vs ORAN~\cite{ranvsopenran}.}
	\label{fig_tutorial:RANvsORAN}
\end{figure}

In Fig.~\ref{fig_tutorial:RANvsORAN}, we illustrate the differences in the designs of RAN and ORAN. In particular, ORAN allows the radio access network to work with hardware from any vendor while traditional RAN requires a specific vendor for its hardware. Similarly, in the software part, software from any vendor can be used with a commercial off-the-shelf server, whereas RAN requires proprietary RAN software from a specific vendor. With this open design, ORAN offers various advantages compared to RAN, such as flexibility, open management, and orchestration, enabling multi-vendor solutions, and being able to use less-expensive third-party hardware and software. With these advantages, ORAN can be an important technology for NTNs as well as NTN-integrated terrestrial networks. More detailed applications of ORAN for NTN and current development both in academia and industry will be presented in Section~\ref{sec:oran}.

\section{Network Slicing-aided NTN }\label{sec:network_slicing} 

Network slicing is a powerful technique in telecommunications, which allows operators to create multiple virtual networks from a shared physical network architecture. Utilizing technologies such as SDN and NFV, the operators can dynamically allocate network resources to each network slice according to its specific requirements in terms of bandwidth, latency, and reliability. In the context of NTN, network slicing enables a single network infrastructure to serve multiple types of applications and users with different demands. For example, remote control for UAVs often requires low-latency communication, whereas satellite internet service might require higher bandwidth to serve millions of users. For both cases, network slicing can divide a single physical NTN into two virtual slices to meet different demands by allocating different resources to each slice. 

\subsection{Satellite Networks} 

Satellite networks, due to their benefits such as worldwide coverage area and fast deployment, can significantly enhance mobile terrestrial networks \cite{drif2021slice}. Several challenges for satellite networks such as delay, throughput, and utilization of resources can be addressed effectively by network slicing \cite{suzhi2019space}. However, those advantages come with several problems such as seamless integration of satellite network into the existing mobile networks \cite{drif2021slice}, load balancing, resource allocation considering satellite constellation topology \cite{kim2021satellite}, and efficient routing methods \cite{guo2023online}. To address these problems, multiple approaches have been proposed.
\subsubsection{Resource Allocation}
Among those problems, resource-allocation-related challenges have been addressed in \cite{bisio2019network,deng2019rl,jinkun2021network, ahmed2018demand}. Specifically, in \cite{bisio2019network}, the authors propose innovative approaches to integrate network slicing into railway communication. These approaches aim to optimize network resource allocation by ensuring that latency requirements are met for each class of service in a network. First, a mathematical approach, namely Queuing Theory (QT) method, uses a nonlinear constrained optimization to minimize the difference between the latency. In addition, a Neural Network (NN) approach, namely NN total model, utilizes delays obtained through simulations as input dataset to produce the control variables. Since this approach approximates the required latency by simple fitting, it could lead to inefficient and inaccurate results. To overcome this drawback, the authors propose an NN-divided model \cite{bisio2019network} to approximate the latency of satellite and terrestrial networks using two separate neural networks. This approach solves an optimization problem similar to that of the QT-based method, except that it minimizes an expected mean latency. Simulation results show that these approaches can reduce the latency by roughly 25\%-60\%. With a focus on the dynamic allocation of radio resources, in \cite{deng2019rl}, the authors propose an RL-based slicing strategy, namely dynamic radio resource slicing strategy (Sat-RRSlice), to serve the LEO networks. Particularly, the network slicing problem is modeled as a Semi-Markov Decision Process (SMDP), including network states, actions, state dynamics, and rewards. The states of each slice denote the resource allocation and utilization at a time slot. Moreover, the actions, taken by a resource manager on each slice at a time slot, indicate whether a resource unit is allocated to a slice or not. Furthermore, the reward function is used to calculate the reward, defined by the network's total utility, for taking a certain action. Using Q-learning, optimal Q-values are learned iteratively based on information obtained from a wireless environment over discrete time slots. Simulation results show that, compared with the static slicing strategy \cite{deng2019rl}, Sat-RRSlice can increase the resource utilization rate and sum utility (i.e., the total benefit gained from communication) by at least 18.5\% and 9\%, respectively. 

Taking another approach, in \cite{jinkun2021network}, the authors design a network-slicing architecture for the LEO satellite network with a focus on increasing flexibility and resource utilization. In this architecture, network slicing is managed through Network Slice Selection Assistance Information (NSSAI) and Network Slice Selection Policy (NSSP). The NSSAI, stored in each device, is specifically configured for each device while the NSSP is the set of rules that the network management component must follow. Moreover, two types of network slices are defined: shared slices used by all users for common functions (e.g., storing user data and authenticating users), and dedicated slices for providing personalized service. Experimental results show that a network implemented with this architecture can effectively handle multimedia communications over IP networks. The authors in \cite{ahmed2018demand} design a framework for SDN/NFV-enabled satellite ground segment systems to enable on-demand network slicing. This framework, namely OnDReAMS, has a Service Orchestrator (SO) responsible for managing the life cycle of network slices, e.g., instantiation, maintenance, and termination. Besides, an NFV Manager is used to handle the instantiation, modification, and termination of the VNFs. Moreover, a novel component, namely Satellite Network Slice Descriptor (SNSD), describes the characteristics of the slice as requested by the customer, allowing the SO to set up the slice with flexibility. For resource allocation, the slicing problem is solved using a Mixed Integer Linear Program (MILP) model. A time-window-based online algorithm is then developed to handle the on-demand aspect of the solution, which solves the MILP at the beginning of each time window with updated slice resources. Simulation results show that OnDReAMS can reduce the average QoS violation by approximately 10\%-50\% compared with a baseline approach.
\subsubsection{Routing}
Besides resource allocation, another important aspect of slice-aware satellite networks is routing, which is the focus of \cite{guo2023online} and \cite{sec_3kim2023satellite}. Particularly, in \cite{guo2023online}, a routing strategy for real-time applications over satellite networks with Virtual Functions (VFs) deployed on satellites is proposed. Particularly, the authors develop a VF constrained simple path algorithm to find the shortest path by searching for paths while checking their delay and function requirements iteratively. To overcome the drawbacks of this algorithm (i.e., sub-optimality, instability, and non-scalability), a second algorithm is developed, namely VF-aware shortest path algorithm (VFSP). In this algorithm, the problem of finding the best path in an SN is divided into two sub-problems: finding the shortest path from the source to each functional satellite and from each functional satellite to the destination. By leveraging the unique property of Dijkstra's algorithm \cite{dijkstra1959note}, these sub-problems are solved efficiently by running the algorithm twice, once forward from the source and once from the destination, eventually joining these paths to find the optimal solution. Simulation results show that both algorithms have a run time of approximately 100 times faster than that of the integer linear programming (ILP) approach. Moreover, compared to the KSP-based approach, the VFSP method can improve the acceptance ratio by at least 5\%. Another approach for routing, i.e., link-embedding methods, is presented in \cite{sec_3kim2023satellite}. Specifically, the authors propose a network slicing planning scheme for satellite networks, considering the mobility of LEO satellites and the handover of virtual networks. To this end, two link-embedding methods are developed. The first method is an algorithm that aims to find the shortest path while ignoring the links with limited capacity, thereby significantly reducing the propagation delay. The second method locates the largest link in the neighborhood and updates the path weight to the found path's minimum link capacity, thereby improving the link stability. Simulation results show that the proposed methods can improve the data throughput by up to 21\%.
\subsubsection{Satellite Edge Computing}
\begin{figure}[!]
	\centering
	\includegraphics[width=\columnwidth]{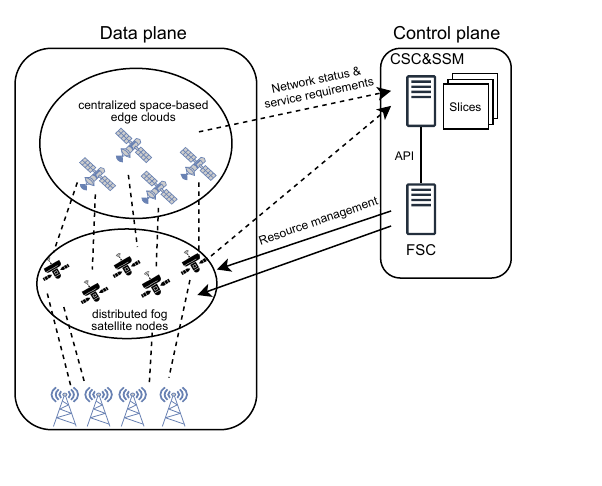}
	\caption{Space edge cloud satellite network architecture~\cite{suzhi2019space}.}
	\label{SpaceBasedSatelliteArchitecture}
\end{figure}

Network slicing applications to satellite edge computing (SatEC) architectures are discussed in \cite{kim2021satellite} and \cite{suzhi2019space}. Particularly, in \cite{kim2021satellite}, the authors present an IoT-supportable SatEC architecture to use satellites for 6G IoT services efficiently. To that end, the authors propose solutions to address two problems: balancing the trade-offs between latency and power and managing network resource allocation. For the first problem, a multi-objective optimization problem is formulated considering latency, computational power, and transmission power attenuation. A Satellite Edge Multi-objective Tabu Search (SE-MOTS) \cite{tang2018multiobjective} algorithm is used to find a Pareto-optimal point, which presents the best trade-off among the three factors considering the dynamic satellite topology and service requirements. For the second aspect, a sliced SatEC optimization problem, formulated as a normalized weighted sum of three objective functions, is used to schedule the tasks with different demands. A golden-section method~\cite{benavoli2009fibonacci} is used to solve this aspect with low computational resources. Taking another approach, in \cite{suzhi2019space}, the authors design a system architecture that combines space-based edge computing and network slicing for space-based network resource management, as illustrated in Fig.~\ref{SpaceBasedSatelliteArchitecture}. In this architecture, edge computing nodes are included due to their benefits of being closer to the data sources and their ability to process data locally. These nodes, namely distributed fog satellite nodes and centralized space-based edge clouds, are responsible for computing tasks with low and high complexity, respectively. In the control plane of the network, the authors introduce the two-layer controller, Fog SDN controller (FSC) and Cloud SDN controller \& Slicing Manager (CSC\&SM). Particularly, FSC abstracts fog satellite nodes and manages their resources while CSC\&SM optimizes resource management of network slices with a global view of the network. Moreover, a 5G Satellite Network Slice Management(5G SNSM) \cite{suzhi2019space} is proposed to perform flexible network slice allocation based on QoS requirements. Simulation results show that the proposed architecture can improve the end-to-end delay and jitter by at least 7\% and 50\%, respectively.

\subsubsection{Slice Admission Control}
Slice admission control is another important aspect of network slicing that should be considered simultaneously with resource allocation. In \cite{sec3_bai2022network}, a two-stage approach is developed to jointly optimize resource allocation and slice admission control in LEO satellite networks. To this end, the authors first formulate a robust optimization problem to optimize slice admission, taking into account the uncertainty in the network state and user distribution. Based on the admission decisions in the first stage, the resource allocation strategies are optimized in the second stage to guarantee QoS for users. Since the original optimization problem is non-convex, auxiliary variables are introduced to transform the problem into a convex form, thereby significantly reducing the problem's complexity. Simulation results show that the proposed approach can improve the user QoS by 7.2\% compared to the non-robust approach.

\subsection{Integrated Satellite-Terrestrial Networks}

Integrated Satellite-Terrestrial Networks (ISTNs) utilize satellite communication to extend network coverage and reduce the dependence on terrestrial infrastructure, thereby offering services to users with higher efficiency. Network slicing can bring several advantages to ISTNs such as efficiency in resource sharing, differentiated QoS provisioning \cite{mendoza2021sdn}, and network flexibility as well as scalability. However, multiple challenges also arise from this application, including network awareness requirements and the highly dynamic environments~\cite{mendoza2021sdn}, QoS reduction due to terrestrial obstacles \cite{minardisast}, limited resource constraint, and service provider incentives. 
\subsubsection{Resource Allocation}
Innovative approaches are proposed to overcome those challenges. To address the resource allocation problem in slice-aware ISTN, optimizing the VNE \cite{fischer2013virtual} algorithm is a promising solution. To that end, the authors in \cite{minardisast} design a framework, namely Slice-Aware VNE for Satellite-Terrestrial (SAST-VNE) to improve the implementation of the VNE algorithm in ISTNs. In this framework, problematic network slices, i.e., those that do not meet Key Performance Indicators (KPIs) or will undergo satellite handover, are inserted into a queue for analysis. In case a new slice is created, the embedding is computed using the DViNE algorithm \cite{maity2022d}. For problematic network slices, if they have high priority, i.e., low tolerated latency, each affected link will be remapped with the shortest-path algorithm. On the contrary, if they have lower priority, the authors propose an iterative algorithm, namely SAST-VNE, to balance the network load and minimize the migration cost. Simulation results show that during handovers, SAST-VNE can reduce the average node migrations and average link migration by roughly 25\% and 6\%, respectively. Another implementation of VNE for ISTNs is presented in \cite{mendoza2021sdn}. Particularly, the authors design a testbed to validate the feasibility of integrating non-GEO satellite constellations and implementing VNE algorithms in highly dynamic network conditions. This testbed consists of a dynamic satellite-terrestrial network emulated in Mininet \cite{mendoza2021sdn}, an external Ryu SDN \cite{mendoza2021sdn} controller and a VNE algorithm script. First, the Mininet emulator is used to build an OpenFlow-based substrate network, enabling the Ryu SDN controller to manage the flow of packets in the network. The Ryu SDN controller also uses a Traffic Engineering (TE) application to create paths for each VN, set rate limits, gather network statistics, read the network topology, and process changes to the network topology in real-time. Furthermore, the VNE algorithm script, implemented in Matlab using an Integer Linear Programming formulation with a load-balancing objective function, handles required resources for virtual networks and updates the network configuration to the Ryu SDN controller. Simulation results show that the testbed can handle many scenarios and changes in the network topology such as adding new VNs, dynamic changes in the network layout, and handling of network failures.  

\begin{figure}[!]
	\centering
	\includegraphics[width=\columnwidth]{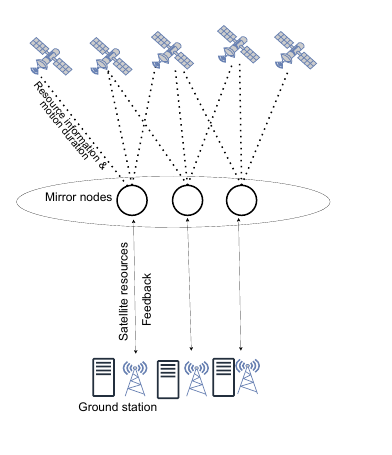}
	\caption{Mirror node model~\cite{kong2022dynamic}.}
	\label{MirrorNodeModel}
\end{figure}

Moreover, approaches to optimize resource allocation are presented in \cite{kong2022dynamic} and \cite{jiang2023multi}. Particularly, the authors in \cite{kong2022dynamic} propose a dynamic slicing strategy for ISTNs. The proposed strategy uses mirror nodes, instead of virtual nodes, to store motion duration and resource information of satellites. Based on that, the service forwarding path and slice resources can be adjusted, as illustrated in Fig.~\ref{MirrorNodeModel}. The mirror nodes also communicate the satellite resources with ground stations and receive feedback, i.e., which satellites to be chosen for the slicing process. An optimization model is then formulated to maximize the long-term average slicing performance of the network with a constraint of computing, storage and bandwidth resources. Moreover, a resource reserved in slices and adaptive adjustment between slices algorithm (R2A2) is developed to solve the optimization problem. Simulation results show that, compared to a DQN approach, R2A2 can increase resource utilization and slice satisfaction by roughly 33\% and 30\%, respectively. With a focus on the admission decision for a slice request, the authors in \cite{jiang2023multi} propose a game-theory-based solution for resource allocation in ISTNs. To solve the challenges related to limited resource constraints, slice admission control is formulated involving different combinations of service providers and users. Specifically, service providers and users form a multi-sided market to exchange resources. To participate in the network, both sides consider their prices, i.e., network resources for the providers and the cost for the users. An auction, designed with the Multi-Sided Ascending-Price Auction Mechanism \cite{jiang2023multi}, is performed by increasing the price for both sides based on supply and demand until a balance is reached. Eventually, the final prices are set, and the allocation of resources to users is determined. Experimental results show that, compared to a baseline approach, i.e., second-price auction \cite{jiang2023multi}, the ascending-price auction can increase the bandwidth per user, the admission ratio, and the gain from trade per user by up to 25\%, 33\%, and 20\%, respectively. 

A novel approach for network slice orchestration is presented in \cite{papa2021cost}. Particularly, the authors present a system model for the deployment of network slicing for aircrafts using satellites. This model aims to enhance resource management in ISTNs by incorporating the concept of slice collaboration, i.e., defining network slices based on their willingness to share user traffic statistics with the infrastructure provider). To this end, a Mixed Integer Non Linear Programming (MINLP) model is developed to maximize the resources, i.e., cache and backhaul resources) while satisfying slice constraints. Moreover, a pricing model \cite{papa2021cost} is proposed to increase the chances of uncooperative slices, i.e., slices that are unwilling to share user traffic statistics, being served. Simulation results show that the MINLP model can increase the selection probability of uncooperative slices by 33\% and cooperative slicing can accommodate 200\% more services compared to that of uncooperative slicing. Taking another approach, the authors in \cite{drif2021slice} propose a slice-aware NTN architecture to enable the seamless integration between NTN and terrestrial networks. To this end, the authors propose an end-to-end slicing model which treats the NTN as a slice-aware link in the terrestrial network. Based on the proposed model, a functional architecture fully compliant with the 3GPP standard is developed to allow the NTN to be integrated with the terrestrial network. The architecture consists of three main segments: the user segment with the satellite terminal, the ground segment with satellite gateways and network control components, and the space segment with one or more satellites. The Satellite Terminal (ST) serves as the User Equipment (UE) in 5G networks, communicating with the space segment, which then relays messages to the satellite gateway. Satellite gateways are the central entities that manage resource allocation, authentication, and data processing. Simulation results show that the proposed slice-aware scheme can achieve better trip time, packet error rate, and jitter compared to those of the cases without network slicing.     
\subsubsection{User Assignment}
In addition to routing, several frameworks have been proposed to enhance network performance by optimizing user assignment in \cite{sec3_hendaoui2020cognitive}, \cite{wu2021learning}, and \cite{hendaoui2021leveraging}. Particularly, in \cite{sec3_hendaoui2020cognitive}, the authors propose a network slicing scheme for hybrid satellite-terrestrial networks, aiming to improve reliability and reduce the video traffic offload. To this end, the authors develop a scheduling strategy that assigns the users to three different types of networks, e.g., only satellite, only terrestrial, and hybrid satellite-terrestrial, based on Channel Quality Indicator (CQI) and QoS requirements. Simulation results show that the proposed scheduling strategy can improve the network throughput by up to 48\% compared to the case where only the satellite network is employed. Differently, with a focus on space-terrestrial integrated vehicular networks (STIVN), the authors in \cite{wu2021learning} propose a scheme for network slicing to support both delay-tolerant services (DTSs) and delay-sensitive services (DSSs). This scheme aims to solve the problem of scheduling and resource slicing in STIVN, which involves allocating spectrum resources to slices, determining bandwidth allocation, and user assignment for each vehicle. Specifically, the system cost (i.e., DSS requirement violation, DTS delay, and slice reconfiguration) is taken into account in two subproblems: a resource slicing subproblem in large-timescale, and a resource scheduling subproblem in a smaller timescale. To solve those subproblems, a two-layered RL-based scheme is developed. In the first layer, namely resource slicing layer, spectrum-resources are pre-allocated to slices using a proximal policy optimization (PPO)-based RL algorithm \cite{wu2021learning}. In the second layer, namely resource scheduling layer, depending on network conditions and service requirements, spectrum resources are assigned to vehicles using match-based algorithms. Simulation results show that the proposed approach can reduce the overall system cost up to roughly 72.57\%, compared to that of a baseline approach. In another approach, the authors in \cite{hendaoui2021leveraging} propose a network slicing scheme to improve their existing resource management approach in satellite-LTE networks. In the previous approach, an adaptive hybrid satellite-LTE downlink scheduler (H-MUDoS) \cite{zangar2019leveraging} determines if users can be served through satellite network or ground-based stations. However, simulation results show that the decrease in QoS of the satellite network affects the performance of the entire hybrid network. In the current approach, the network is separated into isolation slices, each with its own scheduling strategy assigned by the scheduler. Simulation results show that the decrease in QoS of a slice does not affect other slices. 
\subsubsection{Routing}
Next, approaches for optimizing routing are presented in \cite{rodrigues2022network} and \cite{sec3_kak2021towards}. Particularly, in \cite{rodrigues2022network}, the authors propose a novel framework to optimize resource distribution for network slicing management in ISTNs. In this framework, a hybrid approach combining ML and Ant Colony Optimization, is implemented to associate a new metric, i.e., a cost metric, to any route in the network. In this case, when a route is chosen more frequently, its cost is increased and vice versa. With this mechanism, the network can adapt to changes in user demands and effectively allocate resources. Experimental results show that the proposed framework increases the user acceptance ratio up to roughly 15\% under different numbers of users and 20\% under different numbers of servers. Taking another approach, an automatic network slicing framework for ISTNs is presented in \cite{sec3_kak2021towards}. The proposed framework aims to find the satellite-gateway assignments and resource allocation to optimize the overall utility, e.g., throughput and latency, of network slices. To that end, a Voronoi tessellation-based topology construction mechanism is proposed to map the satellite constellations to equivalent network topologies. Based on that, an MILP is formulated. To reduce the complexity of the MILP problem, the optimization problem is decomposed into a resource allocation and a satellite-gateway assignment sub-problems. An online iterative algorithm is then developed to solve the two sub-problems. Simulation results show that the proposed framework can increase the slice admittance rate by up to 23\% and reduce the control traffic by 76\% compared to the standard SDN approach.

\subsection{Space-Air-Ground Integrated Networks} 

Space-Air-Ground Integrated Networks (SAGINs) aim to achieve full network coverage and ubiquitous services by integrating terrestrial networks, satellite networks, and aerial networks\cite{yin2022cybertwin,hou2022edge}. While applying network slicing to SAGINs is an effective solution for efficient usage of network resources, this technique faces new challenges such as complex slice orchestration in multi-domain networks, resource optimization considering UAVs position, and dispatching cost. To address those challenges, multiple innovative schemes have been proposed.
\subsubsection{Resource Allocation}
For instance, as the main concern of network slicing application, resource allocation is discussed in \cite{zhou2023multi,lyu2021service,wu2022ai,yang2020research}. Particularly, in \cite{zhou2023multi}, the authors propose a framework for integrating network slicing to a SAGIN, which establishes three types of RAN slices, i.e., high-throughput, low-delay, and wide-coverage. In particular, a non-scalar multi-objective optimization problem (MOOP) is formulated to jointly optimize throughput, service delay, and coverage area. Moreover, a Central and Distributed Multi-agent Deep Deterministic Policy Gradient (CDMADDPG) algorithm is developed to solve the problem. This CDMADDPG algorithm uses a centralized unit to determine the optimal positions for the virtual UAVs (vUAVs) and the most suitable subchannels as well as power resources among the slices. Then, intra-slice resource sharing is arranged by virtual base stations, vUAVs, or virtual LEO satellites, depending on the distributed units. Eventually, near-Pareto optimal solutions can be found. Simulation results show that, compared to a baseline approach, the proposed framework can improve throughput and delay by up to 10\% and 50\%, respectively. With a focus on dynamically slicing spectrum resource in SAGINs an online control framework is proposed in \cite{lyu2021service}. Here, the proposed framework aims to adapt to varying vehicular environments and achieve isolated service provisioning, i.e., each type of service is processed in an independent queue. To that end, the authors propose a workflow of dynamic slicing consisting of four steps, i.e., request admission, request scheduling, UAV dispatching, and resource slicing, to ensure that services with different QoS are adequately served. Based on that, a Lyapunov-based approach is proposed to maximize the system revenue and minimize a time-averaged penalty while stabilizing the system. This approach aims to minimize time-averaged queue backlogs of all services, i.e., the average number of unprocessed requests in a queue over a period. Simulation results show that the proposed framework can increase the throughput by approximately 26\%. 

\begin{figure}[!]
	\centering
	\includegraphics[width=\columnwidth]{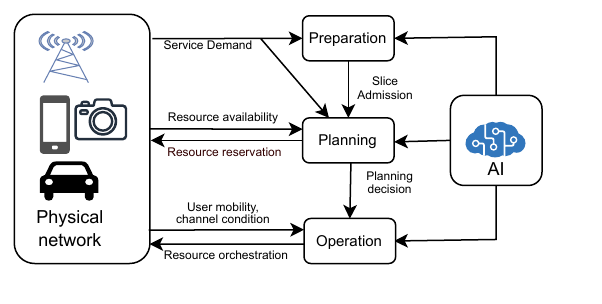}
	\caption{AI based network slicing solution~\cite{wu2022ai}.}
	\label{AIBasedNetworkSlicingSolution}
\end{figure}

Taking another approach utilizing AI, the authors in \cite{wu2022ai} introduce an AI-enabled network slicing architecture for 6G networks, aiming to enable intelligent network management and facilitate emerging AI services. Particularly, the architecture has two main characteristics: AI for slicing, i.e., using AI to manage multiple network slices with strict QoS requirements, and slicing for AI, i.e., creating special slices for AI services. It is shown in \cite{wu2022ai} that AI can be used to support different phases of the network slicing process such as preparation (e.g., service demand prediction and slice admission), planning (e.g., VNF placement and resource reservation), and operation (e.g., resource orchestration and Radio Access Technology selection), as illustrated in Fig.~\ref{AIBasedNetworkSlicingSolution}. Moreover, in the proposed architecture, network slices are tailored to support three stages of AI services (including data collection, model training and model inference) with their corresponding required QoS. Simulation results show that a deep deterministic policy gradient (DDPG)-based network slicing solution has a system cost (a weighted sum of resource reservation cost, slice reconfiguration cost, and delay requirement violation
penalty) 15\% lower than that of myopic resource reservation \cite{wu2022ai}.

Considering the current status of power communication network development, the authors in \cite{yang2020research} propose a SAGIN slicing architecture. Specifically, the proposed architecture consists of four layers, i.e., service request, slicing management and scheduling, virtualized resource, and infrastructure layer. The service request layer receives requests from users to define requirements and build service slices. The slicing management and scheduling layer provides the management interface for users, controls the scheduling management, stores the data required by the platform environment and transmits the corresponding data to the lower functional components. The virtualized resource layer utilizes NFV to virtualize the physical resources and allocates sufficient virtual resources to different slices. The infrastructure layer provides the physical resources. It is shown in \cite{yang2020research} that this architecture can manage network resource allocation efficiently to meet different demands of the power grid communication service.

\subsubsection{Slice Configuration and Routing}
Together with resource allocation, multiple aspects are considered to enhance network performance, including slice configuration, Service Level Agreement (SLA) decomposition, and routing. A multi-domain network slicing framework is proposed in \cite{esmat2023leons}, which aims to jointly optimize all the three aspects. It is shown in \cite{esmat2023leons} that the proposed framework can handle multiple slice configurations in satellite-terrestrial edge computing networks (STECNs). Particularly, a Markov process is modeled to track the probability of satisfying the SLA per configuration. Moreover, an index-based slice configuration policy, based on restless multi-armed bandits (RMABs) \cite{lorenzo2021autonomous} \cite{liu2010indexability}, is defined to solve the multi-domain slicing problem. Based on the slice configuration policy and the slice/resource availability, the SLA is decomposed. The routing and resource allocation optimization problem is solved by each domain controller with consideration of the slice/resource availability. Simulation results show that the proposed framework can achieve three to six times higher rewards than those of baseline approaches in terms of user requirements in the SLA and the energy consumption. It is also shown in \cite{esmat2023leons} that the proposed framework can achieve optimal configuration under different network conditions, e.g., outage and delay.

In \cite{hu2023space}, the authors propose a scheme to integrate network slicing to SAGINs with native AI. In particular, under the assumption that all network entities are software-defined and can load multiple network functions and take any required roles, all network entities are classified into three classes, i.e., end entity, routing entity, and management as well as orchestration entity. End entities are communication sources and destinations. Routing entities are responsible for access control, traffic routing, and QoS management. Management and orchestration entities are responsible for entity management, network slicing management, and resource orchestration. A pre-defined operating entity \cite{hu2023space}, aware of network characteristics, adopts the Distributed Weighted Classification Method (DWCM) to assign roles to all network entities. Experimental results show that, compared to a baseline approach \cite{hu2023space}, the proposed scheme can increase the delivery ratio and the overall data rate by roughly 4\% and 28\%, respectively and reduce the end-to-end delay and routing overhead by approximately 46\% and 6\%, respectively. 

\textit{Summary:} In this section, we have discussed multiple approaches for network slicing integration in NTNs, e.g., satellite networks, ISTNs, and SAGINs. Specifically, network slicing is an effective solution to enhance NTNs with better delay, throughput, network resource utilization, network flexibility, and scalability. However, multiple challenges arise when applying network slicing to NTNs such as resource allocation, routing, load balancing, slice admission control, and SLA decomposition. Among those challenges, resource allocation is the most noticeable concern. To address resource allocation, one approach is to formulate an optimization problem, which can be solved by AI/ML-based methods, mathematic algorithms, scheduling strategies, or innovative mechanisms. Additionally, optimizing VNE is effective in enhancing resource allocation. Another challenge to network slicing is routing, which can be solved by shortest-path approaches. Although the proposed approaches can effectively address various challenges of network slicing in NTN, there are still open issues that stem from the complexity of network slicing optimization problems. Particularly, these problems often include numerous constraints due to resource limitations and various platforms, e.g., space, air, and ground, resulting in high complexity. As a result, most of the proposed approaches do not consider network slicing holistically or are based on simplified assumptions. Therefore, additional effort is needed to develop effective and holistic solutions for network slicing in NTNs, especially the integration of NTNs and terrestrial networks. The works surveyed in this section are summarized in Table~\ref{TableNetworkSlicingReview}.

\begin{table*}[]
	\centering
	\caption{Summary of network slicing approaches in NTN.}
	\label{TableNetworkSlicingReview}
	\begin{tabular}{|l|l|l|l|}
		\hline
		Reference                         & Network Type                                             & Problem                                          & Technique                                        \\ \hline
		\cite{bisio2019network}           & \multicolumn{1}{c|}{\multirow{9}{*}{Satellite Network}}                                    & Resource allocation                             & NN-based model                                   \\ \cline{1-1} \cline{3-4} 	
		\cite{deng2019rl}                 & \multicolumn{1}{c|}{}                                    & Resource allocation                             & RL-based slicing strategy                        \\ \cline{1-1} \cline{3-4} 
		\cite{jinkun2021network}          & \multicolumn{1}{c|}{}                                    & Resource allocation                             & Managing slice through NSSAI and NSSP      \\ \cline{1-1} \cline{3-4}       
		\cite{ahmed2018demand}            & \multicolumn{1}{c|}{}                                    & Resource allocation                             & MILP                       \\ \cline{1-1} \cline{3-4}
		\cite{guo2023online}              & \multicolumn{1}{c|}{}                                    & Routing                                         & Dijkstra's algorithm                             \\ \cline{1-1} \cline{3-4} 
		\cite{sec_3kim2023satellite}      & \multicolumn{1}{c|}{}                                    & Routing                                         & Develop new shortest path algorithm              \\ \cline{1-1} \cline{3-4}
		\cite{kim2021satellite}           & \multicolumn{1}{c|}{}                                    & Satellite edge computing             & SE-MOTS, golden-section method                   \\ \cline{1-1} \cline{3-4} 
		\cite{suzhi2019space}              &  & Satellite edge computing     & 5G SNSM                        \\ \cline{1-1} \cline{3-4} 
		
		\cite{sec3_bai2022network}        & \multicolumn{1}{c|}{}                                    & Slice admission control    & Simplifying robust optimization problem using auxiliary variables        
		\\ \hline
		
		\cite{minardisast}                &          \multirow{10}{*}{ISTN}                                                & Resource allocation                                & SAST-VNE                                         \\ \cline{1-1} \cline{3-4}                
		\cite{mendoza2021sdn}             &                                    & Resource allocation                                  & Testbed design                                                   \\ \cline{1-1} \cline{3-4} 
		
		\cite{kong2022dynamic}            &                                                          & Resource allocation                             & R2A2                                             \\ \cline{1-1} \cline{3-4} 
		\cite{jiang2023multi}             &                                                          & Resource allocation                             & Multi-Sided Ascending-Price Auction Mechanism    \\ \cline{1-1} \cline{3-4} 
		\cite{papa2021cost}               &                                                          & Resource allocation                             & MINLP        \\
		\cline{1-1} \cline{3-4} 
		\cite{drif2021slice}              &  & Resource allocation     & End-to-end slicing model
		\\ \cline{1-1} \cline{3-4} 
		
		\cite{sec3_hendaoui2020cognitive} &                                                          & User assignment             & Hybrid scheduling strategy                       \\ \cline{1-1} \cline{3-4} 
		\cite{wu2021learning}             &                                                          & User assignment                             & PPO-based RL algorithm, matched-based algorithms \\ \cline{1-1} \cline{3-4} 
		\cite{hendaoui2021leveraging}     &                                                          & User assignment                             & Smart scheduler         \\ \cline{1-1} \cline{3-4} 
		
		\cite{rodrigues2022network}       &                                                          & Routing                    & ACO                                              \\ \cline{1-1} \cline{3-4} 
		\cite{sec3_kak2021towards}        &                                                          & Routing                    & MILP                                             \\ \hline

		\cite{zhou2023multi}              & \multirow{6}{*}{SAGIN}                                   & Resource allocation                             & CDMADDPG                                         \\ \cline{1-1} \cline{3-4} 
		\cite{lyu2021service}             &                                                          & Resource allocation                             & Lyapunov-based algorithm                         \\ \cline{1-1} \cline{3-4} 
		\cite{wu2022ai}                   &                                                          & Resource allocation                             & Managing slice by AI                             \\ \cline{1-1} \cline{3-4} 
		\cite{yang2020research}           &                                                          & Resource allocation                             & Scheduling                        \\ \cline{1-1} \cline{3-4} 
		\cite{esmat2023leons}             &                                                          & Slice configuration and routing & RMABs                                            \\ \cline{1-1} \cline{3-4} 
		
		\cite{hu2023space}                &                                                          & Slice configuration                  & DWCM                                             \\ \hline
	\end{tabular}
\end{table*}



\section{AI/ML-aided NTN }\label{sec:aiml}

As people continue to explore and establish more practical applications throughout the Earth, there exists a growing need for advanced technologies to support NTN. AI/ML-aided NTN offers a guarantee for significant benefits in this area, from improving communication efficiency for IoT systems to enabling autonomous UAV operations in huge and remote areas. Specifically, AI/ML can increase efficiency and reliability to optimize communication protocols using NTN. They can also enable the autonomous operation of numerous wireless applications, thereby reducing the need for constant human intervention. Finally, AI/ML can facilitate new scientific discoveries and insights by enabling more efficient and targeted data analysis using NTN. As a result, NTN can generate vast quantities of data, and AI/ML may help to extract valuable insights and enable better-informed decisions in a variety of practical applications, as discussed in the following.



\subsection{IoT}

\begin{figure}[t]
	\centering
	\includegraphics[scale=0.4]{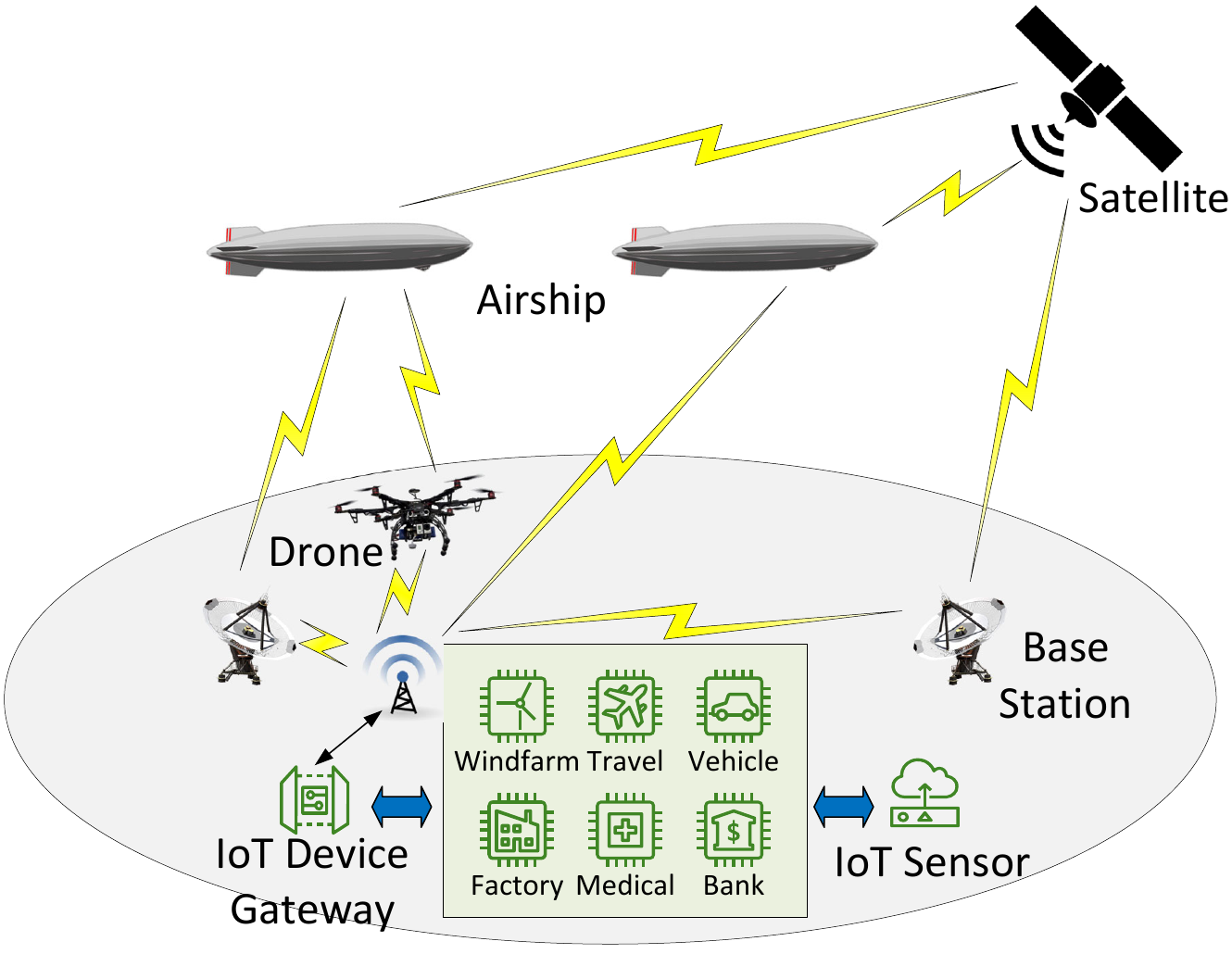}
	\caption{The integrated air-ground and NTN for IoT systems.}
	\vspace{-0.5em}
	\label{fig:iot_1}
\end{figure}

The utilization of AI and ML in NTN has the potential to revolutionize IoT systems by improving data collection and enabling extensive communication for sensors located in remote and broad areas. NTN may increase radio coverage, provide monitoring, and incorporate sensing services to remote locations by utilizing satellites, airships, and aircraft as illustrated in Fig.~\ref{fig:iot_1}. This can lead to significant improvements in data collection from areas that were previously inaccessible, allowing for more comprehensive and accurate insights. 

In~\cite{sec5_michailidis2020ai}, the authors investigate the potential use of DL for NTN-based industrial IoT (IIoT) services, i.e., spaceborne-based and airborne-based intelligent IIoT systems. This research describes how diverse DL-based algorithms have been used and tested for NTN-based IIoT applications based on the optimization objective, the available power, and processing resources. Here, DL is considered the most effective AI algorithm when it comes to end-to-end optimization. Specifically, for spaceborne applications, an ensemble DNN-based optimization can be utilized for resource allocation while a deep-Q network can be used for energy consumption and processing latency in Satellite-IoT (S-IoT) networks. Additionally, S-IoT edge computing and DL methods can accelerate data processing and transmission as well as enhance bandwidth utilization. Since the practical viability of conventional DL is contingent on the availability of a large amount of sensor data and sufficient computing power, DRL with lower complexity can be used. For airborne applications, an IoT network with three layers for online large data processing based on MEC is proposed with UAVs as edge servers, Lyapunov optimization, DRL algorithm, and CNN Q-networks. In this case, the CNN Q-network, which enables action reward prediction, is trained using the UAVs' views of the surrounding environment, and the DRL can efficiently optimize the path planning of the UAVs.

To provide more benefits of using ML/AI-aided NTN, the authors in~\cite{sec5_vaezi2022cellular} categorize the NTN into integrated UAV-IoT networks and integrated S-IoT networks. For integrated UAV-IoT networks, an RL method can be applied to build the data collection trajectory of a UAV from IoT devices with the aim to increase the UAV's flight duration. Using a double deep Q-network strategy, an effective path to maximize collected IoT data under flight time and obstacle avoidance limitations can be designed. Then, to increase the IoT network's energy efficiency, a DRL strategy is employed to optimize the channel and power allocation of IoT devices in the uplink communication. Meanwhile, a DRL approach can be adopted in the integrated S-IoT networks. In this case, an energy-efficient channel allocation mechanism for LEO S-IoT network can be implemented. The DRL approach can also be employed to address the complexity of energy cost and latency minimization influenced by IoT user association and resource allocation. 

Despite that NTNs, especially the ISTNs, can provide wide coverage for IoT systems, their links still suffer from high dynamics and latency. To address the problem, several ML/DL approaches are surveyed in~\cite{sec5_zhu2021integrated}, aiming at optimizing the routing strategy update. Specifically, traffic patterns of the integrated network are learned using CNN, resulting in routing paths that balance traffic. Then, a deep Q-learning-based method is proposed to reduce the routing delay in the integrated network by combining both satellite and terrestrial users' networking, caching, and computing resources. Additionally, the ISTN has considerable spectrum-sharing potential, and thus two spectrum-sharing systems based on SVM and CNN are presented. Here, the intelligent spectrum sharing reduces interference and increases spectrum efficiency compared to traditional approaches. As the satellite covers a large area, it is important for IoT devices to be accurately positioned to ensure that the service stays up and runs in the integrated network.

\subsection{Mobile Services}

\begin{figure}[!t]
	\centering
	\includegraphics[scale=0.3]{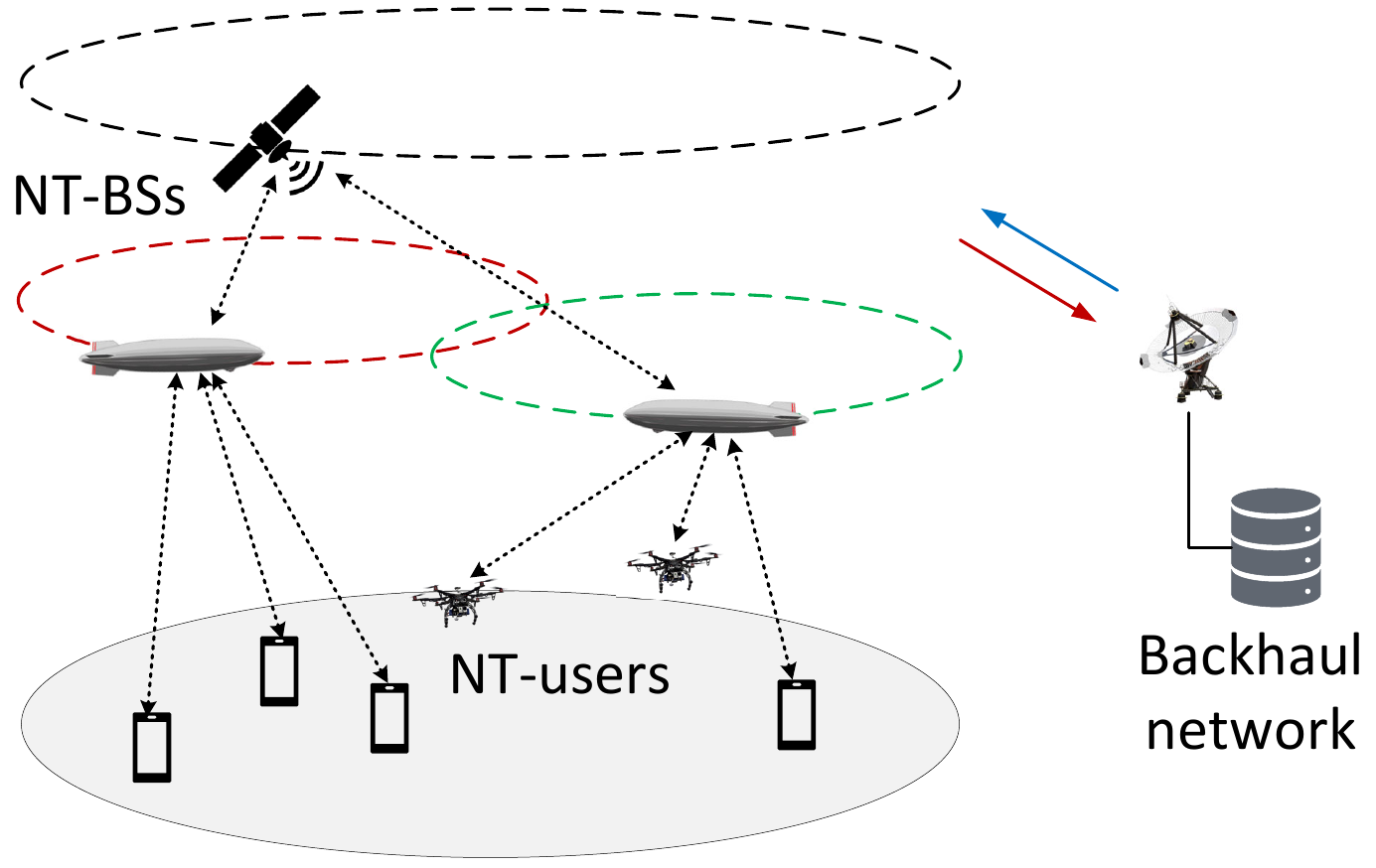}
	\caption{NTN with moving NT-BSs and NT-users.}
	\vspace{-0.5em}
	\label{fig:mobile_1}
\end{figure} 

Since NTN consists of space-borne base stations (BSs), e.g., satellites, and airborne BSs, e.g., UAVs and aircraft, they may lead to a dynamic and non-stationary environment. As such, non-terrestrial users (NT-users) on the ground need to predict the mobility of those non-terrestrial BSs (NT-BSs) autonomously. Likewise, NT-users may move frequently, e.g., drones, which makes the trajectory estimation from the NT-BSs side challenging (as illustrated in Fig.~\ref{fig:mobile_1}). 

To cope with those issues, the work in~\cite{sec5_cao2021deep} and~\cite{sec5_lien2022autonomous} propose new ML/AI-aided NTN, aiming at supporting the activities of NT-BSs and NT-users. In~\cite{sec5_cao2021deep}, the authors introduce an NT-users-driven DRL approach for handover and throughput optimization in NTN-based multi-user access control. Specifically, a centralized agent on the backhaul side of NT-BSs trains DQN parameters with a sigmoid activation function. The trained DQN is then used by each NT-user with slow mobility, e.g., smartphone user, to make his/her own access decisions. The proposed technique allows each NT-user to intelligently access a suitable NT-BS to improve long-term system throughput and reduce the NT-BSs' handovers. Compared with other benchmark methods, e.g., RSS-based and Q-learning algorithms, the proposed framework is superior in terms of the long-term throughput and the handover numbers by 6 times and 8\%, respectively. 

The above work is then extended with the existence of NT-users with medium/high mobility, i.e., drones in low-altitude or high-altitude, by the authors in~\cite{sec5_lien2022autonomous}. In this case, each NT-BS should autonomously forecast NT-users' trajectories and the likelihood of their presence at any site. For that, instead of using the aforementioned complex DRL method, this work proposes novel RL systems in which each of many NT-BSs independently calculates deployment trajectories to maximize the access of NT-users. For the deployment trajectory, multiple NT-BSs can apply \emph{k-step} state reduction (SR) distributive Q-learning to optimize the autonomous trajectory. Via simulation results, the suggested approaches outperform Q-learning, maximal SINR, and distributive DRL in terms of the average number of served NT-users up to 47\%.

\subsection{Network Management}

One of the NTN communication systems, i.e., the satellite communication networks, typically has a challenge in terms of system capacity, in which the resources of most satellites are usually underutilized. To address this challenge, collaboration among different satellite systems can be one of the most effective solutions to enhance resource utilization. Currently, each satellite has an unconnected system architecture and dedicated resource utilization. For that, ML/AI approaches can be used to cope with the above limitations by developing intercommunication frameworks among different satellite systems. 

\begin{figure}[t]
	\centering
	\includegraphics[scale=0.3]{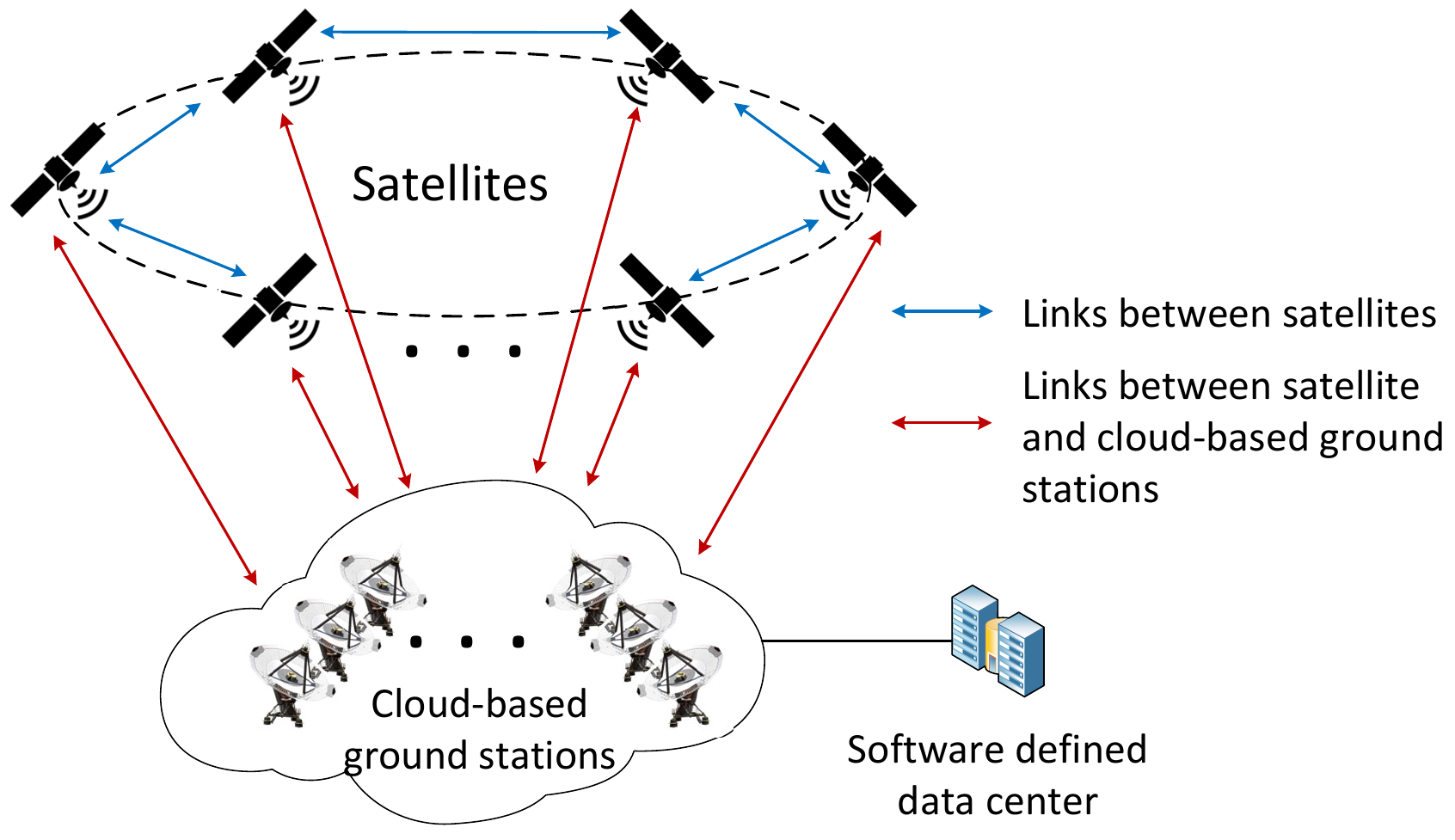}
	\caption{The communication between satellites and ground stations using software-defined network and virtualization.}
	\vspace{-0.5em}
	\label{fig:netman_1}
\end{figure}

The authors in~\cite{sec5_deng2019next} propose a resource management framework in heterogeneous satellite networks. In the framework, SDN and virtualization methods in the data center are used to manage and combine disparate resources as shown in Fig.~\ref{fig:netman_1}. To obtain the optimal resource utilization, a DRL approach that uses the Markov decision process (MDP) and integrates DL for inference capability as well as RL for decision-making potential is then applied. In this work, the state space contains service and resource states, whereas the action state includes all actions in which an agent provides resources to the services. Additionally, the reward may include spectral efficiency, bandwidth, throughput, and power efficiency. Here, each satellite acts as a smart multi-agent that can perform distributed data processing and transmit information between satellites and/or ground stations in the cloud system. This reduces the workload of the data center and enhances the efficiency of communication because of shorter transmission paths. The SDN-based integrated networks for resource allocation are also proposed in~\cite{sec5_qiu2019deep}. The purpose of this integrated network is to monitor networks, computation resources as well as caches, and orchestrate them all simultaneously. They use the Markov decision process for the resource allocation optimization problem, and deep Q-learning approach for the problem solving. The simulation results reveal that the proposed scheme can achieve the highest expected utility per resource up to 10 times compared with other baseline methods.

Then, the work in~\cite{sec5_darwish2021vision} presents a novel framework of "self-evolving networks (SENs)," which employs AI through ML algorithms, e.g., federated learning and online learning, to fully automate and intelligently evolve future integrated NTN in terms of network management, communication, computation, and mobility of mobile users. For that, the authors utilize the intelligent vertical heterogeneous network (I-VHetNet) architecture as a model to envision the idea of SEN in future integrated networks. The I-VHetNet design not only combines terrestrial, aerial, and satellite networks, but also includes intelligence, computation, and caching platforms to allow multi-level edge computing. In particular, the SEN engine first employs AI to forecast where additional network capacity and coverage are required based on user movement, behavior, and applications. Utilizing the prediction result, the SEN engine intelligently and automatically sends UAV-BSs or adjusts a HAPS beam to enhance network coverage and increase its capacity to serve customers. It then can choose terrestrial, aerial, or satellite networks to backhaul UAV-BSs based on the QoS requirements. The SEN engine can also execute computational offloading to the best computational level systems, e.g., cloud computing, fog computing, or collaborative computing of mobile users. Afterward, the SEN engine can monitor network performance and user satisfaction. To adapt to dynamic changes and create more accurate automated and intelligent decisions, the SEN engine uses network environment assessment as feedback. From the simulation using three data centers and 300/3000 users, the proposed system can minimize the data offloading and computing delay by 0.05 seconds with 10GB-10TB offloaded data per user.

To further relax the network management and service-oriented resource allocation of the integrated terrestrial and NTN in B5G/6G networks, network slicing can be used through leveraging AI-based approaches. In~\cite{sec5_lei2021dynamic}, dynamic-adaptive AI-enabled network slicing management to deal with dynamic wireless environments is discussed. Specifically, to allow intelligent orchestration of optimization problems in network slicing management, several AI-based approaches can be applied. However, it is worth noting that using conventional DNN, CNN, or DRL methods may suffer from a slow convergence rate, and thus it can deteriorate the learning performance especially when parameters change fast dynamically. To this end, transfer learning and meta-learning can provide a fast response with fewer samples. Both learning methods benefit from transfer/meta knowledge without requiring data training from scratch to solve the new learning problem with less training data. The use of RL then can speed up the convergence in re-training and improve the re-fitting ability. Based on the case study, the proposed transfer and meta-learning framework can minimize the loss and cost of slices up to 0 and 0.7 over time, respectively, in two typical dynamic schemes, i.e., bursty traffic and devices' arrival/departure in slices. A similar network slicing method to guarantee different QoS levels according to users' requirements for eMBB in 5G-satellite networks is proposed in~\cite{sec5_de2020qos}. Particularly, the authors introduce a neural network-based resource allocation optimization problem to satisfy different QoS requirements. Here, the work adopts the neural network with a weighted round-robin scheduler (WRR-NN) since a multi-queue system for packet fetching is utilized, aiming at meeting the delay requirement for average end-to-end packet delivery delay. Based on the comprehensive simulations, this technique can precisely follow system dynamics and satisfy eMBB's service latency and jitter criteria at approximately 0.025s and 10-50ms, respectively. 

The addition of privacy for resource allocation in integrated networks can also be implemented. The authors in~\cite{sec5_li2020distributed} present a distributed federated learning (FL)-based intrusion detection system (IDS) in the integrated terrestrial networks and NTN, aiming at addressing limited satellite network resources and high privacy requirements. As such, the FL is used to allocate resources adequately in each domain and block malicious traffic, including DDoS attacks. Here, the FL only sends the trained model without raw data sharing to preserve privacy. This solution outperforms DL-based IDS for malicious traffic identification rate at 98\%, packet loss at 0\%, and CPU consumption rate at 70\%.

\subsection{Vehicular Networks}

In terms of vehicular networks, the integration between terrestrial and NTN has the potential to bring efficient, reliable, and robust data transmission for medium/high speed or delay-sensitive/delay-tolerant services. This can reduce the dynamic network environment including handover and unstable network connection problems. 

For example, the authors in~\cite{sec5_xu2021deep} apply multipath transmission control protocol (MPTCP) congestion control mechanism to perform data transmission through the integrated networks concurrently for high-speed railway schemes. Nonetheless, due to frequent handover problems, they utilize reference signal received power information according to a DRL approach to boost the goodput (i.e., the amount of successful data received by the recipient within the deadline) performance. Particularly, the integration among gated recurrent unit (GRU), CNN, and deep deterministic policy gradient (DDPG) are used to build a learning model. First, the representation network generates a network state representation for the actor and critic networks to develop congestion control actions. Second, the actor network can derive the congestion control actions according to the observed state. Meanwhile, the critic network can assess congestion control actions performance to further optimize the policy gradients and update the actions. Finally, the experience memory stores congestion control state, action, reward, and next state sequences. Random sampling from saved experience sequences then trains the representation, critic, and actor networks, thereby reducing data reliance. Through simulations using static and high-speed mobile scenarios, the proposed solution can achieve the highest goodput up to 63\% compared with other baseline algorithms.

\begin{figure}[t]
	\centering
	\includegraphics[scale=0.5]{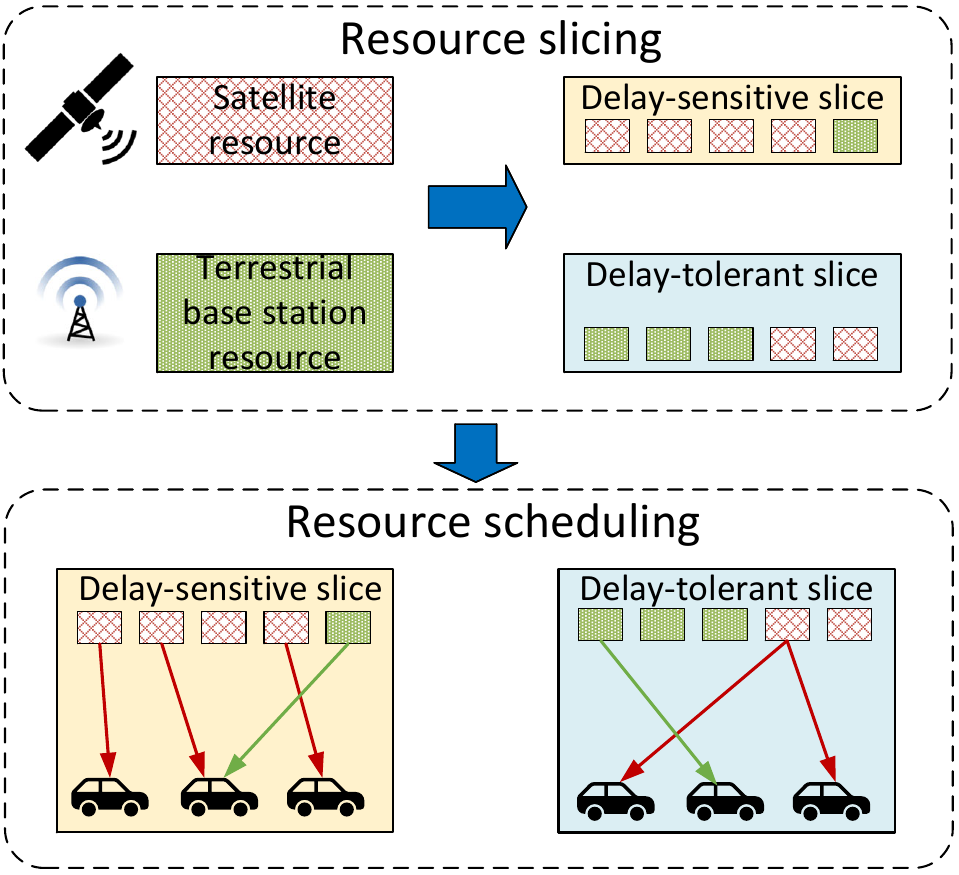}
	\caption{The resource slicing and scheduling in the integrated space-terrestrial vehicular networks.}
	\vspace{-0.5em}
	\label{fig:vehnet_1}
\end{figure}

Next, a joint resource slicing and scheduling problem to minimize the system cost in a long-term scenario in the integrated space-terrestrial vehicular networks is investigated in~\cite{sec5_wu2021learning}. Here, the system cost includes the delay-sensitive service cost, delay-tolerant service cost, and slice configuration cost. To find the solution, a two-layer RL-based approach is used, as illustrated in Fig.~\ref{fig:vehnet_1}. In the resource slicing layer, a proximal policy optimization-based RL method pre-allocates spectrum resources. Meanwhile, matching-based algorithms assign spectrum resources in each slice to each vehicle based on dynamic network circumstances and service requirements in the resource scheduling layer. From the trace-driven experiments, the proposed framework can efficiently minimize the system cost by 98\% while meeting service quality standards, compared with the proportional slicing scenario.

Then, the authors in~\cite{sec5_kim2023intelligent} foresee 6G convergent terrestrial and NTNs of virtual emotion and pandemic prevention from two perspectives that are Red AI for accuracy and Green AI for efficiency. Specifically, the Red AI-enabled 6G virtual emotion approach leveraging DL algorithms can be used to detect specific emotions of humans passing through a specified area using vehicles with high accuracy and low delay. Furthermore, the Red AI-enabled 6G epidemic prevention approach is utilized to provide epidemic services including fast medical item delivery and epidemic prevention map construction using autonomous vehicles with smart devices on the ground. Meanwhile, Green AI-enabled 6G virtual emotion focuses on computation cost reduction in detecting emotion with pre-defined accuracy. For Green AI-enabled 6G epidemic prevention using DL algorithms, it provides the same services as the Red AI however with the minimum data and reduced number of training processes and communications, thereby maximizing the efficiency. All the above are supported by the NTN via 6G communications.

\subsection{UAV}

In the next decade, B5G and 6G wireless networks that include terrestrial networks and NTN will be heavily reliant on UAVs for a variety of purposes \cite{HieuTVTBackscatter}. For that, wireless communication must be reliable, trustworthy, and inexpensive to support such huge UAV deployments. Through unlimited connectivity in 6G, UAVs for commercial use can be used widely at different altitudes from low to high ones, e.g., delivery, surveillance, traffic control, and aerial imaging.

The authors in~\cite{sec5_mozaffari2021toward} investigate antenna tilt deployment optimization of air-to-ground (A2G) network between terrestrial base stations and UAVs using a DL method, aiming at maximizing users' throughput in the air. In particular, bi-DNN is employed to approximate the behavior of the A2G network and decide the optimal network configuration. The optimal solution can be achieved by considering inter-site distance, number of antenna sectors, UAV altitude, base stations' locations, and traffic load. Through the experiment, optimal antenna tilt angles decrease by up to 30 degrees as inter-site distance increases between 20km and 80km to ensure adequate coverage throughout the cell.

In~\cite{sec5_nemati2022non}, UAV-based low altitude platform systems (LAPS) with ML/DL approaches to create a flying ad hoc network (FANET) is discussed. Specifically, Q-learning-based RL can be used to develop autonomous and adaptive packet routing among UAVs. In addition to Q-learning, SVM and logistic regression can support real-time and dynamic resource allocation to provide robust UAV services. The combination of Q-learning and FL, i.e., federated Q-learning can also be utilized to protect the FANET through jamming detection. {\color{black}A case study with ground base station and UAVs in smart farming scenarios using the opportunistic network environment shows that the proposed system can achieve more than 95\% delivery ratio and average latency 50\% lower than those of other delay tolerant networks}.

\begin{figure}[t]
	\centering
	\includegraphics[scale=0.35]{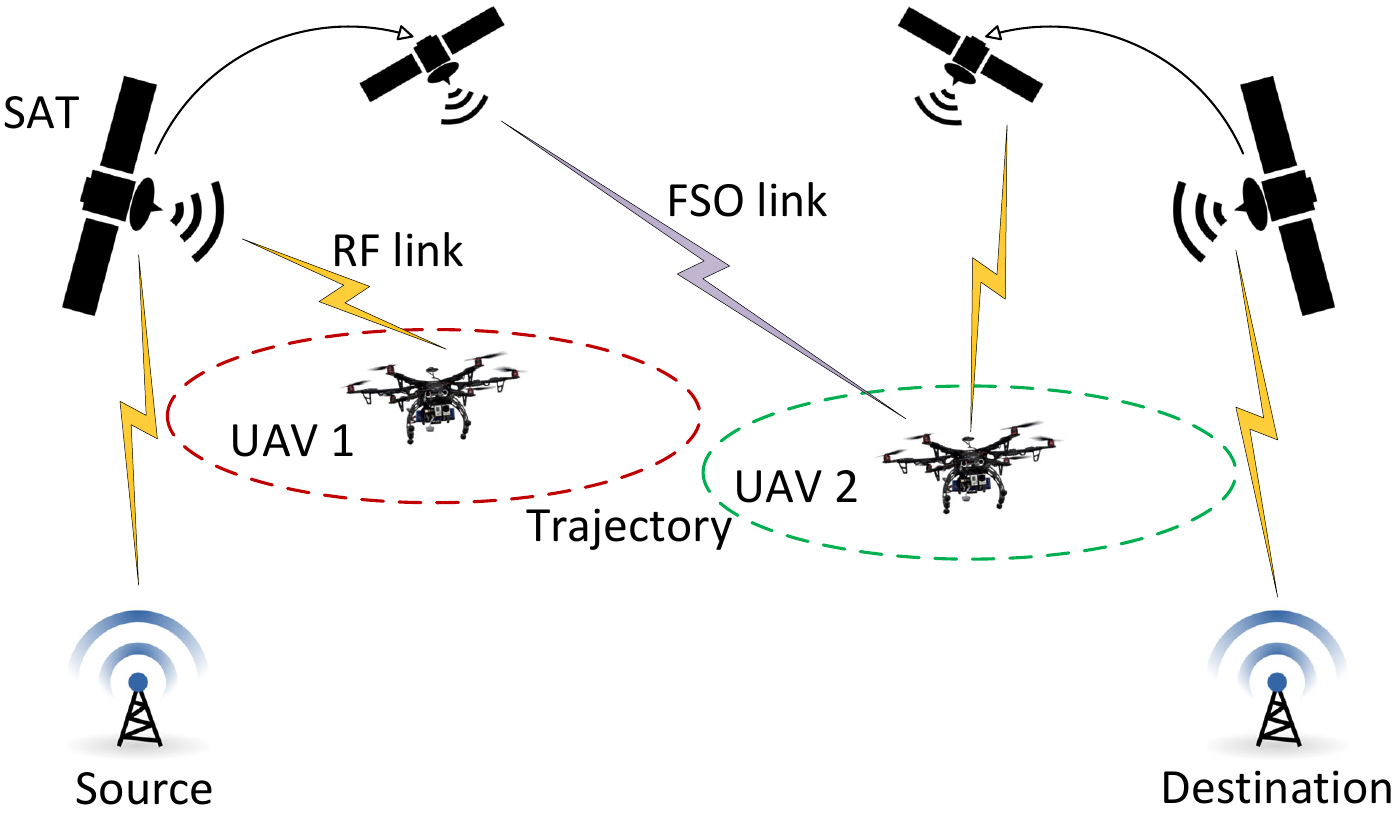}
	\caption{The integration between LEO SATs and UAVs with RF/FSO links for packet forwarding.}
	\vspace{-0.5em}
	\label{fig:uav_1}
\end{figure}

To further optimize packet forwarding between distant ground terminals, the integration between LEO satellites and UAVs to provide seamless relays using radio frequency (RF) or free-space optical (FSO) links is presented in~\cite{sec5_lee2022integrating} (as shown in Fig.~\ref{fig:uav_1}). Here, the authors propose multi-agent DRL to optimize the relationship between orbiting LEO satellites and UAVs' trajectories by maximizing the overall throughput of communication over long distances while reducing the system's energy consumption. The environment includes multi-hop communication, while the state includes LEO satellites' position in two orbital planes, the UAV's position, the link distance for each RF/FSO link, the UAV's energy consumption, and the time slot. The actions contain associations and accelerations, while the reward function optimizes the actions that maximize throughput and minimize energy consumption and distance between UAVs. Through simulations, it is shown that the proposed framework can obtain throughput two times higher than that of a baseline method with fixed ground relays. Furthermore, the energy efficiency can be improved by 2.25 times compared with the baseline method.

\subsection{Maritime Services}

Seamless marine connection service is becoming a reality as a result of recent advances in merging high-capacity and ultra-reliable integrated terrestrial network and NTN technologies. Here, NTN can boost terrestrial system coverage and enable access to marine services in offshore and non-line-of-sight (NLoS) environments. To deal with the rising complexity of controlling these interconnected systems, ML/AI-based approaches can be applied, aiming at achieving the service needs and energy efficiency objectives in diverse marine communication conditions. 

For example, the use of ML/AI-based methods to provide sustainable 6G maritime networks through NTN is investigated in~\cite{sec5_saafi2022ai}. Particularly, the most potential maritime scenarios are first described including maritime search and rescue, intelligent harbor and vessel logistics, on-board entertainment, navigation and fleet management, and shipborne IoT. In this case, distributed intelligence, wherein learning and inference are utilized at several system levels, is required to cope with the dynamic networking. Then, the authors show how ML may improve energy-efficient topology management and scheduling in dynamic marine networks over baseline model-based techniques. For the energy-efficient topology management, a multi-hop wireless network with the aid of NTN is deployed in the area of 100 km$^2$ to reduce energy usage of the network while finding the optimal routing paths for heterogeneous traffic delivery to multiple destinations, e.g., cargo vessels. For that, the DNN-based DL approach is utilized to estimate which connections are absolutely necessary for the best possible configuration as the system's traffic patterns change. As a result, the number of involving connections in the routing and execution time can be minimized. Additionally, the temporal and geographical correlations in traffic may be discovered and multi-objective optimization issues can be supported by employing other promising methods such as LSTM and auto-encoder. For the energy-efficient scheduling, the LSTM approach can be applied to predict channel quality because of its ability to deal with time series problems. Consequently, the channel reporting accuracy can be improved and packet delay can be reduced at a fixed 2 seconds for a user density of more than 2 users/m$^2$.


\subsection{Other Applications}

\begin{figure}[t]
	\centering
	\includegraphics[scale=0.5]{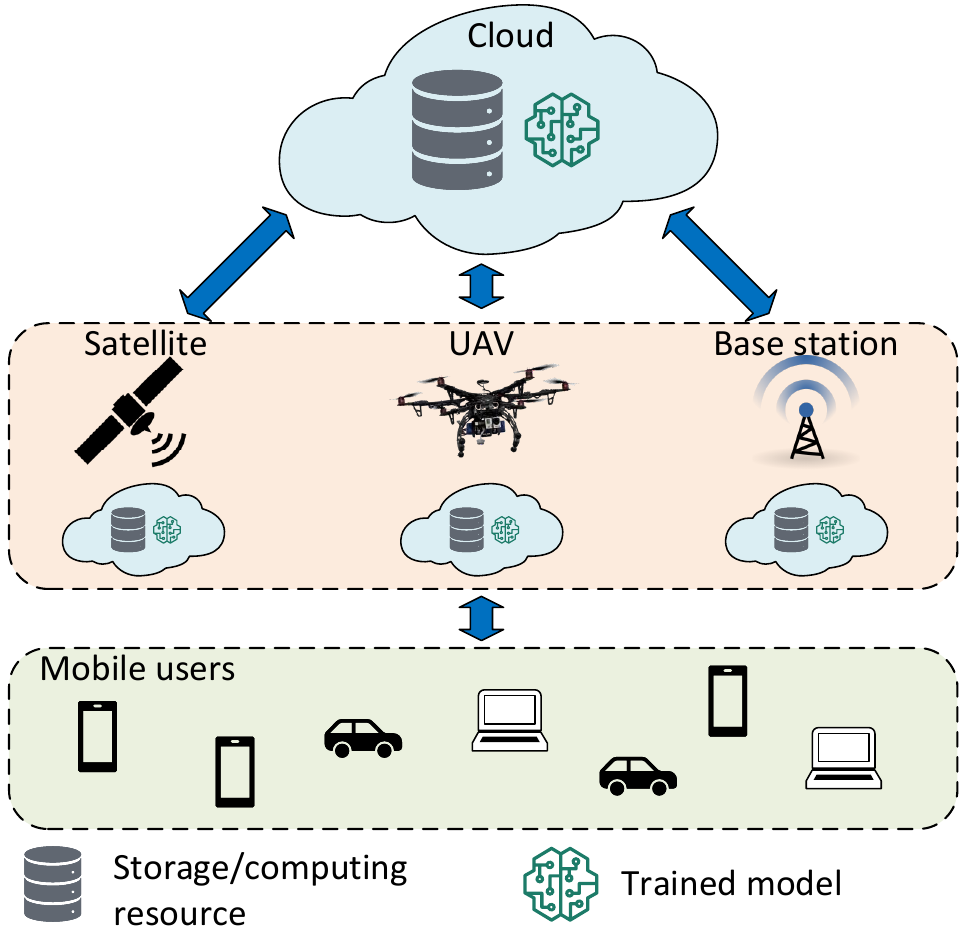}
	\caption{Federated learning using multi-level network architecture, i.e., space-air-ground networks}
	\vspace{-0.5em}
	\label{fig:other_1}
\end{figure}

Aside from all the above applications, ML/AI-aided NTN scenarios have been investigated for other emerging applications. For example, a cybertwin-enabled 6G for SAGIN using FL is discussed in~\cite{sec5_yin2022cybertwin}. Specifically, accounting for non-homogeneity of SAGIN, mobile users that are served by different RANs can offload their local data independently to train them at the respective RANs as shown in Fig.~\ref{fig:other_1}. In this case, the cloud server can work as the trained model aggregator and global model updater. Here, MNIST dataset is used in the cybertwin space of the SAGIN. Using additional helpers from SAGIN as the edge nodes, i.e., LEO satellite for the space network, a UAV for the air network, and a base station for the ground networks, an FL training process can be conducted. From the simulation results, it is shown that the satellite network suffers from the slowest training time and convergence rate due to the large communication delay between the mobile users and the satellite. The convergence gets faster when UAV-based networks and base station-based networks are utilized. Here, the accuracy for SAGIN can reach up to 90\%. To further boost the accuracy, the combination of all the edge nodes and mobile users can be used for the FL training processes.

Due to the unique features of SAGIN, e.g., time-varying connectivity, diverse resources, and complex 3-level network design, an adaptive data transmission mechanism needs to be investigated. In~\cite{sec5_liu2022a}, a DRL-based intelligent adaptive transmission framework for SAGIN is proposed, aiming at maximizing the system throughput while satisfying the packet delay and reliability standards of the traffic flow. Particularly, the authors first formulate a mixed-integer stochastic optimization problem. Then, a re-parameterization method based on a deep deterministic policy gradient (RPDDPG) algorithm is utilized to solve the problem. From the numerical results, the RPDDPG algorithm is effective in improving the throughput and outage probability compared with the relaxation-based DDPG and heuristic algorithms. Likewise, the authors in~\cite{sec5_kato2019optimizing} propose a DL-based approach to enhance the traffic control performance of the SAGIN. Particularly, several GEO, MEO, and LEO satellites as well as hundreds of UAVs are first considered. Then, online training including data collection and the training process using a CNN method is performed. From the trained model, a routing strategy to forward packets to a specific destination and select the optimal traffic path can be executed. From the simulations, the proposed method can significantly boost the network throughput by 9\% over 500 episodes, compared with the conventional routing strategy.

In~\cite{sec5_ortiz2022unsupervised}, a clustering-based users' scheduling problem using their characteristics in high-throughput and multibeam precoded GEO satellite systems is investigated. Particularly, the work considers three clustering algorithms for unsupervised learning including K-means, hierarchical clustering, and self-organization. Each of the algorithms is evaluated as a function of users' feature vectors which contain location and channel information. From the numerical results, it is observed that the channel information of the user improves the clustering and scheduling performance depending on per-beam clusters and the number of multicast users. The numerical results show that the K-means and hierarchical clustering can achieve an average throughput of 0.01Gbps higher than that of the self-organization method for 24 to 32 clusters per beam.

\emph{Summary:} In this section, we have discussed the ML/AI-aided NTN for various emerging applications/services, e.g, IoT systems, mobile services, network management, vehicular networks, UAVs, maritime services, cybertwin, and adaptive data transmission. Particularly, ML/AI approaches have emerged as one of the most potential solutions in dynamic and mobile environments where the NTN radio access network, e.g., satellites, frequently move. Additionally, ML/AI-based NTN can support integrated networks where the deployments of satellites on multiple orbits for space networks, UAVs/drones for aerial networks, and base stations for ground networks exist in such a highly dynamic environment. Here, we can observe from the above discussion that DL, RL, FL, and the combination among them can take part successfully to help NTN implementation for various services. First, the DL method is popular for end-to-end optimization, e.g., resource allocation, channel estimation, and scheduling. In this case, DNN with deep Q-network or the joint DL and RL approaches, i.e., DRL, can be utilized to reduce the complexity of the data training processes. Meanwhile, the use of CNN can be used to optimize the routing strategy of data traffic in the integrated networks. All the aforementioned ML/AI techniques adopt the centralized ML approach in which all radio accesses in the integrated network send their data to a cloud server. Alternatively, the FL approach can be used to provide a decentralized ML approach, where satellites, UAVs, ground base station, and mobile users can train their local data individually and then share the train models with the cloud server. {\color{black}Nonetheless, the existing works of ML/AI for NTN also face several drawbacks. First, ML/AI-aided NTN may increase the complexity of the overall NTN system due to the integration of various ML/AI methods and their compatibility with existing network infrastructure. Here, the next research can focus on simplifying the implementation process without compromising accuracy performance through modular design, protocol standardization, and abstraction layer introduction of ML/AI method in NTN. Second, NTN with ML/AI method can also suffer from large amounts of data that can lead to data storage, processing, and transmission problems. To address this issue, future research can explore techniques for data compression, feature selection, or data augmentation in NTN. Third, the ML/AI-based NTN may introduce computationally-intensive and time-consuming training process which can delay the deployment of NTN solutions. To this end, future research can focus on developing more efficient training algorithms, optimizing hardware accelerators, or exploring distributed training approaches to parallelize the training process. Finally, based on the existing centralized and decentralized ML/AI approaches, future research can investigate the trade-offs between these approaches in terms of privacy, latency, and scalability, and then develop hybrid approaches that combine the benefits of both centralized and decentralized ML/AI in NTN.} The summary of all ML/AI-aided NTN for various applications is presented in Table~\ref{tab:Summary_ml-ai}.




\begin{table*}[]
	\caption{Summary of ML/AI-aided NTN for Various Emerging Applications}
	\renewcommand{\arraystretch}{.5}
	\begin{tabular}{|>{\raggedright\arraybackslash}m{3.8cm}|>{\raggedright\arraybackslash}p{4.8cm}|>{\raggedright\arraybackslash}m{4cm}|>{\raggedright\arraybackslash}m{2.8cm}|}
		\hline
		\multicolumn{1}{>{\centering\arraybackslash}m{3.8cm}|}{\multirow{1}{*}{\textbf{Application}}}&
		\multicolumn{1}{>{\centering\arraybackslash}m{4.8cm}|}{\multirow{1}{*}{\textbf{Service}}} & \multicolumn{1}{>{\centering\arraybackslash}m{4cm}|}{\multirow{1}{*}{\textbf{ML/AI method}}}&
		\multicolumn{1}{>{\centering\arraybackslash}m{2.8cm}|}{\multirow{1}{*}{\textbf{Refs}}}\\
		&&&\\
		\hline 
		\hline 
		\multirow{10}{*}{IoT} & Resource allocation, energy consumption, trajectory optimization & DNN, DRL, CNN & \cite{sec5_michailidis2020ai} \\ \cline{2-4}
		& Power and channel allocation, latency minimization & DRL & \cite{sec5_vaezi2022cellular} \\ \cline{2-4}
		& Routing strategy optimization, spectrum sharing & CNN, DQN, SVM & \cite{sec5_zhu2021integrated} \\
		\hline
		\multirow{4}{*}{Mobile services} & Handover and throughput optimization & DRL & \cite{sec5_cao2021deep} \\ \cline{2-4}
		& Trajectory forecast & SR distributive Q-learning & \cite{sec5_lien2022autonomous} \\
		\hline
		\multirow{10}{*}{Network management} & Resource utilization and allocation & DRL & \cite{sec5_deng2019next, sec5_qiu2019deep} \\ \cline{2-4}
		& Network capacity and coverage prediction & FL, online learning & \cite{sec5_darwish2021vision} \\ \cline{2-4}
		& Network slicing & Transfer and meta learning & \cite{sec5_lei2021dynamic}  \\ \cline{2-4}
		& Network slicing and resource allocation & WR-NN & \cite{sec5_de2020qos}  \\ \cline{2-4}
		& Resource allocation & FL & \cite{sec5_li2020distributed}  \\
		\hline
		\multirow{5}{*}{Vehicular networks} & Congestion control & DRL, CNN & \cite{sec5_xu2021deep} \\ \cline{2-4}
		& Resource slicing and scheduling & RL & \cite{sec5_wu2021learning} \\ \cline{2-4}
		& Virtual emotion and pandemic prevention & DL & \cite{sec5_kim2023intelligent} \\
		\hline
		\multirow{10}{*}{UAVs} & Antenna tilt deployment optimization & Bi-DNN & \cite{sec5_mozaffari2021toward} \\ \cline{2-4}
		& Autonomous and adaptive packet routing, dynamic resource allocation, jamming detection & RL, SVM and logistic regression, Federated RL & \cite{sec5_nemati2022non} \\ \cline{2-4}
		& Packet forwarding & DRL & \cite{sec5_lee2022integrating}  \\ \cline{2-4}
		\hline
		Maritime services & Energy-efficient topology management and scheduling & DNN, LSTM & \cite{sec5_saafi2022ai} \\ 
		\hline
		\multirow{8}{*}{Other applications} & Cybertwin-based image classification & FL & \cite{sec5_yin2022cybertwin} \\ \cline{2-4}
		& Adaptive data transmission & DRL & \cite{sec5_liu2022a} \\ \cline{2-4}
		& Optimal routing path & CNN & \cite{sec5_kato2019optimizing} \\ \cline{2-4}
		& Clustering-based scheduling problem & K-means, hierarchical clustering, and self-organization & \cite{sec5_ortiz2022unsupervised} \\
		\hline
	\end{tabular}
	\label{tab:Summary_ml-ai}
	
\end{table*}

\section{ORAN-aided NTN }\label{sec:oran}



\begin{figure*}[t]
	\centering
	\includegraphics[height=7cm,width = 12cm]{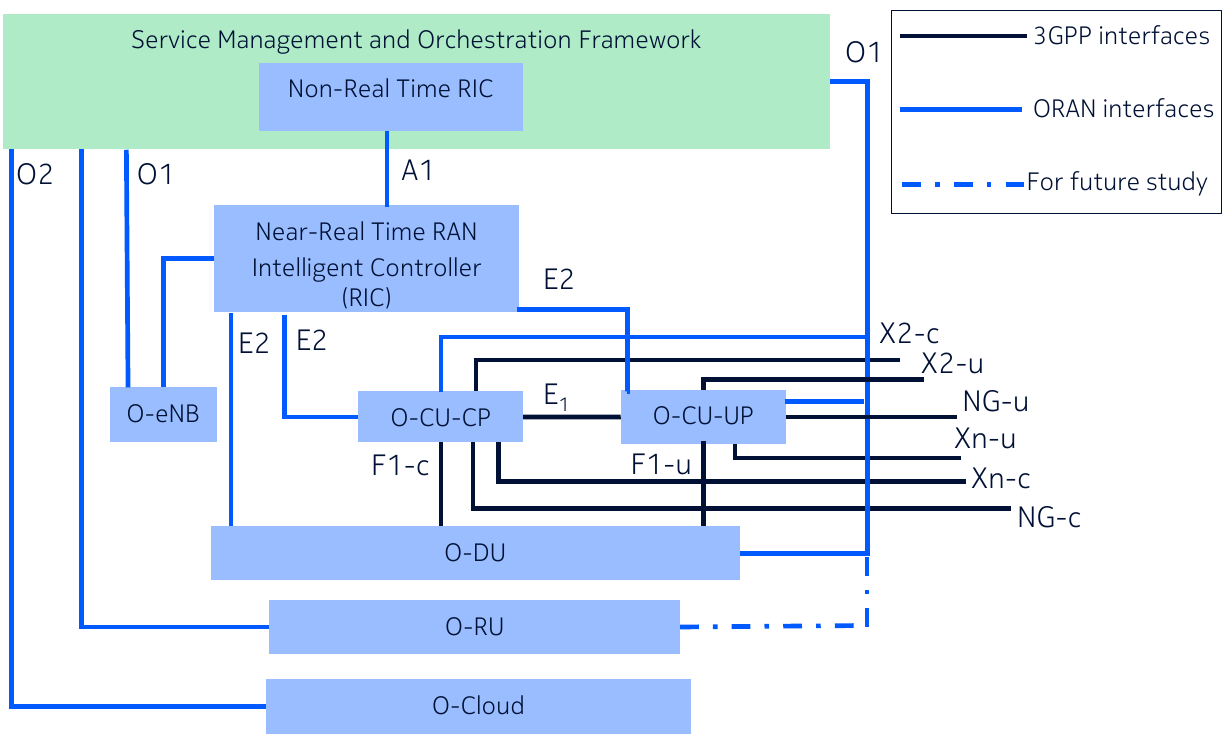}
	\caption{ORAN architecture}
	\label{FigORANArch}
\end{figure*}
\begin{table*}[t]
	\caption{ORAN interface notations}
	\label{Table_ORAN_interface}
	\centering
	\setlength{\tabcolsep}{5pt}
	\begin{tabular}{|>{\centering\arraybackslash} m{2cm} | >{\centering\arraybackslash} m{12cm}| >{\centering\arraybackslash} m{2cm}|}
		\hline \hline
		\textbf{Interfaces} & \textbf{Description} & \textbf{References} \\  \hline\hline
		E2&  This interface establishes a connection between the near-RT RIC and multiple O-CU-CPs, O-CU-UPs, and O-DUs. & \cite{ORAN-1}\\ \hline
		E1&  Interface linking the SMO with O-RAN managed elements & \cite{ORAN-1} \\ \hline O1&  The interface facilitating communication between management entities (NMS/EMS/MANO) and O-RAN managed elements enables operational and managerial functions. This encompasses FCAPS management (fault, configuration, accounting, performance, and security), software management, and file management & \cite{ORAN-1} \\ \hline O2&  Interface between the SMO and the O-Cloud & \cite{ORAN-sli-arc} \\ \hline
		A1&  The interface linking the non-RT RIC and Near-RT RIC facilitates policy-driven guidance of Near-RT operations  & \cite{ORAN-sli-arc} \\ \hline F1-u&  F1 User Plane interface  & \cite{38470} \\ \hline  F1-c&  F1 Control Plane interface  & \cite{38470} \\ \hline
		NG-c&  The next generation (NG) control plane interface is defined between an NG-RAN node and a 5GC  & \cite{38410} \\ \hline NG-u&  The NG user plane (NG-U) interface is established between an NG-RAN node and a UPF. It ensures the non-guaranteed delivery of PDU Session/MBS session user plane PDUs between the NG-RAN node and the user plane function (UPF).  & \cite{38410} \\ \hline Xn-u&  The Xn user plane (Xn-U) interface is established between two NG-RAN nodes. It facilitates the non-guaranteed delivery of user plane PDUs between these two NG-RAN nodes  & \cite{38420} \\ \hline  Xn-c&  Reference point for the control plane protocol interconnecting NG-RAN nodes & \cite{38420} \\ \hline X2-c&  Control plane interface for the exchange of application-level configuration data required for seamless interoperability between eNB and en-gNB & \cite{36423} \\ 
		\\ \hline \color{black}{Open FH M-Plane}& \color{black}{Open fronhaul M-Plane interface which is a management interface controlling O-RU}  & \cite{ORANwg1} \\ 
		\hline \color{black}{NGAP} & \color{black}{Control plan protocol of 5G} & \cite{128413}\\
		\hline \color{black}{GTP} & \color{black}{GTP tunnel to control the UE's data traffic} & \cite{128413} \\
		\hline \color{black}{NR Uu} & \color{black}{Satellite radio interface (SRI) between satellite and UE/NTN gateway.} & \cite{38.821}\\
		\hline \color{black}{NG over SRI} & \color{black}{Satellite radio interface between satellite and NTN gateway (NTNGW) in the regenerative satellite scenario.} & \cite{38.821}\\
		\hline 
		\color{black}{NG } & \color{black}{Next generation interface between satellite and gNB/5G core network (5G CN) in the transparent/regenerative satellite scenario. Besides, it is also the interface between gNB and 5G CN in the transparent satellite case.} & \cite{38.821}\\
		\hline \color{black}{N6 } & \color{black}{It is the interface connecting 5G CN to the data network.} & \cite{38.821}\\
		\hline \hline
	\end{tabular}
	\label{tab2}
\end{table*}

\begin{figure*}[t]
	\centering
	\includegraphics[height=8cm,width = 18cm]{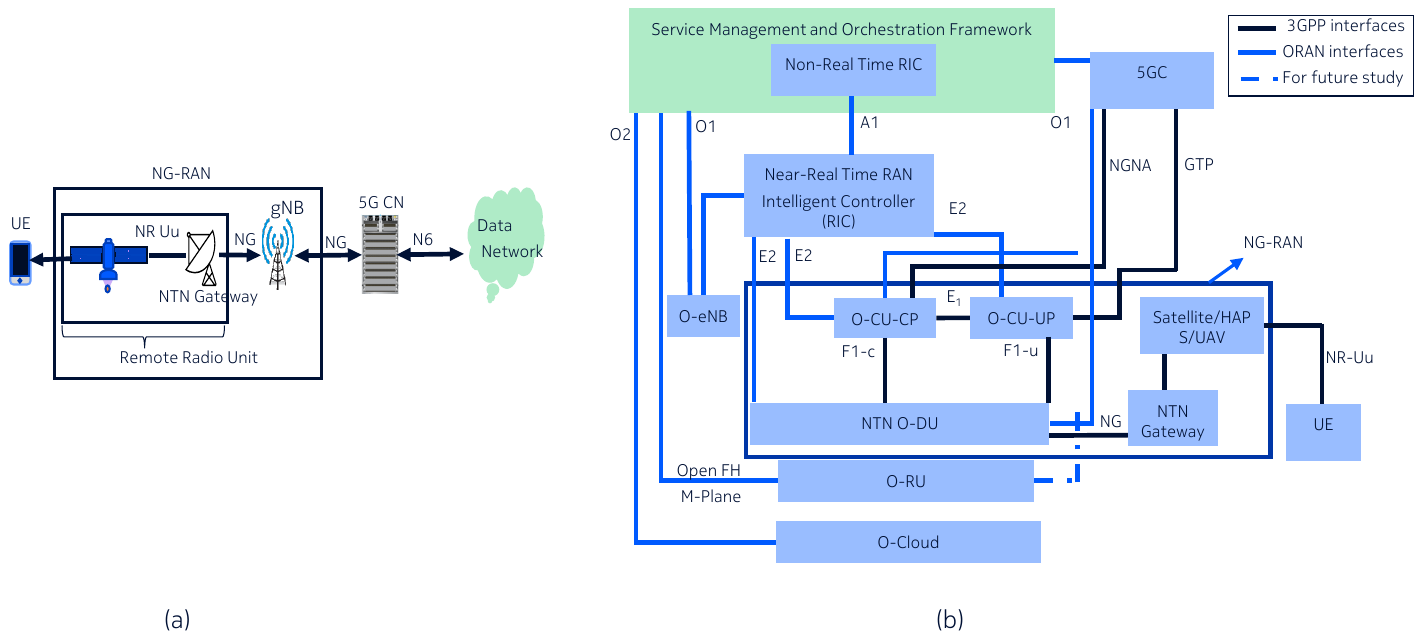}
	\caption{\color{black}{ORAN-aided NTN architecture for the transparent NTN and gNB on Earth.} }
	\label{FigNTNORAN1}
\end{figure*}

\begin{figure*}[t]
	\centering
	\includegraphics[height=8cm,width = 18cm]{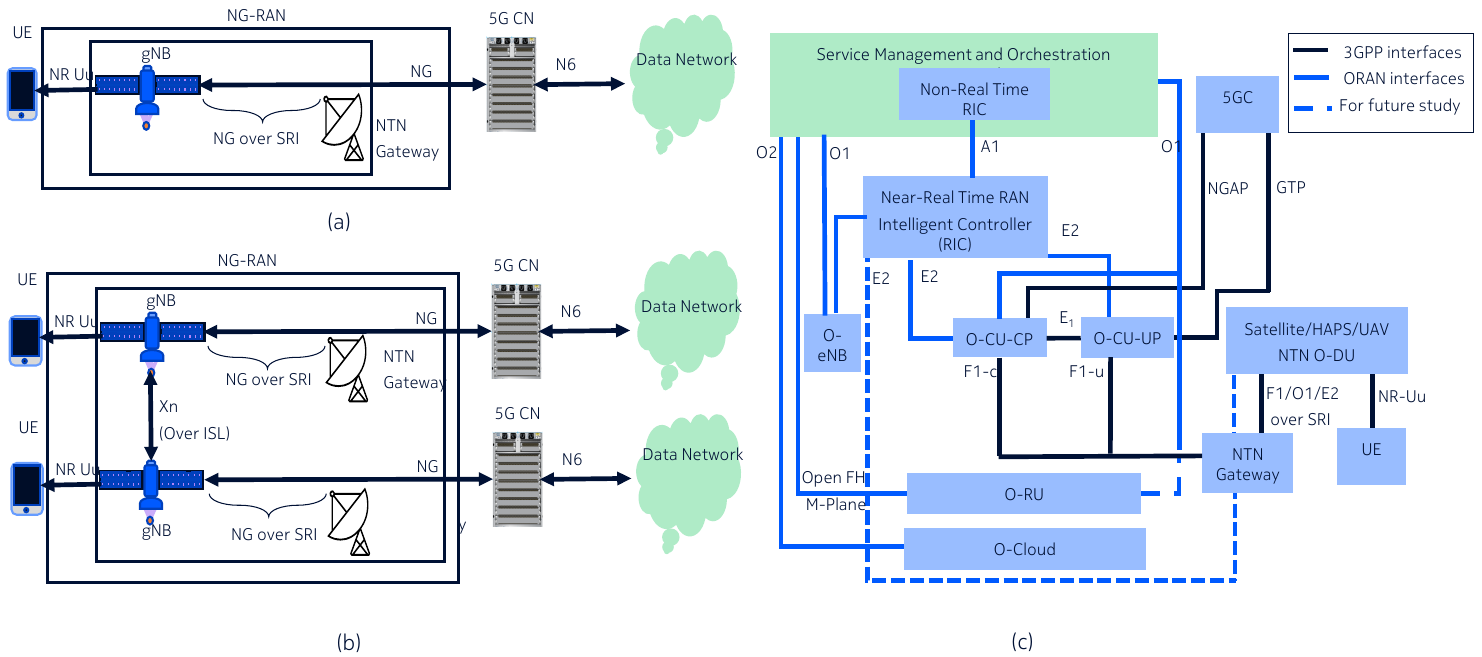}
	\caption{\color{black}{ORAN-aided NTN architecture for the regenerative NTN with gNB onboard.} }
	\label{FigNTNORAN2}
\end{figure*}

\begin{figure*}[t]
	\centering
	\includegraphics[height=8cm,width = 18cm]{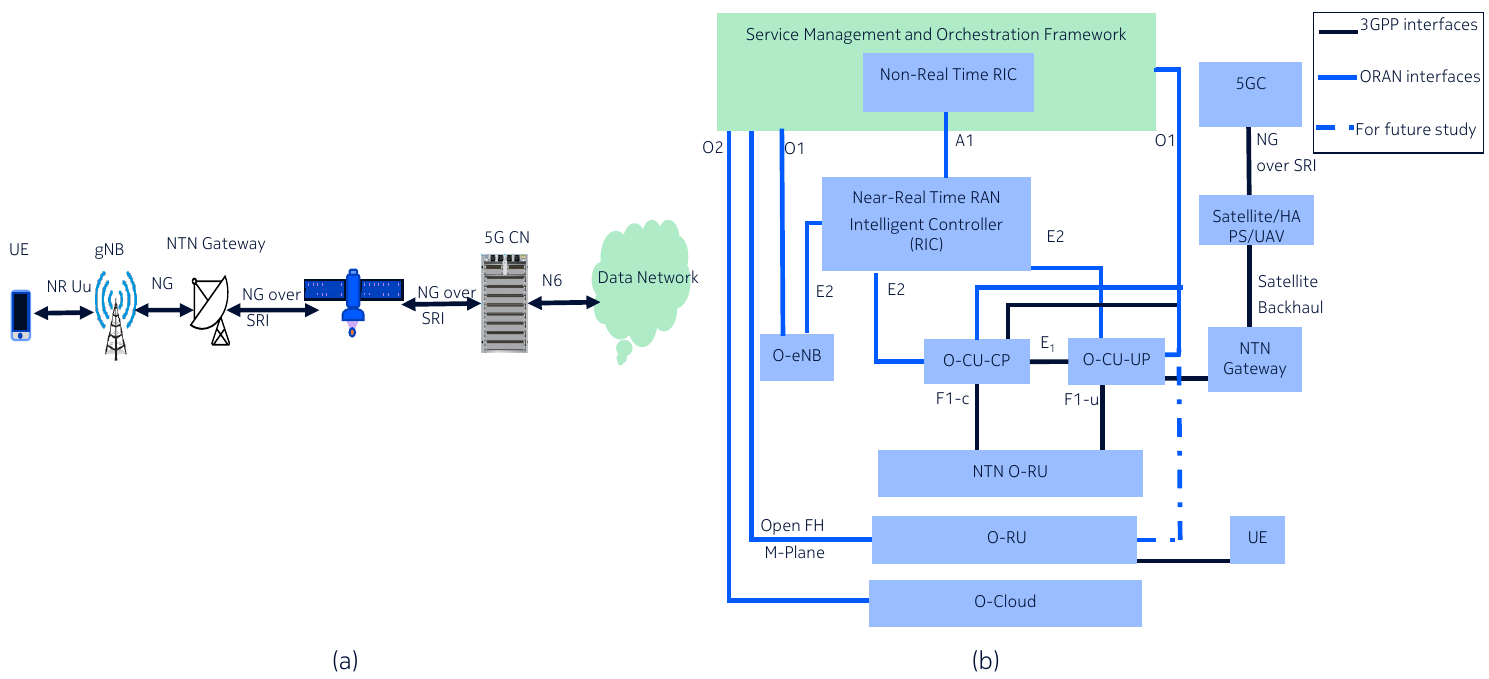}
	\caption{\color{black}{ORAN-aided non-3GPP NTN architecture.} }
	\label{FigNTNORAN3}
\end{figure*}

\begin{figure}[t]
	\centering
	\includegraphics[height=6cm,width = 6cm]{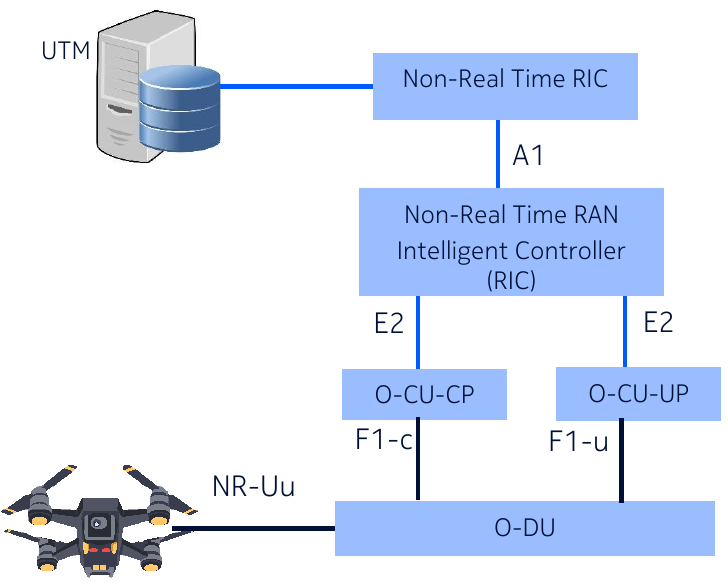}
	\caption{UAV trajectory path design based on ORAN-aided UAV radio resource allocation \cite{ORAN-use-cases}. }
	\label{FigUAVRRA}
\end{figure}
\begin{figure*}[t]
	\centering
	\includegraphics[height=8cm,width = 14cm]{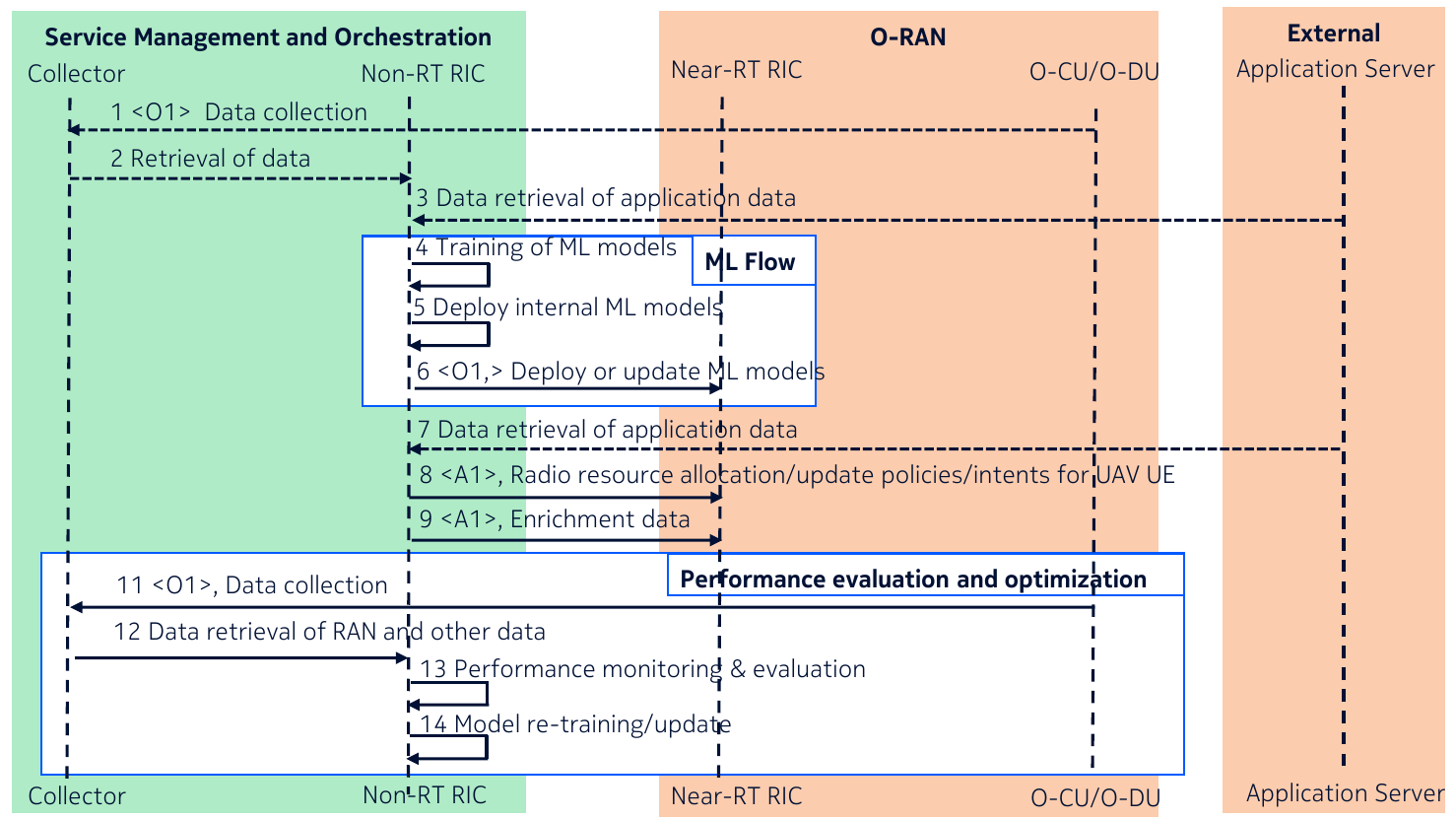}
	\caption{Flow Diagram for UAV trajectory path design based on ORAN-aided UAV radio resource allocation \cite{ORAN-use-cases}. }
	\label{FigUAVRRAWorkflow}
\end{figure*}

As open networks move the Radio Access Network (RAN) to the cloud, an intriguing alternative is venturing into space. The concept of NTN has garnered increasing attention, with numerous experts asserting that this innovative approach holds the potential to extend 6G coverage ubiquitously \cite{CaioNTNORAN}. Nevertheless, there is currently a scarcity of research in this domain. Consequently, this section delves into the potential utilization of ORAN for NTN. {\color{black}{In this section, we describe the 3GPP and non-3GPP-based NTN architectures. Our purpose is to provide a comprehensive overview of the NTN architectures from both industrial and academic perspectives. Subsequently, based on these architectures, we will demonstrate how ORAN connects to the NTN system according to each specific one. In particular, this work provides detailed information about the NTN architecture in both 3GPP and non-3GPP environments. Additionally, it offers insights into components within the ORAN system that can connect to satellites, which were previously not clearly addressed. This comprehensive perspective aids researchers in gaining a complete insights into the subject.  We commence by examining the advantages of ORAN, followed by the presentation of a general ORAN application architecture. Then, we present each NTN architecture and the corresponding ORAN-based NTN architecture. More specifically, we show two main NTN architectures in 3GPP: transparent and regenerative NTN systems in Release 16 of 3GPP\cite{38.821}. Then, we show 1 more NTN architecture in non-3GPP scenario. Based on each NTN system, we detail which components in ORAN can connect to the NTN system, which has not been covered before. This can provide an easy-to-understand perspective for academic researchers instead of reading jargon-filled and confusing industry documents.  }}
\subsection{Benefits of ORAN for 5G and beyond}

ORAN plays a crucial role in 5G and beyond since it provides several benefits as follows:
\begin{itemize}
	\item Cost reduction and network resiliency: ORAN has the potential to reduce costs for operators by opening the RAN ecosystem to multiple vendors. Therefore, operators have more options to select cost-effective solutions suitable to their specific requirements. ORAN also leverages off-the-shelf hardware and cloud-based deployment, thereby lowering costs. This diverse vendor ecosystem not only saves costs but also enhances network resiliency. By using equipment from different vendors, operators can reduce the risk of being dependent on one vendor and better protect their networks against service disruption or potential vulnerabilities.
	\item Scalability, flexibility, and interoperability between vendors: ORAN allows operators to easily scale up/down their networks through disaggregation and hardware and software supply chain decoupling. Hereby, disaggregation refers to the separation of traditional monolithic RAN components such as baseband processing, radio functions, and control functions, into independent and functional entities/modules that can be developed, deployed, and operated independently. For example, the control unit (CU) is responsible for centralized functions, i.e., baseband processing, control plane processing, and network management functions. The distributed unit (DU) is responsible for radio functions such as physical layer processing (PLP), radio frequency processing (RFP), and connection management. By disaggregating the RAN components, ORAN offers flexibility and vendor diversity to the infrastructure. Therefore, the operator can select the best solutions from different vendors for each functional component, promoting competition and innovation in the market.  It also leverages interoperability between different vendors' devices, and thus, it is easier to integrate and upgrade network components.
	\item Innovation, forward-looking, and technology convergence: ORAN encourages innovation by allowing start-ups and niche vendors to introduce their novel solutions and technologies into the RAN domain. It fosters a vibrant ecosystem that helps drive advancements in VR, MEC, AI, digital twins, semantic communications, haptic communications, and network automation. ORAN architecture interfaces can facilitate the integration and interoperability of the above-mentioned technologies.
	\item Security enhancement: With ORAN disaggregated architecture, security checks and verifications can be performed more efficiently. Moreover, operators can analyze and evaluate the security of each component, including CUs and DU. This level of granularity allows for better identification and mitigation of potential vulnerabilities. Another reason for the security enhancement is that ORAN supports network slicing, allowing operators to create virtualized networks dedicated to specific use cases. Each network slice can have its security policies and mechanisms, creating separate security zones. This allows for custom security configurations for different services and applications, ensuring that appropriate security levels are maintained based on the specific requirements of each slice.
\end{itemize}

\subsection{ORAN architecture}
Based on the above-mentioned benefits of ORAN for 5G and beyond, we introduce ORAN architecture as shown in Fig.~\ref{FigORANArch}. Fig.~\ref{FigORANArch} represents different components of ORAN such as service management and orchestration, near-real-time RIC, open evolved NodeB (O-eNB), O-RAN Central Unit-Control Plane (O-CU-CP), Open Distributed Unit (O-DU) and Open Radio Unit (O-RU), and O-Cloud (a cloud computing platform comprising physical infrastructure nodes to host O-RAN functions) \cite{alliance2021ran}. First, O-Cloud is a cloud-computing platform consisting of a set of physical infrastructure components satisfying the ORAN requirements to host related ORAN functions such as O-CU-UP, O-CU-CP, O-DU, and near-RT RIC. Moreover, O-Cloud supports three software components, i.e., virtual machine monitor, operating system (OS), and container runtime. Second, O-eNB enables O-RU and O-DU with an Open Fronthaul Interface (OFI) between them. Third, an Open radio unit (O-RU ) is used to terminate the OFI and LOW-PHY functions (i.e., Fast Fourier Transform (FFT), inverse Fast Fourier Transform (iFFT), and physical random access channel (PRACH) extraction.) of the radio interface (RI) to the UE. Moreover, O-RU also terminates the OF M-Plane Interface (OFMPI) to the Open distributed unit (O-DU) and Service Management and Orchestration (SMO). Fourth, O-DU is used to terminate OFI, F1, and E2 interfaces, together with the radio link control (RLC), medium access control (MAC), and HIGH-PHY functions of the RI to the UE. O-DU also terminates the OFMPI to the O-RU and O1 interface to the SMO. Fifth, the Oran Control Unit User Plane (O-CU-UP) helps to terminate the F1-u, X2-u, S1-u, Xn-u, E1, 01, and NG-u interfaces. Besides, it also terminates the Service Data Adaption Protocol (SDAP) and Packet Data Convergence Protocol (PDCP) protocols. Sixth, the ORAN control unit control plane helps to terminate the X2-c, F1-c, NG-c, Xn-c, E2, O1 interfaces, PDCP protocol, and radio resource control (RRC) protocol. Seventh, the near-real-time (RT) RAN intelligent control (RIC) supports RT control and optimization for E2 functions and actions over the E2 interface with latency between 10 ms to 1 second. Lastly, the SMO helps to manage the O-Clouds and support the orchestration of platform, and workflow management. The following functionalities are supported by SMO, i.e., software management of deployments; deployments and allocated O-Cloud resources; creating, deleting, and associating O-Cloud resources; software management of cloud platform; administration of O-Cloud resources.

In the description above, we provide an overview of each ORAN function but it would be remiss if we did not mention the interface between them. The details of interfaces in ORAN architecture can be described in Table \ref{Table_ORAN_interface}.

\subsection{\color{black}ORAN-aided 3GPP NTN architecture}
{\color{black}In this subsection, we describe about 3GPP-based NTN architectures \cite{38.821}. Then, we outline the specific ORAN components that can interface with it, a perspective not previously presented in detail. In 3GPP Release 16, two primary NTN architectures are listed: transparent and regenerative NTN systems \cite{38.821}. Consequently, we'll illustrate these two NTN architectures alongside their respective ORAN counterparts.}
\subsubsection{\color{black}ORAN-aided NTN architecture for transparent satellite, without gNB onboard}
{\color{black}{Fig. \ref{FigNTNORAN1} a) shows the transparent NTN system whereas the satellite acts as a relay between the NTN ground gateway and the UE \cite{38.821}. More specifically, the satellite plays the roles of RF amplifier and frequency conversion\color{black} \cite{38.821,han2022general}. Besides, the satellite can replicate the radio interface, transmitting signals from the feeder link (connecting the NTNGW and the satellite) to the service link (linking the satellite and the UE), and vice versa. The interfaces in a transparent NTN system can be described as follows:}}

{\color{black}{\begin{itemize}
			\item The UE connects to the satellite through NR Uu interface. Moreover, the NR Uu interface will also be used to connect between satellite and NTNGW and from NTNGW to gNB. 
			\item The gNB connects to the 5G CN through next generation (NG) interface.
			\item N6 interface is used to connect between 5G CN to the data network.    
\end{itemize} }} 

{\color{black}{In Fig.~\ref{FigNTNORAN1} b), we show a scenario with transparent NTN while gnB is assumed to be allocated on the Earth \cite{38.821}. Therefore, it is assumed that the NTN O-DU is located on Earth and is connected to the NTNGW through the next generation (NG) interface \cite{han2022general}. The UE then connects to the satellite through NR-Uu interface \cite{han2022general}. Moreover, the O-CU-CP and O-CU-UP have NGNA and GTP interfaces with 5GC, respectively.}}

\subsubsection{ORAN-aided NTN architecture for regenerative satellite, with gNB onboard} 

{\color{black}{Fig. \ref{FigNTNORAN2} a) and b) present the regenerative NTN systems in which the satellite is equipped as a base station \cite{38.821}. In Fig. \ref{FigNTNORAN2} a), we depict the scenario where only a single satellite connects to the NTNGW. Fig. \ref{FigNTNORAN2} b) presents a generalized version of Fig. \ref{FigNTNORAN2} a), incorporating an inter-satellite link (ISL) interface between satellites. The interfaces in regenerative NTN system can be described as follows:}}

{\color{black}{\begin{itemize}
			\item The UE connects to the satellite through NR-Uu interface. 
			\item The Xn interface, operating over the ISL, facilitates connections between satellites. The ISL can utilize either a radio interface (RI) or an optical interface (OI) and may conform to either 3GPP or non-3GPP standards.
			\item The regenerative satellite connects to the NTNGW through the NG satellite radio interface (SRI), which serves as a transport link. Moreover, the NTNGW is a transport network layer (TNL) node capable of supporting all transport protocols.
			\item N6 interface is used to connect between 5G CN to the data network.    
\end{itemize} }}
\color{black}
In Fig.~\ref{FigNTNORAN2} c), {{we show a novel architecture for the regenerative satellite \cite{campana2023ran}. To the best of our knowledge, this is the first proposed architecture that shows how ORAN can work with NTN systems \cite{campana2023ran}.}} Specifically, the gNB is equipped with a satellite. {\color{black}In this scenario, the NTNGW connects directly to the O-CU-UP and O-CU-UP through the F1-c and F1-u interfaces, respectively.} Besides, the NTNGW also connects with non-real-time RIC, and the in-band controls such as E2, F1-c, and F1-u. {\color{black}Then, the NTNGW connects to the satellite through F1/O1/E2 over SRI interface.} In this case, the satellite is assumed to be equipped with NTN O-DU. The UE also connects to the satellite through NR-Uu as in the transparent satellite scenario.

\subsection{\color{black}ORAN-aided non-3GPP NTN architecture} {\color{black}{In this subsection, we describe one more NTN architecture that is not agreed and listed in 3GPP documents \cite{38.821}. Then, we show in detail which ORAN components can connect to the NTN system. } }

{ \color{black}{In Fig. \ref{FigNTNORAN3} a), we illustrate a scenario in which the gNB cannot directly connect to the 5G CN \cite{drif2021slice}. This may occur if the gNB is deployed in an isolated area without an optical link to the 5G CN, or if the connection is disrupted due to a disaster or accident. The interfaces in a transparent NTN system can be described as follows:}}

{{\color{black}{\begin{itemize}
				\item The UE connects to the gNB through NR Uu interface. 
				\item The gNB connects to the NTNGW through the NG interface.
				\item The satellite connect to the NTNGW and 5G CN through NG over SRI interface.
				\item N6 interface is used to connect between 5G CN to the data network.    
\end{itemize} }} } 

{\color{black} In In Fig. \ref{FigNTNORAN3} a), we detail how ORAN components connect to the satellite. In this scenario, the satellite plays as a relay between NTNGW and the 5GC. Therefore, it provides a satellite backhaul link to the NTNGW. Moreover, the satellite connects to the 5GC through NG over SRI interface. As an illustration, let us consider an example to facilitate understanding. Initially, let's assume there exists an optical link between the 5GC and the gNB. However, in the event of a disaster, this link is destroyed. To expedite the rescue mission in the isolated area, satellites can serve as relays to transmit information from the 5GC to the gNB located in this isolated region.}
\subsection{ORAN-aided UAV trajectory design based on radio resource allocation }
In this section, we describe a case where a UAV trajectory flight path is based on the UAV radio resource allocation, which helps operators fine-tune their radio resource policies in the ORAN architecture \cite{ORAN-use-cases}. Since the location along the trajectory is mainly focused on GUEs, the UAV does not always belong to the main lobe of the ground base station (GBS). Moreover, the side lobes of the GBS antenna lead to the scattered cell association phenomenon, especially in the sky. The cell pattern on the ground is ideally a contiguous area whereas the best cell is usually the cell closest to the UE. When the UE moves upward, the side lobes of the antenna start to show up and the best cell may no longer be the closest one. The connectivity of the cell in this case becomes discrete, particularly at an altitude of 300m or more. 

As shown in Fig. \ref{FigUAVRRA}, the NRT RIC can retrieve essential UAV-related measurements from the network based on UE and SMO reports, as well as UAV trajectory information, aerial load information, and climate information. For example, unmanned traffic management (UTM) is used to build/train relevant AI/ML models deployed in RAN. This could be uplink/downlink interference from the UAVs, UAV detections, and prediction of available radio resources such as bandwidth, frequency, cell, and beam. Based on this information, the NRT RIC can support the building and execution of AI/ML models from Non-real-time RIC. Moreover, the NRT RIC can make radio resource allocation for on-demand coverage of UAVs taking into account the trajectory and radio channel information.

{\color{black}Recently, there are some works that investigate how ORAN meets UAV communications \cite{MitsuiORANUAV,BetaloORANUAV,ChuanORANUAV,LorenzoORANUAV}. In \cite{MitsuiORANUAV}, Mitsui et al. tried to solve the fairness problem from ground UEs to UAV by optimizing the UAV's speed and allocated bandwidth. Particularly, the authors proposed iterative algorithms that run on ORAN architecture to solve the above-mentioned problem. More specifically, they first fixed the allocated bandwidth and optimized the UAV's flying speed. Then, they used the optimized UAV's flying speed to optimize the allocated bandwidth. This algorithm was stopped when it reached the saturation point. In \cite{BetaloORANUAV}, Betalo et al. tried to minimize the UAV's energy consumption by optimizing task scheduling, trajectory design, and resource allocation in wireless sensor networks (WSNs). As the problem is NP-hard, they designed a multi-agent deep reinforcement learning (MADRL) algorithm to solve it, and this algorithm is deployed on the ORAN system. \cite{ChuanORANUAV} is the first work that investigated the benefits of multi-UAV in ORAN architecture. More specifically, the authors aimed to maximize network utility, i.e., data rate and computing resources, by optimizing the UAVs' flight paths and the allocated offloading tasks in ORAN architecture. Different to other works that investigated from the theoretical perspective, Lorenzo et al. \cite{LorenzoORANUAV} conducted an experimental study on drone video streaming applications on ORAN architectures. More specifically, the authors designed a control system for UAV positions and transmission directionality to improve the total network performance, i.e., the UE's uplink throughput. Specifically, they prototyped the proposed solution in a real testbed.     }

\subsection{Flow diagram for ORAN-aided UAV trajectory design based on radio resource allocation }

In Fig.~\ref{FigUAVRRAWorkflow}, we describe a flow diagram for an ORAN-aided UAV trajectory design based on radio resource allocation using ML models \cite{ORAN-use-cases}. In step 1, the collector in the SMO performs data collection from O-CU/O-DU via the O1 interface. Then this data is transmitted to the non-RL RIC in step 2. In step 3, the application server transmits the application data to non-RT RIC. From steps 4 to 6, the ML flow is performed. In particular, non-RT RIC trains and deploys the ML models in steps 4 and 5, respectively. In step 6, non-RT RIC deploys or updates ML models in near-RT-RIC through the O1 interface. In step 7, the application data is transmitted from the application server to non-RT RIC. In steps 8 and 9, non-RT RIC sends radio resource allocation, updated policies, intents, and enrichment data to near-RT RIC via the A1 interface. From steps 11 to 14, the performance evaluation and optimization are performed. In step 11, the collector collects data from O-CU/O-DU and then transmits it to the non-RT RIC in step 12. In step 13, the non-RT RIC executes the performance monitoring and evaluation. Finally, the process of re-training or updating the model is performed in step 14.

\subsection{\color{black}Summary}
{\color{black} In this section, we have discussed multiple approaches for ORAN-aided NTN, i.e., benefits of ORAN for 5G and beyond, ORAN architecture, ORAN-aided 3GPP NTN architecture (ORAN-aided NTN architecture for transparent satellite and ORAN-aided NTN architecture for regenerative satellite), ORAN-aided non-3GPP NTN architecture, ORAN-aided UAV trajectory design based on radio resource allocation. Different to the traditional monolithic RAN architecture, ORAN provides many advantages such as cost reduction, scalability, interoperability between vendors, and disaggregation. It also brings new challenges and opportunities for researchers due to the architecture of ORAN is totally different from the RAN. Some potential research directions of ORAN-aided NTN can be described as follows: }
\begin{itemize}
	\item {\color{black} \textit{Design ORAN-NTN routing path:} Designing ORAN-NTN routing paths to efficiently support the heterogeneous requirements of UEs in each area, with different latency and rate requirements and in a large coverage area, has become essential to satisfy the UE's QoS. This is especially important in emergency/disaster scenarios or when the terrestrial network becomes overloaded. }
	\item {\color{black} \textit{Real time applications:} How to design efficient algorithms in an ORAN-NTN system to support real-time (e.g., latency requirement of less than 1 ms), near-real-time (e.g., latency requirement from 10 ms to 1 second), and non-real-time RIC (e.g., latency requirement greater than 1 second) requirements at UEs is a complex issue. Depending on different data rates and latency requirements at UEs, we need to allocate resources and design routing paths efficiently. This problem becomes tricky in large-scale scenarios in practice, e.g., more than 1000 BSs and hundreds of thousands of UEs. }
	\item {\color{black} \textit{AIML for ORAN control:} In order to utilize ORAN efficiently on a large scale and automatically, AIML can provide efficient solutions by predicting the incoming traffic of the network. Therefore, it offers good solutions to deal with dynamic control and varying traffic requirements. } 
	\item {\color{black} \textit{Security for ORAN-NTN:} It is necessary to design, maintain, develop, and test algorithms to improve the security of the ORAN-aided NTN system, especially since the large distance between NTN and ORAN creates high transmission delays, making it vulnerable to attacks }
	\item {\color{black} \textit{Energy efficiency:} One of the biggest concerns of the operators is how to reduce the energy consumption at the BSs, which contributes to 2-3$\%$ of greenhouse gas emissions and is a major operational expense. Therefore, designing efficient algorithms to operate the ORAN-aided NTN system has become an urgent need.   } 
	
	\item {\color{black} \textit{ORAN-based NTN network slicing:}  How we enable ORAN NTN network slicing to support diverse applications with varying latency and rate requirements. How we design routing paths from NTN through different ORAN components to meet the UE's QoS requirements with limited resources. }
	
	\item {\color{black} \textit{NTN-based ORAN digital twin:} How to enable ORAN NTN digital twin for real-time, near real-time, and non-real-time in different NTN architectures, such as transparent and regenerative NTN modes. }
\end{itemize}
\color{black}
\section{Challenges, Open Issues, and Future Research Directions}\label{sec:challenge}

\subsection{Challenges and Open Issues}

\subsubsection{Complexity of network slicing}
The complexities of constrained optimization problems inherent in network slicing for NTN pose significant challenges. Particularly, these optimization problems are characterized by numerous practical constraints, ranging from bandwidth limitations and latency requirements to security considerations. Moreover, the combination of various platforms in different altitudes, e.g., satellites, UAVs, and base stations, further aggravates the problem complexity. Consequently, to ensure the effectiveness of network slicing in NTN, optimization techniques need to be carefully designed to take into account the unique characteristics of each platform and the dynamic nature of the network environment.  
\subsubsection{Challenges of AI-enabled NTN}
AI techniques in NTN face a wide range of challenges due to the distinctive features of non-terrestrial communication environments. First, the dynamic nature of NTN, influenced by atmospheric conditions and connectivity shifts, hinders the effectiveness of conventional AI techniques with limited domain adaptation capability. Moreover, constrained bandwidth further complicates the deployment of AI models, limiting the transmission of large datasets, which in turn reduces the effectiveness of AI models' training. Non-terrestrial platforms also often have limited power and computational capabilities, which complicates the deployment of AI techniques. Additionally, applications demanding high prediction accuracy, such as autonomous driving, introduce challenges in adapting AI models to the inherent propagation delay of NTN. To address these challenges, techniques such as Transfer Learning (TL)~\cite{nguyen2022transfer} can be employed to improve the domain adaption ability of AI models, as well as to address the limited power and computational issues. Moreover, both FL and TL can be utilized to reduce the data transmission demands. 

\subsubsection{Security}
There is a wide range of security issues in NTNs, ranging from data transmission and signal jamming to unauthorized access. Particularly, due to the vast distance of transmission, NTNs are susceptible to interception and eavesdropping. Moreover, the expansive coverage areas of NTN also make them potential targets for unauthorized access. This also brings forth privacy concerns as sensitive data are transmitted over long distances. Additionally, the reliance on satellite communication introduces challenges related to secure satellite operation, such as satellite commands and cyberattacks. To overcome these problems, promising solutions such as AI-empowered intrusion detection systems, blockchain-based authentication, and privacy-preserving homomorphic encryption should be investigated.

\subsection{Future Research Directions}
\subsubsection{Blockchain for NTN}
Potential applications of blockchain in NTN encompass diverse areas, aiming to enhance security, interoperability, efficiency, and privacy.~\color{black}For example, in~\cite{kumar2023distributed}, a blockchain-based approach is developed for secure data sharing in SAGINs. The proposed approach utilizes session-based authentication and public key cryptography for secure communication between satellites, UAVs, and IoT devices. In this system, a blockchain serves as a decentralized platform to manage authentication processes, leveraging smart contracts to automate and securely execute the authentication protocols between entities.\color{black}

Besides automating the authentication processes, smart contracts can be leveraged to orchestrate network slicing, enabling the automatic execution of agreements among network entities, as well as streamlining the process of allocating and managing resources dynamically.~\color{black}Blockchain can also play a pivotal role in stimulating collaboration and resource-sharing through digital tokens and incentive mechanisms~\cite{nguyen2023metashard}.~\color{black}Particularly, participants can be incentivized to contribute resources, share data, or engage in collaborative efforts through token-based rewards, thereby alleviating resource and data demands in NTN applications.
\subsubsection{FL for NTN}
In NTN, FL can enable collaborative model training across various non-terrestrial platforms. This collaborative learning paradigm brings two significant advantages. First, data can be used to train the models locally, and thus reducing the risks of exposing sensitive data over the network.~\color{black}Moreover, this also reduces the need to transmit data, thereby alleviating the transmission demands~\cite{abdulrahman2020survey}.~\color{black}Nevertheless, the integration of FL in NTN involves challenges related to varying connectivity, platform mobility, and heterogeneous datasets. Specifically, different non-terrestrial platforms may experience intermittent or fluctuating network connections, affecting the synchronization and coordination required for FL processes. Moreover, the mobility of these platforms introduces complexities in maintaining consistent and reliable communication channels for collaborative model training.~\color{black}The transmission of training data for FL, especially over large distances, might also be prone to eavesdropping~\cite{hou2023uav}.~\color{black}Addressing these challenges, therefore, is crucial for unlocking the full potential of FL in NTN and requires more efforts from the academia.
\subsubsection{Generative AI for NTN}
Exploring the integration of Generative Artificial Intelligence (GAI) in NTN is a promising future research direction with huge potential.~\color{black}Generative AI, known for its ability to create synthetic data and generate novel content~\cite{nguyen2024generative}, can be utilized to address several challenges in NTN.~\color{black}For instance, it can be employed to generate high-quality synthetic data, complementing the training and testing of traditional AI techniques in the dynamic and diverse NTN environments. Moreover, GAI can enhance the security of NTN by generating realistic attack scenarios, enabling robust evaluation of network defense mechanisms. Another approach is utilizing GAI to generate network data for simulating various network conditions, thereby aiding network optimization and predicting potential bottlenecks.
\color{black}
\subsubsection{Holographic MIMO and Intelligent Reflecting Surfaces}
The recent advancement of antenna technologies has opened promising research directions that can significantly enhance the performance of NTN. For example, holographic MIMO is an advanced antenna technology that leverages principles from holography to enable highly directional and focused transmissions from a single aperture. This technique enables highly directional transmissions while mitigating interference, thereby improving spectrum utilization and throughput for NTN systems~\cite{deng2022holographic}. Additionally, the concept of intelligent reflecting surfaces (IRS) has emerged as a promising enabling technology for NTNs. IRSs comprise arrays of reconfigurable meta-surfaces that can alter the propagation properties of impinging electromagnetic waves~\cite{zhang2023secrecy}. By strategically deploying IRSs in conjunction with satellites or aerial platforms, NTN providers can optimize signal reflections, extend coverage areas, and enhance the overall quality of service. Moreover, another promising direction could be the combination of IRS and holographic MIMO, e.g., deploying IRSs in strategic locations to intelligently reflect and focus the multiple beams generated by a holographic MIMO transmitter towards intended receivers. 
\color{black}

\section{CONCLUSION}\label{sec:Summary}
In this paper, we have presented a comprehensive survey of NTN for 6G networks from both academic and industry perspectives. Particularly, we have presented an in-depth tutorial on NTN and the enabling technologies including network slicing, AI/ML, and ORAN. Then, we have surveyed state-of-the-art network slicing and AI/ML approaches that are proposed to address various challenges of NTN in the literature. Moreover, we have presented how ORAN has been utilized for NTN from the industry standpoint. Finally, we have discussed the current challenges as well as open issues and introduced promising technologies as future research directions of NTNs.

\bibliographystyle{IEEEtran}
\bibliography{Bibtex/intro,Bibtex/tutorial,Bibtex/network_slicing,Bibtex/oran,Bibtex/aiml,Bibtex/challenge}

\begin{IEEEbiography}[{\includegraphics[width=1in,height=1.25in,clip,keepaspectratio]{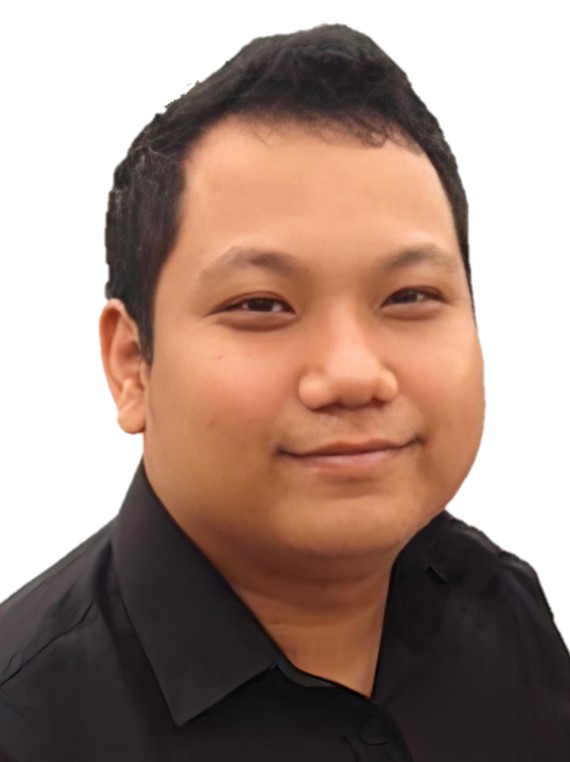}}]{Cong T. Nguyen } is a researcher and lecturer at the Institute of Fundamental and Applied Sciences and the Faculty of Information Technology, Duy Tan University, Vietnam. His research interests include blockchain technology, game theory, optimization, and generative AI. He received his BS degree from Frankfurt University of Applied Sciences, Germany, MS degree from Technical University
Berlin, Germany, and PhD degree from the University of Technology Sydney (UTS), Australia.
\end{IEEEbiography}

\begin{IEEEbiography}[{\includegraphics[width=1in,height=1.25in,clip,keepaspectratio]{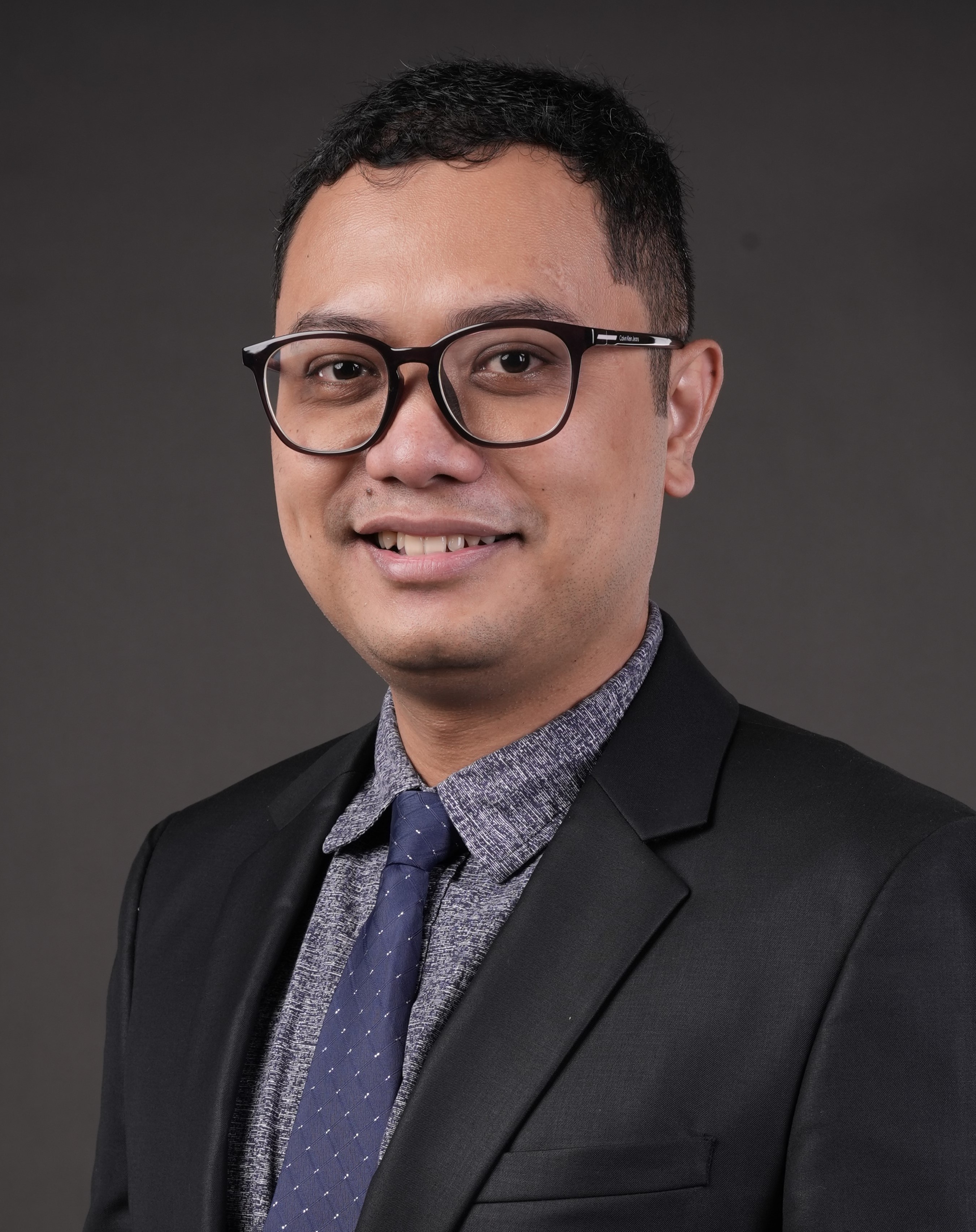}}]{Yuris Mulya Saputra } (Senior Member, IEEE) received the Ph.D degree from the University of Technology Sydney (UTS), Australia in 2022. Currently, he is an Assistant Professor and Deputy Head of the Department of Electrical Engineering and Informatics, Vocational College at the Universitas Gadjah Mada (UGM), Indonesia, and an Adjunct Fellow at the School of Electrical and Data Engineering, UTS. His research interests include mobile edge computing, economic efficiency, IoT, data science, machine learning, and optimization problems for wireless communication networks and applications.
\end{IEEEbiography}

\begin{IEEEbiography}[{\includegraphics[width=1in,height=1.25in,clip,keepaspectratio]{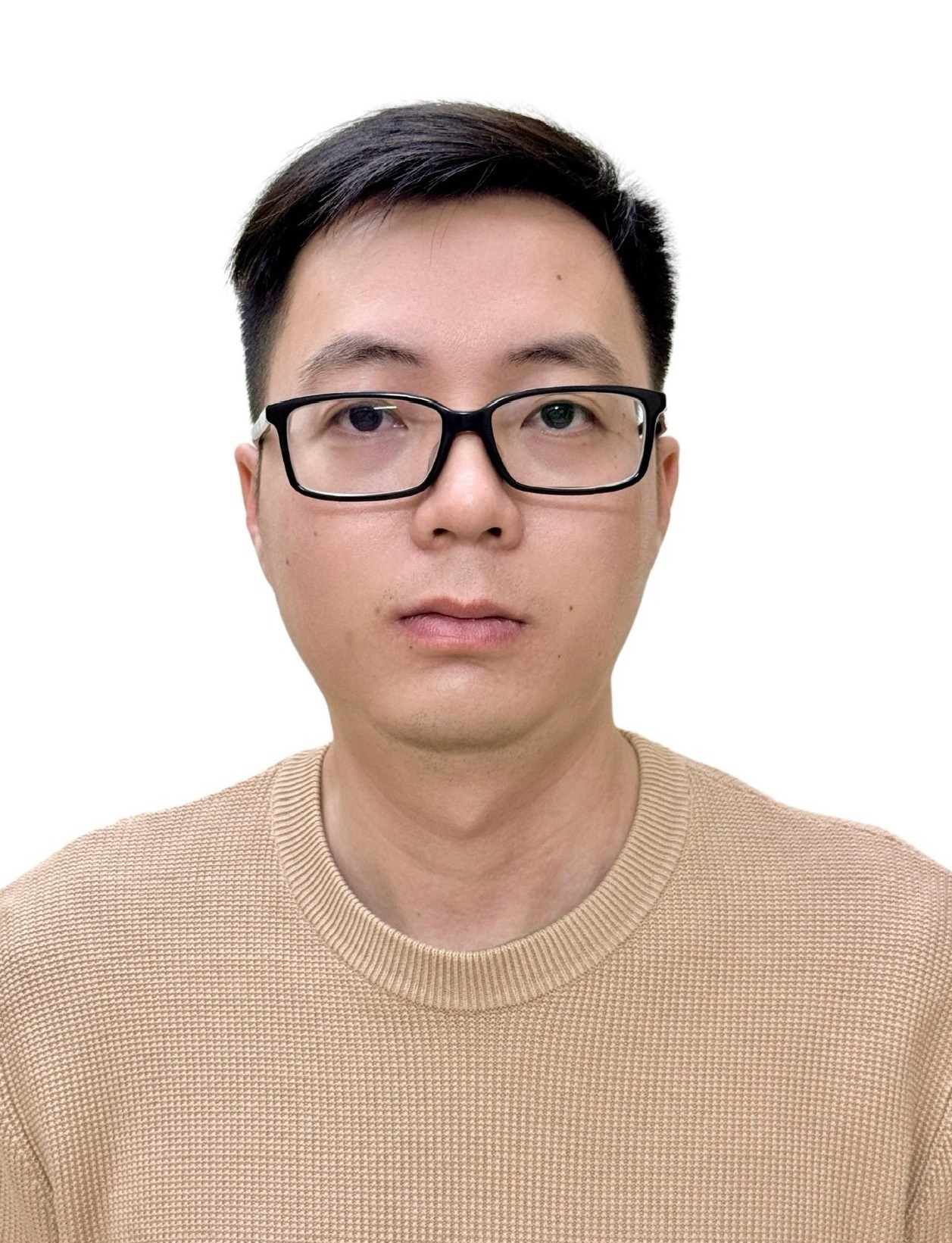}}]{Nguyen Van Huynh } (Member, IEEE) received the Ph.D. Degree in Electrical and Computer Engineering from the University of Technology Sydney (UTS), Australia in 2022. He is currently a Lecturer at the Department of Electrical Engineering and Electronics, University of Liverpool (UoL), United Kingdom. Before joining UoL, he was a Postdoctoral Research Associate in the Department of Electrical and Electronic Engineering, Imperial College London, United Kingdom. His research interests include mobile computing, 5G/6G, IoT, and machine learning.
\end{IEEEbiography}

\begin{IEEEbiography}[{\includegraphics[width=1in,height=1.25in,clip,keepaspectratio]{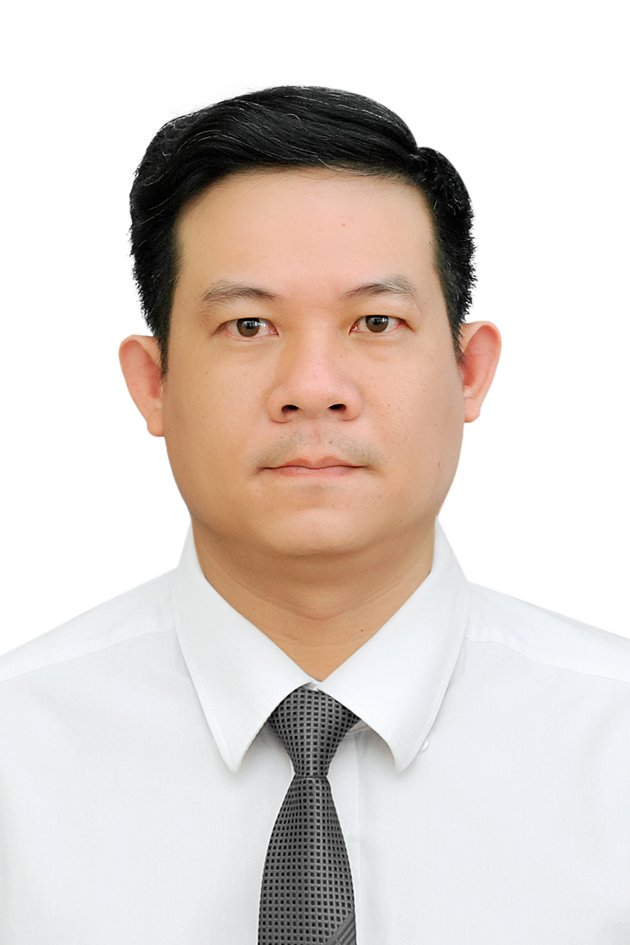}}]{Tan N. Nguyen } (Member, IEEE)  (member IEEE) was born in 1986 in Nha Trang City, Vietnam. He received a BS degree in electronics in 2008 from Ho Chi Minh University of Natural Sciences and an MS degree in telecommunications engineering in 2012 from Vietnam National University. He received a Ph.D. in communications technologies in 2019 from the Faculty of Electrical Engineering and Computer Science at VSB – Technical University of Ostrava, Czech Republic. He joined the Faculty of Electrical and Electronics Engineering of Ton Duc Thang University, Vietnam, in 2013, and since then has been lecturing. He started as the Editor-in-Chief of Advances in Electrical and Electronic Engineering (AEEE) journal in 2023. His major interests are cooperative communications, cognitive radio, signal processing, satellite communication, UAV, and physical layer security.
\end{IEEEbiography}

\begin{IEEEbiography}[{\includegraphics[width=1in,height=1.25in, clip,keepaspectratio]{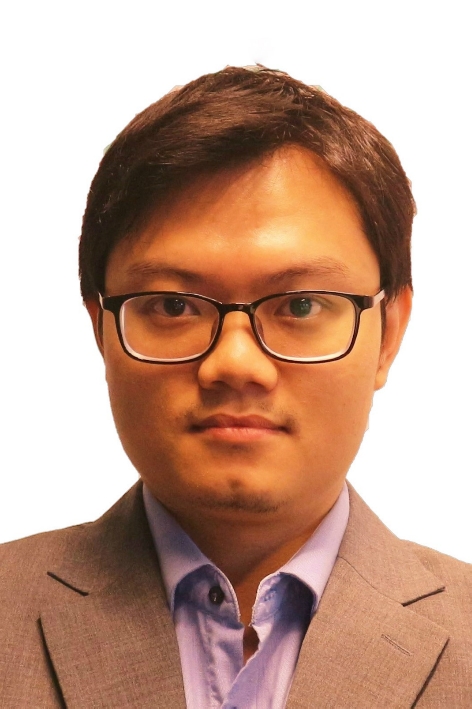}}]{Dinh Thai Hoang } (Senior Member, IEEE) is currently a faculty member at the School of Electrical and Data Engineering, University of Technology Sydney, Australia. He received his Ph.D. in Computer Science and Engineering from the Nanyang Technological University, Singapore 2016. His research interests include emerging wireless communications and networking topics, especially machine learning applications in networking, edge computing, and cybersecurity. He has received several precious awards, including the Australian Research Council Discovery Early Career Researcher Award, IEEE TCSC Award for Excellence in Scalable Computing for Contributions on ``Intelligent Mobile Edge Computing Systems'' (Early Career Researcher), IEEE Asia-Pacific Board (APB) Outstanding Paper Award 2022, and IEEE Communications Society Best Survey Paper Award 2023.  He is currently an Editor of IEEE TMC, IEEE TWC, IEEE TCCN, IEEE TVT, and IEEE COMST.
\end{IEEEbiography}

\begin{IEEEbiography}[{\includegraphics[width=1in,height=1.25in, clip,keepaspectratio]{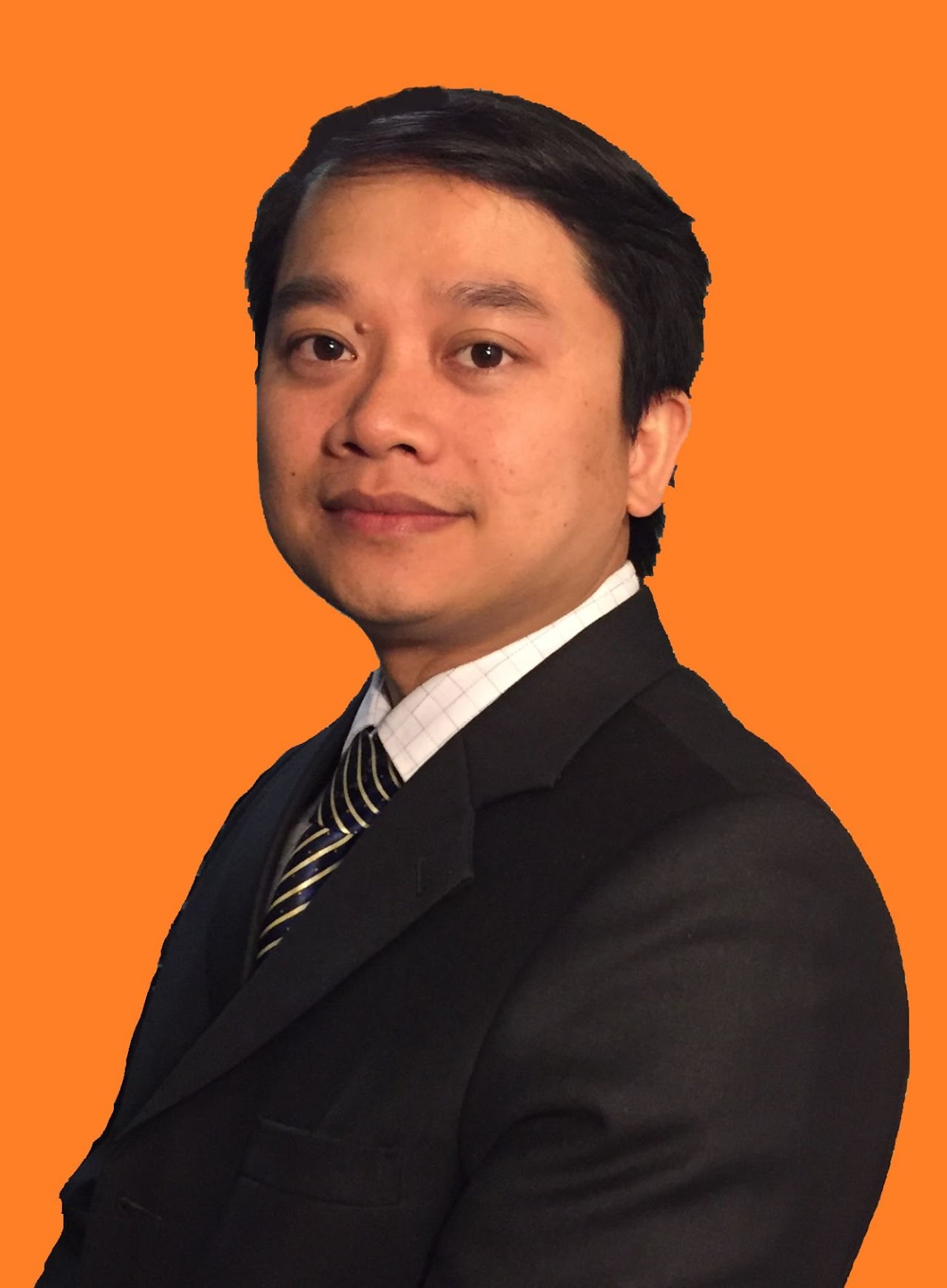}}]{Diep N. Nguyen } (Senior Member, IEEE) is a faculty member of the Faculty of Engineering and Information Technology, University of Technology Sydney (UTS). He received an M.E. and Ph.D. in Electrical and Computer Engineering from the University of California San Diego (UCSD) and The University of Arizona (UA), respectively. Before joining UTS, he was a DECRA Research Fellow at Macquarie University, a member of technical staff at Broadcom (California), ARCON Corporation (Boston), consulting the Federal Administration of Aviation on turning detection of UAVs and aircraft, US Air Force Research Lab on anti-jamming. He has	received several awards from LG Electronics, the University of California, San Diego, The University of Arizona, the US National Science Foundation, Australian Research Council. His recent research interests are in the areas of computer networking, wireless communications, and machine learning applications, with an emphasis on systems' performance and security/privacy. He is an Associate Editor of IEEE Transactions on Mobile Computing, IEEE Access (special issue), and a Senior Member of IEEE.
\end{IEEEbiography}

\begin{IEEEbiography}[{\includegraphics[width=1in,height=1.25in,clip,keepaspectratio]{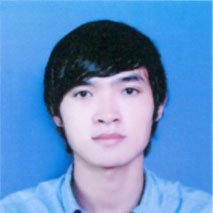}}]{Van-Quan Pham } received a BS degree in Electronic and Telecommunication Engineering from PTIT HCM, Vietnam, in 2013, a Master's Degree in Telecommunication System engineering from the IMT Atlantique, Brest, France, and a Ph.D. degree in Computer Science from the Institut Polytechnique de Paris, France. He is currently a Member of the Technical Staff in the Mobile Networks Systems Department within the Network Systems and Security Research Lab at NOKIA Bell Labs, New Jersey, USA. His research interests include 6G network systems, ORAN, IP/Optical Networking, software-defined networking, Cloud-Native Orchestration, Network as Code, Machine learning operations, E2E slicing architecture, Network Orchestration, and Automation.
\end{IEEEbiography}

\begin{IEEEbiography}[{\includegraphics[width=1in,height=1.25in,clip,keepaspectratio]{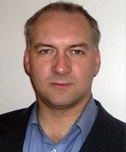}}]{Miroslav Voznak } Miroslav Voznak (M’09-SM’16) received his PhD in telecommunications in 2002 from the Faculty
of Electrical Engineering and Computer Science at VSB – Technical University of Ostrava, and
achieved habilitation in 2009. He was appointed Full Professor in Electronics and Communications Technologies in 2017. His research interests generally focus on ICT, especially on quality of service and
experience, network security, wireless networks, and big data analytics. He has authored and co-authored
over one hundred articles indexed in SCI/SCIE journals. According to the Stanford University study released in 2020, he is one of the World’s Top 2\% of scientists in Networking \& Telecommunications and
Information \& Communications Technologies. He served as a general chair of the 11th IFIP Wireless and Mobile Networking Conference in 2018 and the 24th IEEE/ACM International Symposium on Distributed Simulation and Real Time Applications in 2020. He participated in six projects funded by EU
in programs managed directly by European Commission. Currently, he is a principal investigator in the research project QUANTUM5 funded by NATO, which focuses on the application of quantum cryptography in 5G campus networks.
\end{IEEEbiography}

\begin{IEEEbiography}[{\includegraphics[width=1in,height=1.25in,clip,keepaspectratio]{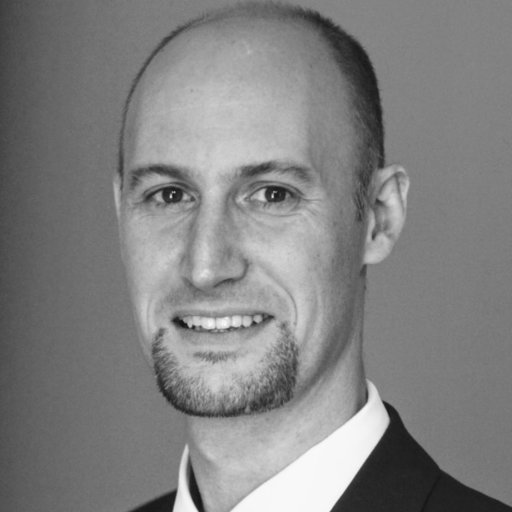}}]
{Symeon Chatzinotas } (MEng, MSc, PhD,
FIEEE) is currently Full Professor / Chief Scientist
I and Head of the research group SIGCOM in the
Interdisciplinary Centre for Security, Reliability
and Trust, University of Luxembourg. In parallel,
he is an Adjunct Professor in the Department of
Electronic Systems, Norwegian University of Science and Technology and a Collaborating Scholar
of the Institute of Informatics $\&$ Telecommunications, National Center for Scientific Research
“Demokritos”. In the past, he has lectured as
Visiting Professor at the University of Parma, Italy and contributed in
numerous R$\&$D projects for the Institute of Telematics and Informatics,
Center of Research and Technology Hellas and Mobile Communications
Research Group, Center of Communication Systems Research, University of Surrey. He has received the M.Eng. in Telecommunications from
Aristotle University of Thessaloniki, Greece and the M.Sc. and Ph.D. in
Electronic Engineering from University of Surrey, UK in 2003, 2006 and
2009 respectively. He has authored more than 700 technical papers in
refereed international journals, conferences and scientific books and has
received numerous awards and recognitions, including the IEEE Fellowship
and an IEEE Distinguished Contributions Award. He is currently in the
editorial board of the IEEE Transactions on Communications, IEEE Open
Journal of Vehicular Technology and the International Journal of Satellite
Communications and Networking.
\end{IEEEbiography}

\begin{IEEEbiography}[{\includegraphics[width=1in,height=1.25in,clip,keepaspectratio]{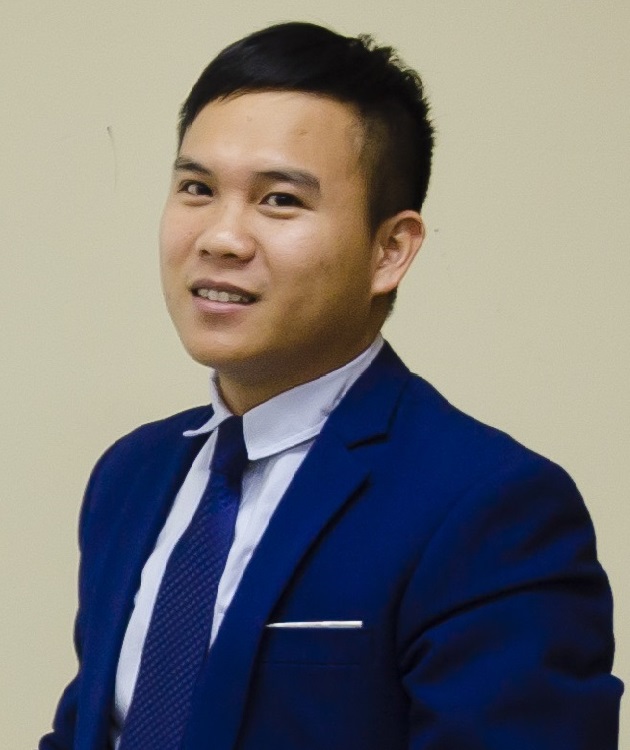}}]{Dinh-Hieu Tran} was born and grew up in Gia Lai, Vietnam. He is currently a senior research specialist 5G+ at Nokia, France. He finished his M.Sc degree in Electronics and Computer Engineering from Hongik University, Korea, in 2017, and the Ph.D. degree at the University of Luxembourg in 2022. His major interests include Non-Terrestrial-Network, Open Radio Access Networks (ORAN), Network Digital Twins, 5G+ for wireless networks. In 2016, he received the Outstanding Hongik Rector Award. He was a co-recipient of the IS3C 2016 best paper award. In 2022, he was nominated for the ``FNR Outstanding Thesis Award" and won the award ``Excellent thesis award 2022 in Doctoral School Science and Engineering".
\end{IEEEbiography}
\end{document}